\documentclass[preprint,prd,aps,showpacs,showkeys,nofootinbib]{revtex4}
\usepackage{amssymb}
\usepackage{amsmath}

\usepackage{graphicx}
\usepackage{subfigure}

\usepackage{float}

\begin{document}


\title{$t\rightarrow c \gamma$ and $t\rightarrow c g$ in Warped Extra Dimensions}

\author{Tie-Jun Gao$^{a,b}$\footnote{email:gao-t-j@foxmail.com}, Tai-Fu Feng$^{a,b}$\footnote{email:fengtf@dlut.edu.cn}, Jian-Bin Chen$^{a,b}$ }

\affiliation{$^a$Department of Physics, Hebei University, Baoding, 071002, China\\
$^b$Department of Physics, Dalian University of Technology, Dalian,
116024, China}

\begin{abstract}

 In this work, we calculate the top quark rare decays $t\rightarrow c\gamma$ and $t\rightarrow cg$  in the
framework where the standard model is embedded in a warped extra
dimension with the custodial symmetry $SU(3)_c\times SU(2)_L\times
SU(2)_R\times U(1)_X\times P_{LR}$. Adopting reasonable assumptions
on the parameter space, we numerically find the branching ratios of
$t\rightarrow c\gamma$ exceeding $10^{-6}$ and that of $t\rightarrow
cg$ exceeding $10^{-5}$ respectively, which can be detected in near
future.

\end{abstract}


 \maketitle

\section{Introduction\label{sec1}}
\indent\indent

Top quark plays a special role in the Standard Model(SM) and holds
great promise in revealing the secret of new physics beyond the SM.
And the study of the rare top quark decays have long been a subject
of intense theoretical and experimental study. Among those rare
processes, the flavor-changing neutral current (FCNC) decays
$t\rightarrow c\gamma$ and $t\rightarrow cg$ deserve special
attention, since the branching ratios(BRs) of those rare processes
are strongly suppressed in the SM.

On the theoretical aspect, the authors of Ref.\cite{ref1} gives a
general expression for the one-loop fermion-neutral boson coupling
keeping all masses and momenta. In the framework of the SM, the BRs
of top quark FCNC
 $t\rightarrow c\gamma$ is of the order $10^{-13}$ and that of $t\rightarrow cg$ is of the order $10^{-11}$
\cite{ref2,ref3}. In  extensions of the SM, the BRs for FCNC top
decays can be orders of magnitude larger. For example in two Higgs
doublet models Br$(t\rightarrow c\gamma)\sim10^{-7} $, Br$(t
\rightarrow cg) \sim10^{-5}$ can be achieved \cite{ref4}, and in
supersymmetric models with R parity conservation  these branching
ratios can reach Br$(t\rightarrow c\gamma)\sim10^{-6} $, Br$(t
\rightarrow cg) \sim10^{-5}$ \cite{ref5,ref6}.
 The authors of  Ref.\cite{ref7} discuss the process $t\rightarrow c\gamma$  in a model with a
single universal extra dimension which is compactified  gauge, and
get the branching fraction Br$(t \rightarrow cg) \sim10^{-10}$. In
Ref.\cite{ref8}, the author computed the BRs for $t\rightarrow
c\gamma$ and $t\rightarrow cg$ in minimal extensions of the SM,
where the additional vector-like up and down quarks singlets break
the unitarity of the $3\times3$ Cabibbo-Kobayashi-Maskawa (CKM)
matrix.

The running LHC is a top-quark factory, and provides a great
opportunity to seek out top-quark rare decays. Given the annual
yield of 80 million $t\bar{t}$ events plus 34 million single-top
events, one may hope to search for rare decays with a branching
fraction as small as $10^{-6}$\cite{ref9}. Then the experimental
observation on LHC will bring a tremendous improvement in our
knowledge of top quark properties.

Models with a warped extra dimension, also called Randall-Sundrum
(RS) models \cite{ref10,ref11}, where all SM fields are allowed to
propagate in the bulk, offer natural solutions to many outstanding
puzzles of contemporary particle physics. In addition to providing a
geometrical solution to the hierarchy problem related to the vast
difference between the Planck scale and the electroweak (EW) scale,
they also allow to naturally generate hierarchies in fermion masses
and weak mixing angles \cite{ref14,ref15}, suppress FCNC
interactions \cite{ref16,ref17,ref18}, construct realistic models of
EW symmetry breaking \cite{ref19,ref20,ref21,ref22,ref23} and
achieve gauge coupling unification \cite{ref24,ref25}.

A necessary condition for direct signals of RS models at the LHC is
the existence of Kaluza-Klein (KK) modes with $\mathcal {O}$(1TeV)
masses. But early studies of EW precision observables
\cite{ref18,ref26} have shown that with the SM gauge group
$SU(2)_L\times U(1)_Y$ in the bulk, the EW precision observables,
for example the experimental data on $S,\;T$ parameters and the
well-measured $Z\bar{b}_Lb_L$
coupling\cite{ref27,ref28,ref29,ref30}, generally require that the
exciting KK modes are heavier than 10 TeV and exceed the reach of
colliders running now. To solve this problem,
literature\cite{ref31,ref32} enlarges the gauge group in the bulk to
$SU(3)_c\times SU(2)_L\times SU(2)_R\times U(1)_X\times P_{LR}$. The
presence of new light KK modes necessary to solve the
$Z\bar{b}_Lb_L$ problem, implies significant contributions to the T
parameter at the one loop level \cite{ref33}. With an appropriate
choice of quark bulk mass parameters, an agreement with the EW
precision data in the presence of light KK modes can be obtained
\cite{ref34,ref35}.Actually, the EW precision observables are
consistent with the light fermion KK modes with masses even below
$1{\rm TeV}$ while the masses of KK gauge bosons are forced to be at
least $2-3{\rm TeV}$ to be consistent with experimental data on the
parameter $S$.

It is well known that all virtual KK excitations contribute their
corrections to theoretical predictions on the physical quantities at
EW scale, and those theoretical corrections should be summed over
infinite KK modes in principle\cite{ref36,ref37,ref38}. In this
paper, we sum over the infinite series of KK modes using the method
in Ref.\cite{ref39}, and analyze the corrections from exciting KK
modes to the top-quark decay $t\rightarrow c\gamma$ and
$t\rightarrow cg$ in the scenario with a warped extra dimension and
the custodial symmetry.

This paper is composed of the sections as follows. In section II, we
present the main ingredients of the SM extension with a warped extra
dimension and the custodial symmetry. And we list some useful
formulas for summing over infinite series of KK modes. In section
III, we present the theoretical calculation on the $t\rightarrow
c\gamma$ and $t\rightarrow cg$ processes. Section IV is devoted to
the numerical analysis and discussion. In section V, we deliver our
conclusions. The relevant nontrivial couplings approached to the
order ${\cal O}(\mu_{EW}^2/\Lambda_{KK}^2)$ are in appendix, where
$\Lambda_{KK}$ denotes the energy scale of low-lying KK excitations
and $\mu_{EW} $ denotes the EW energy scale.

\section{A warped extra dimension with custodial protection and summing over infinite series of KK modes \label{sec2}}
\indent\indent

Models with extra dimensions have attracted great attention in
recent years as they offer new perspectives on challenging problems
in modern physics. It was demonstrated by Randall and Sundrum that a
small but warped extra dimension provides an elegant solution to the
gauge hierarchy problem \cite{ref10,ref40}.
 In the RS scenario, four dimensional (4D)
Minkowskian space-time is embedded into a slice of five dimensional
(5D) anti de-Sitter (ADS$_5$) space with curvature $k$. The fifth
dimension is a $S^1/Z_2\times Z_2^\prime$ orbifold of size $r$
labeled by a coordinate $\phi\in[-\pi,\pi]$, such that the points
$(x^\mu,\phi)$, $(x^\mu,\pi-\phi)$, $(x^\mu,\pi+\phi)$ and
$(x^\mu,-\phi)$ are identified all. The corresponding metric of the
non-factorizable RS geometry is given by
\begin{eqnarray}
&&ds^2=e^{-2\sigma(\phi)}\eta_{\mu\nu}dx^\mu dx^\nu-r^2d\phi^2,
\;\sigma(\phi)=kr|\phi|,
\label{metric}
\end{eqnarray}
where $x^\mu\;(\mu=0,\;1,\;2,\;3)$ denote the coordinates on the 4D
hyper-surfaces of constant $\phi$ with metric
$\eta_{\mu\nu}=(1,-1,-1,-1)$, and $e^\sigma$ is called the warp
factor. Two branes are located on the orbifold fixed points $\phi=0$
and $\phi=\pi/2$, respectively. The brane on $\phi=0$ is called
Planck or ultra-violet (UV) brane, and the brane on $\phi=\pi/2$ is
called TeV or infra-red (IR) brane. The parameters $k$ and $1/r$ are
assumed to be of order the fundamental Planck scale $M_{\rm Pl}$ and
choosing the product $kr\simeq24$, one gets the inverse warp factor
\begin{eqnarray}
&&\epsilon={\Lambda_{\rm IR}\over\Lambda_{\rm UV}}\equiv
e^{-kr\pi/2}\simeq10^{-16},
\label{warped-factor}
\end{eqnarray}
which explains the hierarchy between the EW and Planck scale
naturally. Meanwhile, the warp factor also sets the mass scale for
the  low-lying KK excitations
\begin{eqnarray}
&&\Lambda_{_{KK}}\equiv k\epsilon=ke^{-kr\pi/2}={\cal O}(1{\rm
TeV}).
\label{KK-scale}
\end{eqnarray}
 In the EW sector, we consider an $SU(2)_L\times
SU(2)_R\times U(1)_X\times P_{LR}$ gauge symmetry on a slice of
AdS$_5$, where $P_{LR}$ is the discrete symmetry interchanging the
two SU(2) groups. This means for instance that $g_{5L} = g_{5R}
\equiv g_5$. The gauge group is broken by boundary conditions (BCs)
on the UV brane  to the SM gauge group, i. e.

\begin{eqnarray}
&&SU(2)_L\times SU(2)_R\times U(1)_X\times P_{LR}\xrightarrow[]{UV
brane}
 SU(2)_L\times U(1)_Y.
\label{1}
\end{eqnarray}

This breakdown is achieved by the following assignment of BCs

\begin{eqnarray}
&&W_{L,\mu}^{1,2,3}(++),\;B_\mu(++),\;W_{R,\mu}^{1,2}(-+),\;Z_{X,\mu}(-+),\;\;
(\mu=0,\;1,\;2,\;3),
\nonumber\\
&&W_{L,5}^{1,2,3}(--),\;B_5(--),\;W_{R,5}^{1,2}(+-),\;Z_{X,5}(+-).
\label{BCs-gauge}
\end{eqnarray}
where the first (second) sign is the BC on the UV (IR) brane: $+$
stands for a Neumann BC and $-$ stands for a Dirichlet BC.The third
component of $SU(2)_R$ gauge fields $W_{R,M}^3$ and the $U(1)_X$
gauge field $\tilde{B}_M$ are expressed in terms of the neutral
gauge fields $Z_{X,M}$ and $B_M$ as
\begin{eqnarray}
&&W_{R,M}^3={g_5Z_{X,M}+g_{5X}B_M\over\sqrt{g_5^2+g_{5X}^2}},\;
\tilde{B}_M=-{g_{5X}Z_{X,M}-g_5B_M\over\sqrt{g_5^2+g_{5X}^2}},\;(M=0,\;1,\;2,\;3,\;5),
\label{gauge1}
\end{eqnarray}
where $g_{5X}$ is the 5D gauge coupling of $U(1)_X$.

To further proceed it will be useful to follow \cite{ref41} and
define the fields
\begin{eqnarray}
&&A_A={\sqrt{g_5^2+g_{5X}^2}B_A+g_{5X}W_{L,A}^3\over\sqrt{g_5^2+2g_{5X}^2}},
\nonumber\\
&&Z_{A}={-g_{5X}B_A+\sqrt{g_5^2+g_{5X}^2}W_{L,A}^3\over\sqrt{g_5^2+2g_{5X}^2}},
\nonumber\\
&&W_{L,A}^{\pm}={1\over\sqrt{2}}\Big(W_{L,A}^1\mp iW_{L,A}^2\Big)\;,
\nonumber\\
&&W_{R,A}^{\pm}={1\over\sqrt{2}}\Big(W_{R,A}^1\mp iW_{R,A}^2\Big)\;.
\label{gauge2}
\end{eqnarray}

In order to break down the electroweak symmetry, a Higgs boson is
introduced that is localised either on or near the IR brane,
transforming as a self-dual bidoublet of $SU(2)_L\times SU(2)_R$,
and transforms as a singlet with charge $Y_H=0$ under the gauge
group $U(1)_X$:
\begin{eqnarray}
&&H=\left(\begin{array}{cc}-i\pi^+/\sqrt{2}&-(h^0-i\pi^0)/2\\
(h^0+i\pi^0)/2&i\pi^-/\sqrt{2}
\end{array}\right)\;.
\label{Higgs-bidoublet}
\end{eqnarray}

As regards the matter fields, the quarks of one generation are
embedded into the multiplets\cite{ref39,ref42}:
\begin{eqnarray}
&&Q_L^i=\left(\begin{array}{ll}\chi_{u_L}^i(-+)_{5/3}&q_{u_L}^i(++)_{2/3}\\
\chi_{d_L}^i(-+)_{2/3}&q_{d_L}^i(++)_{-1/3}\end{array}\right)\;,\;\;
Q_{u_R}^i=u_R^i(++)_{2/3}
\nonumber\\
&&\tilde{Q}_{d_R}^i=\left(\begin{array}{l}X_R^i(-+)_{5/3}\\U_R^i(-+)_{2/3}
\\D_R^i(-+)_{-1/3}\end{array}\right)\;,\;\;
Q_{d_R}^i=\left(\begin{array}{l}\tilde{X}_R^i(-+)_{5/3}\\
\tilde{U}_R^i(-+)_{2/3}
\\d_R^i(++)_{-1/3}\end{array}\right)\;,
\label{SM-quarks}
\end{eqnarray}
and the corresponding states of opposite chirality are given by
\begin{eqnarray}
&&Q_R^i=\left(\begin{array}{ll}\chi_{u_R}^i(+-)_{5/3}&q_{u_R}^i(--)_{2/3}\\
\chi_{d_R}^i(+-)_{2/3}&q_{d_R}^i(--)_{-1/3}\end{array}\right)\;,\;\;
Q_{u_L}^i=u_L^i(--)_{2/3}
\nonumber\\
&&\tilde{Q}_{d_L}^i=\left(\begin{array}{l}X_L^i(+-)_{5/3}\\U_L^i(+-)_{2/3}
\\D_L^i(+-)_{-1/3}\end{array}\right)\;,\;\;
Q_{d_L}^i=\left(\begin{array}{l}\tilde{X}_L^i(+-)_{5/3}\\
\tilde{U}_L^i(+-)_{2/3}
\\d_L^i(--)_{-1/3}\end{array}\right)\;.
\label{opposite-quarks}
\end{eqnarray}
Here, $i=1,\;2,\;3$ denotes the index of generation, the $U(1)_X$
charges are all assigned as
\begin{eqnarray}
&&Y_{Q^i}=Y_{u^i}=Y_{Q_d^i}={2\over3}\;.
\label{U1_charges}
\end{eqnarray}

In order to give the kinetic terms of triplets, we redefine the
quarks in triplet as
\begin{eqnarray}
&&\tilde{T}_{Q_R}^i=\left(\begin{array}{c}{1\over\sqrt{2}}(X_R^i+D_R^i)
\\{i\over\sqrt{2}}(X_R^i-D_R^i)\\U_R^i\end{array}\right)\;,\;\;
T_{Q_R}^i=\left(\begin{array}{c}{1\over\sqrt{2}}(\tilde{X}_R^i+d_R^i)
\\{i\over\sqrt{2}}(\tilde{X}_R^i-d_R^i)\\ \tilde{U}_R^i\end{array}\right)\;,
\nonumber\\
&&\tilde{T}_{Q_L}^i=\left(\begin{array}{c}{1\over\sqrt{2}}(X_L^i+D_L^i)
\\{i\over\sqrt{2}}(X_L^i-D_L^i)\\U_L^i\end{array}\right)\;,\;\;
T_{Q_L}^i=\left(\begin{array}{c}{1\over\sqrt{2}}(\tilde{X}_L^i+d_L^i)
\\{i\over\sqrt{2}}(\tilde{X}_L^i-d_L^i)\\ \tilde{U}_L^i\end{array}\right)\;.
\label{redefine-triplet}
\end{eqnarray}

 the Lagrangian we will use is written as

\begin{eqnarray}
&&{\cal L}={\cal L}_{gauge}+{\cal L}_H+{\cal L}_Q+{\cal L}_Y^Q.
\label{L_gauge}
\end{eqnarray}

where
\begin{eqnarray}
&&{\cal L}_{gauge}={\sqrt{\cal G}\over r}{\cal G}^{KM}{\cal
G}^{LN}\Big(-{1\over4}W_{KL}^iW_{MN}^i
-{1\over4}\tilde{W}_{KL}^i\tilde{W}_{MN}^i-{1\over4}\tilde{B}_{KL}\tilde{B}_{MN}
-{1\over4}G_{KL}^aG_{MN}^a\Big)\;.
\label{L_gauge}
\end{eqnarray}
is the Lagrangian for gauge sector, and the corresponding Lagrangian
for Higgs field is written as

\begin{eqnarray}
&&{\cal L}_{H}={\rm Tr}\Big[\Big(D_\mu
\Phi(x)\Big)^\dagger\Big(D^\mu\Phi(x)\Big)\Big] -\mu^2{\rm
Tr}\Big(\Phi^\dagger(x)\Phi(x)\Big) +{\lambda\over2}\Big[{\rm
Tr}\Big(\Phi^\dagger(x)\Phi(x)\Big)\Big]^2\;
\label{IR-Brane-Higgs0}
\end{eqnarray}

with

\begin{eqnarray}
&&D_MH=\partial_MH+\frac{i}{2}g_5(\sum^3_{a=1}W^a_{L,M}\sigma^a)H+\frac{i}{2}g_5H(\sum^3_{a=1}W^a_{R,M}\sigma^a)^T.
\label{IR-Brane-Higgs0}
\end{eqnarray}

The Lagrangian for kinetic terms of quarks can be written as
\begin{eqnarray}
&&{\cal L}_{Q} ={\sqrt{\cal G}\over2r}\sum\limits_{i=1}^3
\Big\{(\overline{Q}^i)_{a_1a_2}iE_A^M\gamma^A\Big[\Big({1\over2}(\partial_M
-\overleftarrow{\partial}_M)+ig_{5s}T^aG^a_M+ig_{5X}Y_{Q^i}\tilde{B}_M\Big)\delta_{a_1b_1}
\delta_{a_2b_2}
\nonumber\\
&&\hspace{1.2cm}
+ig_5({\sigma^{c_1}\over2})_{a_1b_1}W_{L,M}^{c_1}\delta_{a_2b_2}
+ig_5({\sigma^{c_2}\over2})_{a_2b_2}W_{R,M}^{c_2}\delta_{a_1b_1}\Big](Q^i)_{b_1b_2}
\nonumber\\
&&\hspace{1.2cm}
+(\overline{Q}^i)_{a_1a_2}\Big[iE_A^M\gamma^A\omega_M-{\rm
sgn}(\phi)k(c_{_B})_{ij}\Big] (Q^j)_{a_1a_2}
\nonumber\\
&&\hspace{1.2cm}
+\overline{u}^i\Big[iE_A^M\gamma^A\Big({1\over2}(\partial_M
-\overleftarrow{\partial}_M)+ig_{5s}T^aG^a_M+ig_{5X}Y_{u^i}\tilde{B}_M\Big)\delta_{ij}
\nonumber\\
&&\hspace{1.2cm} +iE_A^M\gamma^A\omega_M-{\rm
sgn}(\phi)k(c_{_S})_{ij}\Big]u^j
\nonumber\\
&&\hspace{1.2cm}
+(\overline{\tilde{T}}_Q^i)_{a1}iE_A^M\gamma^A\Big[\Big({1\over2}(\partial_M
-\overleftarrow{\partial}_M)+ig_{5s}T^aG^a_M+ig_{5X}Y_{Q_d^i}\tilde{B}_M\Big)
\delta_{a_1b_1}
\nonumber\\
&&\hspace{1.2cm}
+g_5\varepsilon_{a_1b_1c_1}W_{L,M}^{c_1}\Big](\tilde{T}_Q^i)_{b_1}
+(\overline{\tilde{T}}_Q^i)_{a_1}\Big[iE_A^M\gamma^A\omega_M-{\rm
sgn}(\phi)(\eta_3)_{ij}\Big] (\tilde{T}_Q^j)_{a_1}
\nonumber\\
&&\hspace{1.2cm}
+(\overline{T}_Q^i)_{a_1}iE_A^M\gamma^A\Big[\Big({1\over2}(\partial_M
-\overleftarrow{\partial}_M)+ig_{5s}T^aG^a_M+ig_{5X}Y_{Q_d^i}\tilde{B}_M\Big)
\delta_{a_1b_1}
\nonumber\\
&&\hspace{1.2cm}
+g_5\varepsilon_{a_1b_1c_1}W_{R,M}^{c_1}\Big](T_Q^i)_{b_1}
+(\overline{T}_Q^i)_{a_1}\Big[iE_A^M\gamma^A\omega_M-{\rm
sgn}(\phi)k(c_{_T})_{ij} \Big](T_Q^j)_{a_1}+h.c.\Big\}\;,
\label{kinetic_quark}
\end{eqnarray}

 with
$\gamma^A=(\gamma^\mu,\;-i\gamma^5)$, the inverse vielbein
$E_B^A={\it
diag}(e^{\sigma(\phi)},\;e^{\sigma(\phi)},\;e^{\sigma(\phi)},\;
e^{\sigma(\phi)},{1\over r})$, and the spin connection
$\omega_A=({\rm
sgn}(\phi){i\over2}ke^{-\sigma(\phi)}\gamma_\mu\gamma^5,0)$.
Generally, three bulk mass matrices $c_{_B},\;c_{_S},\;c_{_T}$ are
arbitrarily hermitian $3\times3$ matrices. Furthermore, we also
assume that $c_{_B},\;c_{_S},\;c_{_T}$ are real and diagonal, i.e.
each of them is described by three real parameters. This can always
be obtained through some appropriate field redefinitions.

At last,

\begin{eqnarray}
&&{\cal L}_{Y}^Q= e^{kr\pi/2}\sqrt{-{\cal
G}_{_{IR}}}\sum\limits_{i,j=1}^3\Big\{\sqrt{2}
\lambda_{ij}^{u}\overline{Q}_{a\alpha}^iH_{a\alpha}u^j-2\lambda_{ij}^{d}\Big[
\overline{Q}_{a\alpha}^i(\tau^c)_{ab}(\tilde{T}_d^j)_cH_{b\alpha}
\nonumber\\
&&\hspace{1.2cm}
+\overline{Q}_{a\alpha}^i(\tau^c)_{\alpha\beta}(T_d^j)_cH_{a\beta}\Big]+h.c.\Big\}\;,
\label{Yukawa-quarks}
\end{eqnarray}
is the Lagrangian for Yukawa couplings between quarks and Higgs
field. Here the metric on IR brane ${\cal
G}_{_{IR}}^{\mu\nu}=e^{kr\pi/2}\eta^{\mu\nu}$.

For convenience in our analysis, we define the gauge couplings in 4D
which are related to the 5D gauge couplings via
\begin{eqnarray}
&&g={g_5\over\sqrt{2\pi r}}\;,\nonumber\\
&&g_X={g_{5X}\over\sqrt{2\pi r}}\;.
\label{4D-couplings1}
\end{eqnarray}
Correspondingly, the constant of electromagnetic coupling and
Weinberg angle in 4D are given through
\begin{eqnarray}
&&e={gg_X\over\sqrt{g^2+2g_X^2}}\;,\nonumber\\
&&\sin\theta_{_{\rm W}}={g_X\over\sqrt{g^2+2g_X^2}}\;.
\label{4D-Weinberg-angle}
\end{eqnarray}
In terms of the Weinberg angle $\theta_{_{\rm W}}$ and the constant
of electromagnetic coupling $e$, the gauge couplings in
Eq.(\ref{4D-couplings1}) are written as
\begin{eqnarray}
&&g={e\over\sin\theta_{_{\rm W}}}\;,\nonumber\\
&&g_X={e\over\sqrt{1-2\sin^2\theta_{_{\rm W}}}}\;.
\label{4D-couplings2}
\end{eqnarray}
To discuss the phenomenology at EW scale, we write the KK
decompositions of 5D gauge fields in our notations as
\begin{eqnarray}
&&A_\mu(x,\phi)={1\over\sqrt{r}}\sum\limits_{n=0}^\infty
A_\mu^{(n)}(x) \chi_{_{(++)}}^A(y_{_{(++)}}^{A(n)},t),
\nonumber\\
&&Z_\mu(x,\phi)={1\over\sqrt{r}}\sum\limits_{n=0}^\infty
Z_\mu^{(n)}(x) \chi_{_{(++)}}^Z(y_{_{(++)}}^{Z(n)},t),
\nonumber\\
&&Z_{X,\mu}(x,\phi)={1\over\sqrt{r}}\sum\limits_{n=1}^\infty
Z_{X,\mu}^{(n)}(x) \chi_{_{(-+)}}^{Z_X}(y_{_{(-+)}}^{Z_X(n)},t),
\nonumber\\
&&W_{L,\mu}^\pm(x,\phi)={1\over\sqrt{r}}\sum\limits_{n=0}^\infty
W_{L,\mu}^{\pm(n)}(x) \chi_{_{(++)}}^{W_L}(y_{_{(++)}}^{W_L(n)},t),
\nonumber\\
&&W_{R,\mu}^\pm(x,\phi)={1\over\sqrt{r}}\sum\limits_{n=1}^\infty
W_{R,\mu}^{\pm(n)}(x) \chi_{_{(-+)}}^{W_R}(y_{_{(-+)}}^{W_R(n)},t),
\nonumber\\
&&G_\mu^a(x,\phi)={1\over\sqrt{r}}\sum\limits_{n=0}^\infty
G_\mu^{a,(n)}(x) \chi_{_{(++)}}^g(y_{_{(++)}}^{g(n)},t).
\label{KK-decomposition-Gauge-Higgs}
\end{eqnarray}

Where $y_{_{(++)}}^{G(n)}\;(n=0,\;1,\;\cdots,\;\infty)$,
$G=A,\;Z,\;W_L^\pm,\;g$ denote the roots of equation
$z^2R_{_{(++)}}^{G,\epsilon}(z)\equiv0$ with
\begin{eqnarray}
&&R_{_{(++)}}^{G,\epsilon}(z)=Y_0(z)J_0(z\epsilon)-J_0(z)Y_0(z\epsilon)\;.
\label{root1}
\end{eqnarray}
and $y_{_{(-+)}}^{G(n)}\;(n=1,\;2,\;\cdots,\;\infty)$,
$G=Z_{_X},\;W_R^\pm$ denote the roots of equation
$R_{_{(-+)}}^{G,\epsilon}(z)\equiv0$ with
\begin{eqnarray}
R_{_{(-+)}}^{G,\epsilon}(z)=Y_0(z)J_1(z\epsilon)-J_0(z)Y_1(z\epsilon)\;.
\label{root2}
\end{eqnarray}

Similarly, the KK decompositions of 5D quark fields are written as
\begin{eqnarray}
&&\chi_{u_L}^i(x,\phi)={e^{2\sigma(\phi)}\over\sqrt{r}}\sum\limits_n
\chi_{u_L}^{i,(n)}(x)f_{_{(-+)}}^{L,c_B^i}(y_{_{(\mp\pm)}}^{c_B^i(n)},t),
\;\chi_{d_L}^i(x,\phi)={e^{2\sigma(\phi)}\over\sqrt{r}}\sum\limits_n
\chi_{d_L}^{i,(n)}(x)f_{_{(-+)}}^{L,c_B^i}(y_{_{(\mp\pm)}}^{c_B^i(n)},t),
\nonumber\\
&&q_{u_L}^i(x,\phi)={e^{2\sigma(\phi)}\over\sqrt{r}}\sum\limits_n
q_{u_L}^{i,(n)}(x)f_{_{(++)}}^{L,c_B^i}(y_{_{(\pm\pm)}}^{c_B^i(n)},t),
\;q_{d_L}^i(x,\phi)={e^{2\sigma(\phi)}\over\sqrt{r}}\sum\limits_n
q_{d_L}^{i,(n)}(x)f_{_{(++)}}^{L,c_B^i}(y_{_{(\pm\pm)}}^{c_B^i(n)},t),
\nonumber\\
&&u_R^i(x,\phi)={e^{2\sigma(\phi)}\over\sqrt{r}}\sum\limits_n
u_R^{i,(n)}(x)f_{_{(++)}}^{R,c_S^i}(y_{_{(\mp\mp)}}^{c_S^i(n)},t),
\;X_R^i(x,\phi)={e^{2\sigma(\phi)}\over\sqrt{r}}\sum\limits_n
X_R^{i,(n)}(x)f_{_{(-+)}}^{R,c_T^i}(y_{_{(\pm\mp)}}^{c_T^i(n)},t),
\nonumber\\
&&U_R^i(x,\phi)={e^{2\sigma(\phi)}\over\sqrt{r}}\sum\limits_n
U_R^{i,(n)}(x)f_{_{(-+)}}^{R,c_T^i}(y_{_{(\pm\mp)}}^{c_T^i(n)},t),
\;D_R^i(x,\phi)={e^{2\sigma(\phi)}\over\sqrt{r}}\sum\limits_n
D_R^{i,(n)}(x)f_{_{(-+)}}^{R,c_T^i}(y_{_{(\pm\mp)}}^{c_T^i(n)},t),
\nonumber\\
&&\tilde{X}_R^i(x,\phi)={e^{2\sigma(\phi)}\over\sqrt{r}}\sum\limits_n
\tilde{X}_R^{i,(n)}(x)f_{_{(-+)}}^{R,c_T^i}(y_{_{(\pm\mp)}}^{c_T^i(n)},t),
\;\tilde{U}_R^i(x,\phi)={e^{2\sigma(\phi)}\over\sqrt{r}}\sum\limits_n
\tilde{U}_R^{i,(n)}(x)f_{_{(-+)}}^{R,c_T^i}(y_{_{(\pm\mp)}}^{c_T^i(n)},t),
\nonumber\\
&&d_R^i(x,\phi)={e^{2\sigma(\phi)}\over\sqrt{r}}\sum\limits_n
d_R^{i,(n)}(x)f_{_{(++)}}^{R,c_T^i}(y_{_{(\mp\mp)}}^{c_T^i(n)},t),
\;X_L^i(x,\phi)={e^{2\sigma(\phi)}\over\sqrt{r}}\sum\limits_n
X_L^{i,(n)}(x)f_{_{(+-)}}^{L,c_T^i}(y_{_{(\pm\mp)}}^{c_T^i(n)},t),
\nonumber\\
&&U_L^i(x,\phi)={e^{2\sigma(\phi)}\over\sqrt{r}}\sum\limits_n
U_L^{i,(n)}(x)f_{_{(+-)}}^{L,c_T^i}(y_{_{(\pm\mp)}}^{c_T^i(n)},t),
\;D_L^i(x,\phi)={e^{2\sigma(\phi)}\over\sqrt{r}}\sum\limits_n
D_L^{i,(n)}(x)f_{_{(+-)}}^{L,c_T^i}(y_{_{(\pm\mp)}}^{c_T^i(n)},t),
\nonumber\\
&&\tilde{X}_L^i(x,\phi)={e^{2\sigma(\phi)}\over\sqrt{r}}\sum\limits_n
\tilde{X}_L^{i,(n)}(x)f_{_{(+-)}}^{L,c_T^i}(y_{_{(\pm\mp)}}^{c_T^i(n)},t),
\;\tilde{U}_L^i(x,\phi)={e^{2\sigma(\phi)}\over\sqrt{r}}\sum\limits_n
\tilde{U}_L^{i,(n)}(x)f_{_{(+-)}}^{L,c_T^i}(y_{_{(\pm\mp)}}^{c_T^i(n)},t),
\nonumber\\
&&d_L^i(x,\phi)={e^{2\sigma(\phi)}\over\sqrt{r}}\sum\limits_n
d_L^{i,(n)}(x)f_{_{(--)}}^{L,c_T^i}(y_{_{(\mp\mp)}}^{c_T^i(n)},t),
\;u_L^i(x,\phi)={e^{2\sigma(\phi)}\over\sqrt{r}}\sum\limits_n
u_L^{i,(n)}(x)f_{_{(--)}}^{L,c_S^i}(y_{_{(\mp\mp)}}^{c_S^i(n)},t),
\nonumber\\
&&\chi_{u_R}^i(x,\phi)={e^{2\sigma(\phi)}\over\sqrt{r}}\sum\limits_n
\chi_{u_R}^{i,(n)}(x)f_{_{(+-)}}^{R,c_B^i}(y_{_{(\mp\pm)}}^{c_B^i(n)},t),
\;\chi_{d_R}^i(x,\phi)={e^{2\sigma(\phi)}\over\sqrt{r}}\sum\limits_n
\chi_{d_R}^{i,(n)}(x)f_{_{(+-)}}^{R,c_B^i}(y_{_{(\mp\pm)}}^{c_B^i(n)},t),
\nonumber\\
&&q_{u_R}^i(x,\phi)={e^{2\sigma(\phi)}\over\sqrt{r}}\sum\limits_n
q_{u_R}^{i,(n)}(x)f_{_{(--)}}^{R,c_B^i}(y_{_{(\pm\pm)}}^{c_B^i(n)},t),
\;q_{d_R}^i(x,\phi)={e^{2\sigma(\phi)}\over\sqrt{r}}\sum\limits_n
q_{d_R}^{i,(n)}(x)f_{_{(--)}}^{R,c_B^i}(y_{_{(\pm\pm)}}^{c_B^i(n)},t).\nonumber\\
\label{KK-decomposition-quark}
\end{eqnarray}

In Eq.(\ref{KK-decomposition-quark}), the eigenvalues
$y_{_{(\pm\pm)}}^{c(n)}\;(n\ge1)$ satisfy the equation
$R_{_{(\pm\pm)}}^{c,\epsilon}(z)\equiv0$,
$y_{_{(\mp\mp)}}^{c(n)}\;(n\ge1)$ satisfy the equation
$R_{_{(\mp\mp)}}^{c,\epsilon}(z)\equiv0$,
$y_{_{(\pm\mp)}}^{c(n)}\;(n\ge1)$ satisfy the equation
$R_{_{(\pm\mp)}}^{c,\epsilon}(z)\equiv0$, and the eigenvalues
$y_{_{(\mp\pm)}}^{c(n)}\;(n\ge1)$ satisfy the equation
$R_{_{(\mp\pm)}}^{c,\epsilon}(z)\equiv0$, respectively. Here, the
concrete expressions of
$R_{_{(\pm\pm)}}^{c,\epsilon}(z),\;R_{_{(\pm\mp)}}^{c,\epsilon}(z)
\;,R_{_{(\mp\pm)}}^{c,\epsilon}(z),\;R_{_{(\mp\mp)}}^{c,\epsilon}(z)$
are
\begin{eqnarray}
&&R_{_{(\pm\pm)}}^{c,\epsilon}(z)=\left\{\begin{array}{ll}Y_{N}(z)J_{N}(z\epsilon)-J_{N}(z)Y_{N}(z\epsilon),
&c=N+{1\over2}\\
J_{-c+{1\over2}}(z)J_{c-{1\over2}}(z\epsilon)-J_{c-{1\over2}}(z)J_{-c+{1\over2}}(z\epsilon),
&c\neq N+{1\over2}\end{array}\right.\;,
\nonumber\\
&&R_{_{(\pm\mp)}}^{c,\epsilon}(z)=\left\{\begin{array}{ll}J_{N+1}(z)Y_{N}(z\epsilon)-Y_{N+1}(z)J_{N}(z\epsilon),
&c=N+{1\over2}\\
J_{c+{1\over2}}(z)J_{-c+{1\over2}}(z\epsilon)+J_{-c-{1\over2}}(z)J_{c-{1\over2}}(z\epsilon),
&c\neq N+{1\over2}\end{array}\right.\;,
\nonumber\\
&&R_{_{(\mp\pm)}}^{c,\epsilon}(z)=\left\{\begin{array}{ll}Y_{N}(z)J_{N+1}(z\epsilon)-J_{N}(z)Y_{N+1}(z\epsilon),
&c=N+{1\over2}\\
J_{-c+{1\over2}}(z)J_{c+{1\over2}}(z\epsilon)+J_{c-{1\over2}}(z)J_{-c-{1\over2}}(z\epsilon),
&c\neq N+{1\over2}\end{array}\right.\;,
\nonumber\\
&&R_{_{(\mp\mp)}}^{c,\epsilon}(z)=\left\{\begin{array}{ll}J_{N+1}(z)Y_{N+1}(z\epsilon)
-Y_{N+1}(z)J_{N+1}(z\epsilon),&c=N+{1\over2}\\
J_{c+{1\over2}}(z)J_{-c-{1\over2}}(z\epsilon)-J_{-c-{1\over2}}(z)J_{c+{1\over2}}(z\epsilon),
&c\neq N+{1\over2}\end{array}\right.\;.
\label{root-quark}
\end{eqnarray}

 Since the radiative corrections from all virtual KK
modes to the physics quantities at electroweak scale should be
summed over in principle in order to obtain the theoretical
predictions in extensions of the SM with a warped or universal extra
dimension. In the Ref.\cite{ref39}, the author verify some lemmas on
the eigenvalues of KK modes, and sum over the infinite series of KK
modes using the residue theorem. Here we list some useful results:

For gauge fields with $(++)$ BCs:

\begin{eqnarray}
&&\Sigma_{_{(++)}}^{G}(t,t^\prime)=\sum\limits_{n=1}^\infty
{\Big[\chi_{_{(++)}}^G(y_{_{(++)}}^{G(n)},t)\Big]
\Big[\chi_{_{(++)}}^G(y_{_{(++)}}^{G(n)},t^\prime)\Big]\over
\Big[y_{_{(++)}}^{G(n)}\Big]^2}
\nonumber\\
&&\hspace{2.0cm}= {1\over8\pi}\Big\{t^2(2\ln t-1)+t^{\prime2}(2\ln
t^\prime-1)
\nonumber\\
&&\hspace{2.5cm}
-2\ln\epsilon\Big[t^2\theta(t^\prime-t)+t^{\prime2}\theta(t-t^\prime)\Big]
-{1-\epsilon^2\over\ln\epsilon}\Big\}\;,
\label{propagator-8}
\end{eqnarray}
which is coincided with Eq.(34) exactly in Ref.\cite{ref28}.

Gauge fields with $(-+)$ BCs

\begin{eqnarray}
&&\Sigma_{_{(-+)}}^{G}(t,t^\prime)
=\sum\limits_{n=1}^\infty{\Big[\chi_{_{(-+)}}^G(y_{_{(-+)}}^{G(n)},t)\Big]
\Big[\chi_{_{(-+)}}^G(y_{_{(-+)}}^{G(n)},t^\prime)\Big]\over
\Big[y_{_{(-+)}}^{G(n)}\Big]^2}
\nonumber\\
&&\hspace{2.0cm}=
{1\over4\pi}\bigg({-\epsilon\ln\epsilon\over1-4\epsilon^2+3\epsilon^4
-4\epsilon^4\ln\epsilon}\bigg)^{1/2}\bigg\{\theta(t-t^\prime)
\sqrt{t^{\prime}}(t^{\prime2}-\epsilon^2)
\nonumber\\
&&\hspace{2.5cm}\times
\bigg[1-{\epsilon^2\over4}+2\epsilon^2\ln\epsilon-{1-9\epsilon^4+8\epsilon^6+6\epsilon^2\ln\epsilon
-12\epsilon^4\ln\epsilon-6\epsilon^6\ln\epsilon\over6(1-4\epsilon^2+3\epsilon^4
-4\epsilon^4\ln\epsilon)}
\nonumber\\
&&\hspace{2.6cm} -{\epsilon^2t^{\prime2}(\ln
t^{\prime}-\ln\epsilon)\over t^{\prime2}-\epsilon^2}
+{t^{\prime2}\over4}-{t^2\over2}(1-2\ln t) +{1-\epsilon^2\over2\ln
\epsilon}\bigg] +(t\leftrightarrow t^\prime)\bigg\}\;.
\label{propagator-9b}
\end{eqnarray}

For the left-handed fields with $(++)$ BCs

\begin{eqnarray}
&&\Sigma_{_{(\pm\pm)}}^{L,c}(t,t^\prime)
=\sum\limits_{n=1}^\infty{\Big[f_{_{(++)}}^{L,c}(y_{_{(\pm\pm)}}^{c(n)},t)\Big]
\Big[f_{_{(++)}}^{L,c}(y_{_{(\pm\pm)}}^{c(n)},t^\prime)\Big]\over
\Big[y_{_{(\pm\pm)}}^{c(n)}\Big]^2}
\nonumber\\
&&\hspace{-0.5cm}=
-{(1-2c)\epsilon\ln\epsilon\over(1-\epsilon^{1-2c})}{(tt^\prime)^{-c}\over4\pi}
\Big\{{2(1-2c)(1-\epsilon^{3-2c})\over(3-2c)(1+2c)(1-\epsilon^{1-2c})}+{t^2+t^{\prime2}\over1-2c}
\nonumber\\
&&-{2\over(1-2c)(1+2c)}
\Big[\theta(t^\prime-t)\Big(t^{\prime1+2c}+\epsilon^{1-2c}t^{1+2c}\Big)
\nonumber\\
&&+\theta(t-t^\prime)\Big(t^{1+2c}+\epsilon^{1-2c}t^{\prime1+2c}\Big)\Big]\Big\}\;.
\label{propagator-18}
\end{eqnarray}

When the left-handed fermions satisfy $(--)$BCs,

\begin{eqnarray}
&&\Sigma_{_{(\mp\mp)}}^{L,c}(t,t^\prime)
=\sum\limits_{n=1}^\infty{\Big[f_{_{(--)}}^{L,c}(y_{_{(\mp\mp)}}^{c(n)},t)\Big]
\Big[f_{_{(--)}}^{L,c}(y_{_{(\mp\mp)}}^{c(n)},t^\prime)\Big]\over
\Big[y_{_{(\mp\mp)}}^{c(n)}\Big]^2}
\nonumber\\
&&\hspace{-0.5cm}=
-{(1-2c)(3+2c)\epsilon^{3/2+c}\ln\epsilon\over[\zeta_{_{(uv)}}^{(-)}(c,\epsilon)
\zeta_{_{(ir)}}^{(-)}(c,\epsilon)]^{1/2}}{(tt^\prime)^{1+c}\over4\pi}
\bigg\{\theta(t-t^\prime)(1-t^{-1-2c})(t^{\prime-1-2c}-\epsilon^{-1-2c})
\nonumber\\&&\times
\bigg[{1\over1-\epsilon^{1+2c}}\bigg({1-\epsilon^{3+2c}\over3+2c}
+{\epsilon^2-\epsilon^{1+2c}\over1-2c}\bigg)
+{\epsilon^{3+2c}+\epsilon^{1-2c}-2\epsilon^2-\epsilon^4\over[\zeta_{_{(ir)}}^{(-)}(c,\epsilon)]}
\nonumber\\&&
-{3(1+2c)^2\over(3-2c)(5+2c)[\zeta_{_{(ir)}}^{(-)}(c,\epsilon)]}
+{(1-2c)\epsilon^{5+2c}\over(5+2c)[\zeta_{_{(ir)}}^{(-)}(c,\epsilon)]}
+{(3+2c)\epsilon^{3-2c}\over(3-2c)[\zeta_{_{(ir)}}^{(-)}(c,\epsilon)]}
\nonumber\\&&
-{1+2\epsilon^2-\epsilon^{1-2c}-\epsilon^{3+2c}\over[\zeta_{_{(uv)}}^{(-)}(c,\epsilon)]}
-{3(1+2c)^2\epsilon^4\over(3-2c)(5+2c)[\zeta_{_{(uv)}}^{(-)}(c,\epsilon)]}
+{(1-2c)\epsilon^{-1-2c}\over(5+2c)[\zeta_{_{(uv)}}^{(-)}(c,\epsilon)]}
\nonumber\\&&
+{(3+2c)\epsilon^{1+2c}\over(3-2c)[\zeta_{_{(uv)}}^{(-)}(c,\epsilon)]}+{1-t^{1-2c}\over(1-2c)(1-t^{-1-2c})}
+{t^2-t^{-1-2c}\over(3+2c)(1-t^{-1-2c})} \nonumber\\&&
+{t^{\prime1-2c}-\epsilon^{1-2c}\over(1-2c)(t^{\prime-1-2c}-\epsilon^{-1-2c})}
+{\epsilon^2t^{\prime-1-2c}-\epsilon^{-1-2c}t^{\prime2}\over(3+2c)(t^{\prime-1-2c}-\epsilon^{-1-2c})}\bigg]
+(t\leftrightarrow t^{\prime})\bigg\}\;.
\label{propagator-20}
\end{eqnarray}
For the left-handed fields with $(+-)$BCs

\begin{eqnarray}
&&\Sigma_{_{(\pm\mp)}}^{L,c}(t,t^\prime)
=\sum\limits_{n=1}^\infty{\Big[f_{_{(+-)}}^{L,c}(y_{_{(\pm\mp)}}^{c(n)},t)\Big]
\Big[f_{_{(+-)}}^{L,c}(y_{_{(\pm\mp)}}^{c(n)},t^\prime)\Big]\over
\Big[y_{_{(\pm\mp)}}^{c(n)}\Big]^2}
\nonumber\\
&&\hspace{-0.5cm}=
-{\epsilon\ln\epsilon\over8\pi}\bigg({(1-2c)^2(3+2c)\over(1-\epsilon^{1-2c})
[\zeta_{_{(ir)}}^{(-)}(c,\epsilon)]}\bigg)^{1/2}
\bigg\{\theta(t-t^\prime)t^{1+c}t^{\prime-c}|1-t^{-1-2c}|
\nonumber\\&&\times
\bigg[{2\over1-2c}+{2\epsilon^2\over1+2c}-{4\epsilon^{1-2c}\over(1-2c)(1+2c)}
+{\epsilon^{3+2c}+\epsilon^{1-2c}-2\epsilon^2-\epsilon^4\over[\zeta_{_{(ir)}}^{(-)}(c,\epsilon)]}
\nonumber\\&&
-{3(1+2c)^2\over(3-2c)(5+2c)[\zeta_{_{(ir)}}^{(-)}(c,\epsilon)]}
+{(1-2c)\epsilon^{5+2c}\over(5+2c)[\zeta_{_{(ir)}}^{(-)}(c,\epsilon)]}
+{(3+2c)\epsilon^{3-2c}\over(3-2c)[\zeta_{_{(ir)}}^{(-)}(c,\epsilon)]}
\nonumber\\&& +{(3-2c)\epsilon^{1-2c}-2(1-2c)\epsilon^{3-2c}\over
(3-2c)(1+2c)(1-\epsilon^{1-2c})}
+{1-t^{1-2c}\over(1-2c)(1-t^{-1-2c})}
+{t^2-t^{-1-2c}\over(3+2c)(1-t^{-1-2c})} \nonumber\\&&
+{t^{\prime2}\over1-2c}-{2\epsilon^{1-2c}t^{\prime1+2c}\over(1-2c)(1+2c)}\bigg]
+(t\leftrightarrow t^{\prime})\bigg\}\;.
\label{propagator-22}
\end{eqnarray}
Left-handed fermions with $(-+)$BCs

\begin{eqnarray}
&&\Sigma_{_{(\mp\pm)}}^{L,c}(t,t^\prime)
=\sum\limits_{n=1}^\infty{\Big[f_{_{(-+)}}^{L,c}(y_{_{(\mp\pm)}}^{c(n)},t)\Big]
\Big[f_{_{(-+)}}^{L,c}(y_{_{(\mp\pm)}}^{c(n)},t^\prime)\Big]\over
\Big[y_{_{(\mp\pm)}}^{c(n)}\Big]^2}
\nonumber\\
&&\hspace{-0.5cm}=
-{\epsilon^{3/2+c}\ln\epsilon\over8\pi}\bigg({(1-2c)^2(3+2c)\over(1-\epsilon^{1-2c})
[\zeta_{_{(uv)}}^{(-)}(c,\epsilon)]}\bigg)^{1/2}
\bigg\{\theta(t-t^\prime)t^{-c}t^{\prime1+c}|t^{\prime-1-2c}-\epsilon^{-1-2c}|
\nonumber\\&&\times
\bigg[{2\over1+2c}+{2\epsilon^2\over1-2c}-{4\epsilon^{1+2c}\over(1-2c)(1+2c)}
+{2(1-2c)+(1+2c)\epsilon^{3-2c}-(3-2c)\epsilon^2\over(3-2c)(1+2c)(1-\epsilon^{1-2c})}
\nonumber\\&&
-{1+2\epsilon^2-\epsilon^{1-2c}-\epsilon^{3+2c}\over[\zeta_{_{(uv)}}^{(-)}(c,\epsilon)]}
-{3(1+2c)^2\epsilon^4\over(3-2c)(5+2c)[\zeta_{_{(uv)}}^{(-)}(c,\epsilon)]}
+{(1-2c)\epsilon^{-1-2c}\over(5+2c)[\zeta_{_{(uv)}}^{(-)}(c,\epsilon)]}
\nonumber\\&&
+{(3+2c)\epsilon^{1+2c}\over(3-2c)[\zeta_{_{(uv)}}^{(-)}(c,\epsilon)]}
+{t^2\over1-2c}-{2t^{1+2c}\over(1-2c)(1+2c)}
+{t^{\prime1-2c}-\epsilon^{1-2c}\over(1-2c)(t^{\prime-1-2c}-\epsilon^{-1-2c})}
\nonumber\\&&
+{\epsilon^2t^{\prime-1-2c}-\epsilon^{-1-2c}t^{\prime2}\over(3+2c)(t^{\prime-1-2c}-\epsilon^{-1-2c})}
\bigg]+(t\leftrightarrow t^{\prime})\bigg\}\;.
\label{propagator-24}
\end{eqnarray}

with
\begin{eqnarray}
&&\zeta_{_{(uv)}}^{(-)}(c,\epsilon)=(3+2c)\epsilon^{1+2c}
+(1-2c)\epsilon^{-1-2c}-(1+2c)^2\epsilon^2-(1-2c)(3+2c)\;,
\nonumber\\
&&\zeta_{_{(ir)}}^{(-)}(c,\epsilon)=-(3+2c)\epsilon^{1-2c}
-(1-2c)\epsilon^{3+2c}+(1+2c)^2+(1-2c)(3+2c)\epsilon^2\;.
\label{propagator-19a}
\end{eqnarray}

Similarly,  for the right-handed fields, one analogously has
\begin{eqnarray}
&&\Sigma_{_{(\mp\mp)}}^{R,c}(t,t^\prime)
=\sum\limits_{n=1}^\infty{\Big[f_{_{(++)}}^{R,c}(y_{_{(\mp\mp)}}^{c(n)},t)\Big]
\Big[f_{_{(++)}}^{R,c}(y_{_{(\mp\mp)}}^{c(n)},t^\prime)\Big]\over
\Big[y_{_{(\mp\mp)}}^{c(n)}\Big]^2}=\Sigma_{_{(\pm\pm)}}^{L,-c}(t,t^\prime)\;,
\nonumber\\
&&\Sigma_{_{(\pm\pm)}}^{R,c}(t,t^\prime)
=\sum\limits_{n=1}^\infty{\Big[f_{_{(--)}}^{R,c}(y_{_{(\mp\mp)}}^{c(n)},t)\Big]
\Big[f_{_{(--)}}^{R,c}(y_{_{(\pm\pm)}}^{c(n)},t^\prime)\Big]\over
\Big[y_{_{(\pm\pm)}}^{c(n)}\Big]^2}=\Sigma_{_{(\mp\mp)}}^{L,-c}(t,t^\prime)\;,
\nonumber\\
&&\Sigma_{_{(\mp\pm)}}^{R,c}(t,t^\prime)
=\sum\limits_{n=1}^\infty{\Big[f_{_{(+-)}}^{R,c}(y_{_{(\mp\mp)}}^{c(n)},t)\Big]
\Big[f_{_{(+-)}}^{R,c}(y_{_{(\mp\pm)}}^{c(n)},t^\prime)\Big]\over
\Big[y_{_{(\mp\pm)}}^{c(n)}\Big]^2}=\Sigma_{_{(\pm\mp)}}^{L,-c}(t,t^\prime)\;,
\nonumber\\
&&\Sigma_{_{(\pm\mp)}}^{R,c}(t,t^\prime)
=\sum\limits_{n=1}^\infty{\Big[f_{_{(-+)}}^{R,c}(y_{_{(\pm\mp)}}^{c(n)},t)\Big]
\Big[f_{_{(-+)}}^{R,c}(y_{_{(\pm\mp)}}^{c(n)},t^\prime)\Big]\over
\Big[y_{_{(\pm\mp)}}^{c(n)}\Big]^2}=\Sigma_{_{(\mp\pm)}}^{L,-c}(t,t^\prime)\;.
\label{propagator-191a}
\end{eqnarray}

\section{The theoretical calculation on the  $t\rightarrow c\gamma$ and $t\rightarrow cg$ processes  \label{sec2}}
\indent\indent

 In this section, we present
one-loop radiative corrections to the rare decay $t\rightarrow
c\gamma$ and  $t\rightarrow cg$  in the extension of the SM with a
warped extra dimension and the custodial symmetry.

The infinite dimensional column vectors for quarks in the chirality
basis as\cite{ref42}
\begin{eqnarray}
&&\Psi_L(5/3)=\Big(\chi_{u_L}^{i(n)}(-+),X_{L}^{i(n)}(+-),\tilde{X}_L^{i(n)}(+-),\cdots\Big)^T\;,
\nonumber\\
&&\Psi_R(5/3)=\Big(\chi_{u_R}^{i(n)}(+-),X_{R}^{i(n)}(-+),\tilde{X}_R^{i(n)}(-+),\cdots\Big)^T\;,
\nonumber\\
&&\Psi_L(2/3)=\Big(q_{_{u_L}}^{i(0)}(++),\cdots,q_{_{u_L}}^{i(n)}(++),
U_L^{i(n)}(+-),\tilde{U}_L^{i(n)}(+-),\chi_{d_L}^{i(n)}(-+),u_L^{i(n)}(--),\cdots\Big)^T\;,
\nonumber\\
&&\Psi_R(2/3)=\Big(u_R^{i(0)}(++),\cdots,q_{_{u_R}}^{i(n)}(--),
U_R^{i(n)}(-+),\tilde{U}_R^{i(n)}(-+),\chi_{d_R}^{i(n)}(+-),u_R^{i(n)}(++),\cdots\Big)^T\;,
\nonumber\\
&&\Psi_L(-1/3)=\Big(q_{_{d_L}}^{i(0)}(++),\cdots,q_{_{d_L}}^{i(n)}(++),
D_L^{i(n)}(+-),d_L^{i(n)}(--),\cdots\Big)^T\;,
\nonumber\\
&&\Psi_R(-1/3)=\Big(d_R^{i(0)}(++),\cdots,q_{_{d_R}}^{i(n)}(--),
D_R^{i(n)}(-+),d_R^{i(n)}(++),\cdots\Big)^T\;,
\label{mass1}
\end{eqnarray}
where the flavor index $i=1,\;2,\;3$ runs over the three quark
generations, and $n=1,\;2,\;\cdots,\;\infty$ is the index of KK
exciting modes, the signs in parentheses denote the BCs satisfied by
corresponding fields on UV and IR branes respectively.

We can write the mass eigenstates of charged $5/3$, $2/3$ and $-1/3$
quarks as
\begin{eqnarray}
&&H_{\alpha,L}=\Big[{\cal H}_L^\dagger\Psi_L(5/3)\Big]_\alpha\;,
H_{\alpha,R}=\Big[{\cal H}_R^\dagger\Psi_R(5/3)\Big]_\alpha\;,
\nonumber\\
&&U_{\alpha,L}=\Big[{\cal U}_L^\dagger\Psi_L(2/3)\Big]_\alpha\;,
U_{\alpha,R}=\Big[{\cal U}_R^\dagger\Psi_R(2/3)\Big]_\alpha\;,
\nonumber\\
&&D_{\alpha,L}=\Big[{\cal D}_L^\dagger\Psi_L(-1/3)\Big]_\alpha\;,
D_{\alpha,R}=\Big[{\cal D}_R^\dagger\Psi_R(-1/3)\Big]_\alpha\;.
\label{mass2}
\end{eqnarray}
Here, the charged $2/3$ quarks $U_1,\;U_2,\;U_3$ are identified as
up-type quarks $u,\;c,\;t$, and the charged $-1/3$ quarks
$D_1,\;D_2,\;D_3$ are identified as the down-type quarks $d,\;s,\;b$
in the SM, respectively, and $H_{\alpha,L}, H_{\alpha,R}$ are the
charged $5/3$ quarks, which not exist in the SM.

Similarly, we could express interaction eigenstates of charged and
neutral electroweak gauge bosons in linear combination of the mass
eigenstates as
\begin{eqnarray}
&&W_L^{(0)\pm}=\Big({\cal
Z}_W\Big)_{0,0}W^\pm+\sum\limits_{\alpha=1}^\infty \Big({\cal
Z}_W\Big)_{0,\alpha}W_{H_\alpha}^\pm\;,
\nonumber\\
&&W_L^{(n)\pm}=\Big({\cal
Z}_W\Big)_{2n-1,0}W^\pm+\sum\limits_{\alpha=1}^\infty \Big({\cal
Z}_W\Big)_{2n-1,\alpha}W_{H_\alpha}^\pm\;,
\nonumber\\
&&W_R^{(n)\pm}=\Big({\cal
Z}_W\Big)_{2n,0}W^\pm+\sum\limits_{\alpha=1}^\infty \Big({\cal
Z}_W\Big)_{2n,\alpha}W_{H_\alpha}^\pm\;,
\nonumber\\
&&Z^{(0)}=\Big({\cal Z}_Z\Big)_{0,0}Z+\sum\limits_{\alpha=1}^\infty
\Big({\cal Z}_Z\Big)_{0,\alpha}Z_{H_\alpha}\;,
\nonumber\\
&&Z^{(n)}=\Big({\cal
Z}_Z\Big)_{2n-1,0}Z+\sum\limits_{\alpha=1}^\infty \Big({\cal
Z}_Z\Big)_{2n-1,\alpha}Z_{H_\alpha}\;,
\nonumber\\
&&Z_X^{(n)}=\Big({\cal
Z}_Z\Big)_{2n,0}Z+\sum\limits_{\alpha=1}^\infty \Big({\cal
Z}_Z\Big)_{2n,\alpha}Z_{H_\alpha}\;,
\label{mass3}
\end{eqnarray}
in which ${\cal Z}_{W},\;{\cal Z}_{Z}$ respectively denote the
mixing matrices for charged as well as neutral electroweak gauge
bosons, and $Z,W^\pm$ are identified as the corresponding gauge
bosons in the SM.

The relevant Feynman diagrams are draw in Fig.1 when we adopt the
background gauge\cite{ref43}

\begin{figure}\small
  \centering
   \includegraphics[width=4in]{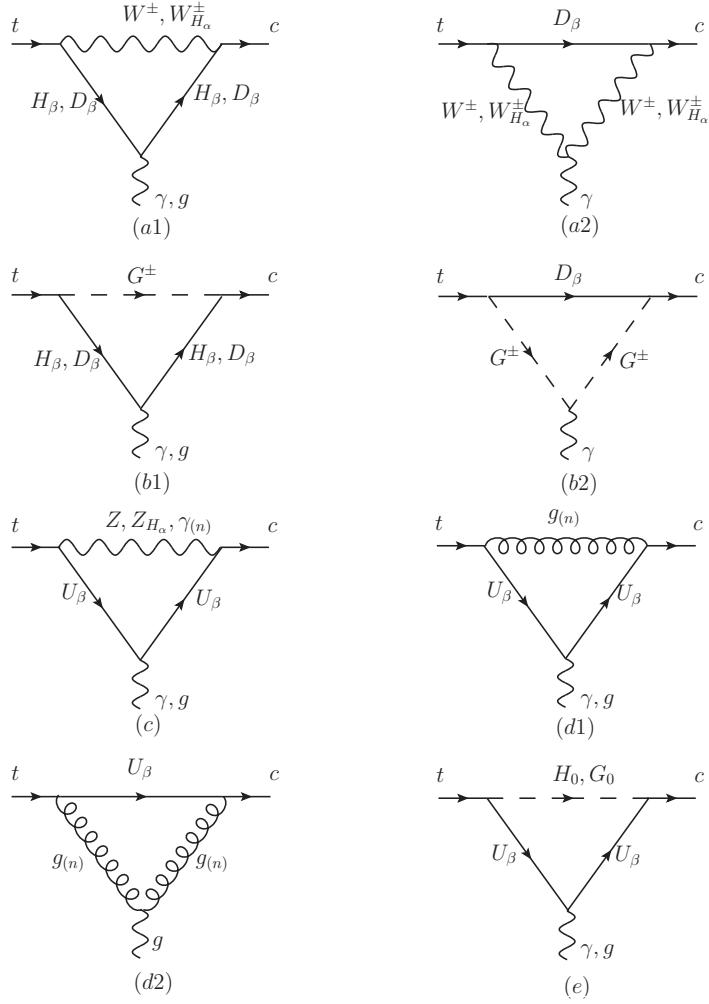}
     \caption{The Feynman diagrams contributing to the
$t\rightarrow c\gamma$ and $t\rightarrow cg$ decay in warped extra
dimension with custodial symmetry. Where
$Z,\;W^\pm,\;H_0,\;G_0,\;G^\pm,\;U_i=u_i,\;D_i=d_i \;(i=1,\;2,\;3)$
denote the normally neutral and charged gauge bosons, neutral Higgs,
neutral and charged Goldstones, and three generation up- and
down-type quarks,
$Z_{H_\alpha},\;W^\pm_{H_\alpha},\;\gamma_{(n)},\;g_{(n)},\;U_{(3+\beta)},\;D_{(3+\beta)},\;H_{(3+\beta)}\;
\;(n,\;\alpha,\;\beta=1,\;2,\;\cdots,\;\infty)$ denote those heavy
gauge bosons together with charge $2/3$, $-1/3$ and  $5/3$ quarks,
respectively.}
    \label{fige3c}
\end{figure}

We could see that the FCNC transitions are mediated by Higgs, the KK
excitations of gluon, photo, and neutral electroweak gauge bosons
besides the charged electroweak gauge bosons $W^{\pm}$ together with
their KK partners.

In order to evaluate there diagrams, and sum over the infinite
series of KK modes, we divided them into two cases: When  all the
propagators are SM particles, such as
$Z,\;W^\pm,\;H_0,\;G_0,\;G^\pm,\;U_i=u_i,\;D_i=d_i \;(i=1,\;2,\;3)$,
we use the general method in \cite{ref1}.

And when the propagator is the KK mode, such as
$Z_{H_\alpha},\;W^\pm_{H_\alpha},\;\gamma_{(n)},\;g_{(n)},\;U_{(3+\beta)},\;D_{(3+\beta)},\;H_{(3+\beta)}\;
\;(n,\;\alpha,\;\beta=1,\;2,\;\cdots,\;\infty)$,   since the mass of
the KK mode is large then the top quark, we could use the effective
Hamilton method\cite{ref39} and expand the amplitudes to the order
$m_t^2/\Lambda_{KK}^2$ for simplicity.

\subsection{The first case}
 When  all the propagators are SM particles, the
effective flavor changing current for $t\rightarrow c \gamma,g$ can
be written in conserved form

\begin{eqnarray}
&&J_\mu ^{\gamma, g} = \bar c(p)[{q^2}{\gamma _\mu }({F_L^{\gamma, g}}{P_L} +
{F_R^{\gamma, g}}{P_R}) + i{\sigma _{\mu \nu }}{q_\nu }({m_c}{F_{TL}^{\gamma, g}}{P_L} +
{m_t}{F_{TR}^{\gamma, g}}{P_R})]t(p') \label{eq1}
\end{eqnarray}

where $p'$ is the momentum of the  initial top quark and $p$ is the
momentum of the final state charm quark, $q_\mu=p-p'$, and we have
dropped the $q_\mu$ term because it does not contribute to the
processes.

In the equations below, $m_i\:(i=1,2,3)$ is the mass of the SM three
generation up- or down-type quarks, $m_W$ and $m_Z$ are the mass of
the SM neutral and charged gauge bosons, $m_{G^\pm}$, $m_{H_0/G_0}$
are the mass of $G^\pm$ and $G_{H_0}, G_{G_0}$ respectively,
$\text{C}_{ij}$ are the coefficients of the Lorentz-covariant tensors in the 3-point standard scalar Passarino-Veltman
integrals(Eq.(4.7) in Ref. \cite{ref431}), and it could be calculated by using 'LoopTools'. And the relevant coefficients $\xi^{L,R} $
and $ \eta^{L,R}$ are the couplings between bosons and quarks, we
approximate them to the order ${\cal O}(\upsilon^2/\Lambda_{KK}^2)$
and present in Appendix.

\subsubsection{Fig.(a1) and (a2) with SM propagators}

In Fig. (a1)and (a2), when one-loop diagrams are composed by the
zero mode of charge gauge bosons $W^\pm$ and charged $-1/3$ quarks
$d_i \;(i=1,\;2,\;3)$, $F_{TL}^{\gamma, g}$ and $F_{TR}^{\gamma, g}$ are formulated as
\begin{eqnarray}
&&F_{TL}^{\gamma(a1)}=\frac{{i{e^3}{Q^F}}}{{16{\pi ^2}{m_c}s_W^2}}
\Big(2 (\xi _{W^ \pm }^{R(-1/3)})_{c,i }^\dag {(\xi _{W^ \pm
}^{L(-1/3)})_{i ,t}} m_{i} \text{C}_1+(\xi _{W^ \pm
}^{R(-1/3)})_{c,i }^\dag {(\xi _{W^ \pm }^{R(-1/3)})_{i ,t}} m_t
\text{C}_{12}\nonumber\\
&&\hspace{2cm}-(\xi _{W^ \pm }^{L(-1/3)})_{c,i }^\dag {(\xi _{W^ \pm
}^{L(-1/3)})_{i ,t}} m_c
   (\text{C}_1-\text{C}_{11}+\text{C}_{12})\Big)\nonumber\\
&&F_{TR}^{\gamma(a1)}=\frac{{i{e^3}{Q^F}}}{{16{\pi ^2}{m_t}s_W^2}}
\Big(2 (\xi _{W^ \pm }^{L(-1/3)})_{c,i }^\dag {(\xi _{W^ \pm
}^{R(-1/3)})_{i ,t}} m_{i} \text{C}_1+(\xi _{W^ \pm
}^{L(-1/3)})_{c,i }^\dag {(\xi _{W^ \pm }^{L(-1/3)})_{i ,t}} m_t
\text{C}_{12}\nonumber\\
&&\hspace{2cm}-(\xi _{W^ \pm }^{R(-1/3)})_{c,i }^\dag {(\xi _{W^ \pm
}^{R(-1/3)})_{i ,t}} m_c
   (\text{C}_1-\text{C}_{11}+\text{C}_{12})\Big)\nonumber\\\label{eq1}
\end{eqnarray}

with
$\text{C}_{ij}=\text{C}_{ij}(p^2,(2p-p')^2,(p-p')^2,m_{i}^2,m_W^2,m_{i}^2)$

\begin{eqnarray}
&&F_{TL}^{\gamma(a2)}=-\frac{{i{e^3}}}{{16{\pi ^2}{m_c}s_W^2}}\Big(2
(\xi _{W^ \pm }^{R(-1/3)})_{c,i }^\dag {(\xi _{W^ \pm
}^{L(-1/3)})_{i ,t}} (\text{C}_1+\text{C}_2)
m_{i}\nonumber\\
&&\hspace{2cm}+(\xi _{W^ \pm }^{L(-1/3)})_{c,i }^\dag {(\xi _{W^ \pm
}^{L(-1/3)})_{i ,t}} (\text{C}_{11}+\text{C}_{12}-\text{C}_{2})
   m_c\nonumber\\
&&\hspace{2cm}+(\xi _{W^ \pm
}^{R(-1/3)})_{c,i }^\dag {(\xi _{W^ \pm }^{R(-1/3)})_{i ,t}} (-\text{C}_1+\text{C}_{12}+\text{C}_{22}) m_t\Big)\nonumber\\
&&F_{TR}^{\gamma(a2)}=-\frac{{i{e^3}}}{{16{\pi ^2}{m_t}s_W^2}}\Big(2
(\xi _{W^ \pm }^{L(-1/3)})_{c,i }^\dag {(\xi _{W^ \pm
}^{R(-1/3)})_{i ,t}} (\text{C}_1+\text{C}_2)
m_{i}\nonumber\\
&&\hspace{2cm}+(\xi _{W^ \pm }^{R(-1/3)})_{c,i }^\dag {(\xi _{W^ \pm
}^{R(-1/3)})_{i ,t}} (\text{C}_{11}+\text{C}_{12}-\text{C}_2)
   m_c\nonumber\\
&&\hspace{2cm}+(\xi _{W^ \pm }^{L(-1/3)})_{c,i }^\dag {(\xi _{W^ \pm
}^{L(-1/3)})_{i ,t}} (-\text{C}_1+\text{C}_{12}+\text{C}_{22})
m_t\Big)\label{eq1}
\end{eqnarray}

with
$\text{C}_{ij}=\text{C}_{ij}(p^2,(p-p')^2,p'^2,m_{i}^2,m_W^2,m_W^2)$

and

\begin{eqnarray}
&&F_{TL}^{g(a1)}=\frac{{i{e^2}{g_S}{T^a}}}{{16{\pi ^2}{m_c}s_W^2}}
\Big(2 (\xi _{W^ \pm }^{R(-1/3)})_{c,i
}^\dag {(\xi _{W^ \pm }^{L(-1/3)})_{i ,t}} m_{i} \text{C}_1\nonumber\\
&&\hspace{2cm}+(\xi _{W^ \pm }^{R(-1/3)})_{c,i }^\dag {(\xi _{W^ \pm
}^{R(-1/3)})_{i ,t}} m_t \text{C}_{12}-(\xi _{W^ \pm
}^{L(-1/3)})_{c,i }^\dag {(\xi _{W^ \pm }^{L(-1/3)})_{i ,t}} m_c
   (\text{C}_1-\text{C}_{11}+\text{C}_{12})\Big)\nonumber\\
&&F_{TR}^{g(a1)}=\frac{{i{e^2}{g_S}{T^a}}}{{16{\pi
^2}{m_t}s_W^2}}\Big(2 (\xi _{W^ \pm }^{L(-1/3)})_{c,i
}^\dag {(\xi _{W^ \pm }^{R(-1/3)})_{i ,t}} m_{i} \text{C}_1\nonumber\\
&&\hspace{2cm}+(\xi _{W^ \pm }^{L(-1/3)})_{c,i }^\dag {(\xi _{W^ \pm
}^{L(-1/3)})_{i ,t}} m_t \text{C}_{12}-(\xi _{W^ \pm
}^{R(-1/3)})_{c,i }^\dag {(\xi _{W^ \pm }^{R(-1/3)})_{i ,t}} m_c
   (\text{C}_1-\text{C}_{11}+\text{C}_{12})\Big)\nonumber\\\label{eq1}
\end{eqnarray}

with
$\text{C}_{ij}=\text{C}_{ij}(p^2,(2p-p')^2,(p-p')^2,m_{i}^2,m_W^2,m_{i}^2)$

Using $\xi^{L,R} $ in appendix, we can approximate the relevant
coefficients above as
\begin{eqnarray}
&&
(\xi _{{W^ \pm }}^{L(-1/3)})_{c,i }^\dag {(\xi _{{W^ \pm }}^{L(-1/3)})_{i ,t}} = \sum\limits_{i = 1}^3 ( V_{CKM}^{(0)}{)_{ci}}(V_{CKM}^{(0)})_{it}^\dag  + \sum\limits_{i = 1}^3 ( \Upsilon _{i,1}^{(a)}{)_{ct}} + O(\frac{{{\upsilon ^4}}}{{\Lambda _{KK}^4}})\;,\nonumber\\
&&
(\xi _{{W^ \pm }}^{L(-1/3)})_{c,i }^\dag {(\xi _{{W^ \pm }}^{R(-1/3)})_{i ,t}} = \frac{{{\upsilon ^2}}}{{2\Lambda _{KK}^2}}\sum\limits_{i = 1}^3 {{{(V_{CKM}^{(0)})}_{ci}}(\Delta _{{W^ \pm }}^R)_{it}^\dag }  + O(\frac{{{\upsilon ^4}}}{{\Lambda _{KK}^4}})\;,\nonumber\\
&&
(\xi _{{W^ \pm }}^{R(-1/3)})_{c,i }^\dag {(\xi _{{W^ \pm }}^{L(-1/3)})_{i ,t}} = \frac{{{\upsilon ^2}}}{{2\Lambda _{KK}^2}}\sum\limits_{i = 1}^3 {{{(\Delta _{{W^ \pm }}^R)}_{ci}}(V_{CKM}^{(0)})_{it}^\dag }  + O(\frac{{{\upsilon ^4}}}{{\Lambda _{KK}^4}})\;,\nonumber\\
\label{eq1}
\end{eqnarray}
here we define the short-cut notation

\begin{eqnarray}
&&
{(\Upsilon _{i,1}^{(a)})_{ct}} = {(V_{CKM}^{(0)})_{ci}}(V_{CKM}^{(0)}\delta Z_L^d)_{it}^\dag  + {(V_{CKM}^{(0)}\delta Z_L^d)_{ci}}(V_{CKM}^{(0)})_{it}^\dag  + {(V_{CKM}^{(0)})_{ci}}(\delta Z_L^{u\dag }V_{CKM}^{(0)})_{it}^\dag \nonumber\\
&&\hspace{1.5cm}
 + {(\delta Z_L^{u\dag }V_{CKM}^{(0)})_{ci}}(V_{CKM}^{(0)})_{it}^\dag  - \frac{{{\upsilon ^2}}}{{4\Lambda _{KK}^2}}[{(V_{CKM}^{(0)})_{ci}}{(\Delta _{{W^ \pm }}^L)_{it}} + (\Delta _{{W^ \pm }}^L)_{ci}^\dag (V_{CKM}^{(0)})_{it}^\dag ]\;.
\label{eq1}
\end{eqnarray}

where  $\delta Z_L^u, \delta Z_L^d$ are the correct to the mixing
matrices, which are presented in the equation (154) and (160) of
Ref. \cite{ref39}. And $(\xi _{{W^ \pm }}^{L(-1/3)})_{c,i }^\dag
{(\xi _{{W^ \pm }}^{L(-1/3)})_{i ,t}} = \sum\limits_{i = 1}^3 (
V_{CKM}^{(0)}{)_{ci}}(V_{CKM}^{(0)})_{it}^\dag$ is just the SM case

\subsubsection{Fig.(b1) and (b2) with SM propagators}

The contributions to $F_{TL}^{\gamma, g}$ and $F_{TR}^{\gamma, g}$
of diagrams (b1) and (b2) with the zero mode of charge Goldstone and
charge $-1/3$ quarks and charge Higgs $G^\pm$ are
\begin{eqnarray}
&&F_{TL}^{\gamma(b1)}=\frac{{i{e}{Q^F}}}{{16{\pi
^2}{m_c}}}\Big((\eta _{G^ \pm }^{L(-1/3)})_{c,i }^\dag {(\eta _{G^
\pm }^{L(-1/3)})_{i ,t}} (\text{C}_0+\text{C}_1)
m_{i}\nonumber\\
&&\hspace{2cm}+(\eta _{G^ \pm }^{R(-1/3)})_{c,i }^\dag {(\eta _{G^
\pm }^{L(-1/3)})_{i ,t}} (\text{C}_{12}-\text{C}_{11}) m_c-(\eta
_{G^ \pm }^{L(-1/3)})_{c,i }^\dag {(\eta _{G^ \pm }^{R(-1/3)})_{i
,t}}
   (\text{C}_{1}+\text{C}_{12}) m_t\Big)\nonumber\\
&&F_{TR}^{\gamma(b1)}=\frac{{i{e}{Q^F}}}{{16{\pi
^2}{m_t}}}\Big((\eta _{G^ \pm }^{R(-1/3)})_{c,i }^\dag {(\eta _{G^
\pm }^{R(-1/3)})_{i ,t}}
(\text{C}_{0}+\text{C}_{1}) m_{i}\nonumber\\
&&\hspace{2cm}+(\eta _{G^ \pm }^{L(-1/3)})_{c,i }^\dag {(\eta _{G^
\pm }^{R(-1/3)})_{i ,t}} (\text{C}_{12}-\text{C}_{11}) m_c-(\eta
_{G^ \pm }^{R(-1/3)})_{c,i }^\dag {(\eta _{G^ \pm }^{L(-1/3)})_{i
,t}}
   (\text{C}_{1}+\text{C}_{12}) m_t\Big)\nonumber\\\label{eq1}
\end{eqnarray}

with
$\text{C}_{ij}=\text{C}_{ij}(p^2,(2p-p')^2,(p-p')^2,m_{i}^2,m_{G^\pm}^2,m_{i}^2)$

\begin{eqnarray}
&&F_{TL}^{\gamma(b2)}=\frac{{i{e}}}{{16{\pi ^2}{m_c}}}\Big((\eta
_{G^ \pm }^{L(-1/3)})_{c,i }^\dag {(\eta _{G^ \pm }^{L(-1/3)})_{i
,t}}
(\text{C}_{0}+\text{C}_{1}+\text{C}_{2}) m_{i}\nonumber\\
&&\hspace{2cm}-(\eta _{G^ \pm }^{R(-1/3)})_{c,i }^\dag {(\eta _{G^
\pm }^{L(-1/3)})_{i ,t}} (\text{C}_{1}+\text{C}_{11}+\text{C}_{12})
   m_c\nonumber\\
&&\hspace{2cm}-(\eta _{G^ \pm
}^{L(-1/3)})_{c,i }^\dag {(\eta _{G^ \pm }^{R(-1/3)})_{i ,t}} (\text{C}_{12}+\text{C}_{2}+\text{C}_{22}) m_t\Big)\nonumber\\
&&F_{TR}^{\gamma(b2)}=\frac{{i{e}}}{{16{\pi ^2}{m_t}}}\Big((\eta
_{G^ \pm }^{R(-1/3)})_{c,i }^\dag {(\eta _{G^ \pm }^{R(-1/3)})_{i
,t}}
(\text{C}_{0}+\text{C}_{1}+\text{C}_{2}) m_{i}\nonumber\\
&&\hspace{2cm}-(\eta _{G^ \pm }^{L(-1/3)})_{c,i }^\dag {(\eta _{G^
\pm }^{R(-1/3)})_{i ,t}} (\text{C}_{1}+\text{C}_{11}+\text{C}_{12})
   m_c\nonumber\\
&&\hspace{2cm}-(\eta _{G^ \pm }^{R(-1/3)})_{c,i }^\dag {(\eta _{G^
\pm }^{L(-1/3)})_{i ,t}} (\text{C}_{12}+\text{C}_{2}+\text{C}_{22})
m_t\Big)\label{eq1}
\end{eqnarray}

with
$\text{C}_{ij}=\text{C}_{ij}(p^2,(p-p')^2,p'^2,m_{i}^2,m_{G^\pm}^2,m_{G^\pm}^2)$

and

\begin{eqnarray}
&&F_{TL}^{g(b1)}=\frac{{i{g_S}{T^a}}}{{16{\pi ^2}{m_c}}}\Big((\eta
_{G^ \pm }^{L(-1/3)})_{c,i }^\dag {(\eta _{G^ \pm }^{L(-1/3)})_{i
,t}} (\text{C}_0+\text{C}_1)
m_{i}\nonumber\\
&&\hspace{2cm}+(\eta _{G^ \pm }^{R(-1/3)})_{c,i }^\dag {(\eta _{G^
\pm }^{L(-1/3)})_{i ,t}} (\text{C}_{12}-\text{C}_{11}) m_c-(\eta
_{G^ \pm }^{L(-1/3)})_{c,i }^\dag {(\eta _{G^ \pm }^{R(-1/3)})_{i
,t}}
   (\text{C}_{1}+\text{C}_{12}) m_t\Big)\nonumber\\
&&F_{TR}^{g(b1)}=\frac{{i{g_S}{T^a}}}{{16{\pi ^2}{m_t}}}\Big((\eta
_{G^ \pm }^{R(-1/3)})_{c,i }^\dag {(\eta _{G^ \pm }^{R(-1/3)})_{i
,t}}
(\text{C}_{0}+\text{C}_{1}) m_{i}\nonumber\\
&&\hspace{2cm}+(\eta _{G^ \pm }^{L(-1/3)})_{c,i }^\dag {(\eta _{G^
\pm }^{R(-1/3)})_{i ,t}} (\text{C}_{12}-\text{C}_{11}) m_c-(\eta
_{G^ \pm }^{R(-1/3)})_{c,i }^\dag {(\eta _{G^ \pm }^{L(-1/3)})_{i
,t}}
   (\text{C}_{1}+\text{C}_{12}) m_t\Big)\nonumber\\\label{eq1}
\end{eqnarray}

with
$\text{C}_{ij}=\text{C}_{ij}(p^2,(2p-p')^2,(p-p')^2,m_{i}^2,m_{G^\pm}^2,m_{i}^2)$

Using $\eta^{L,R} $ in appendix, the relevant coefficients above can be approximate to
the order ${\cal O}(\upsilon^2/\Lambda_{KK}^2)$ as

\begin{eqnarray}
&&
(\eta _{{G^ \pm }}^{L(-1/3)})_{c,i }^\dag {(\eta _{{G^ \pm }}^{L(-1/3)})_{i ,t}} = \frac{{{e^2}}}{{2s_{\rm{w}}^2}}\sum\limits_{i = 1}^3 ( V_{CKM}^{(0)}{)_{ci}}(V_{CKM}^{(0)})_{it}^\dag \frac{{{m_t}{m_c}}}{{m_W^2}} + \frac{{{e^2}}}{{2s_{\rm{w}}^2}}\sum\limits_{i = 1}^3 {{{(\Upsilon _{i,1}^{(b)})}_{ct}}} + O(\frac{{{\upsilon ^4}}}{{\Lambda _{KK}^4}})\;,\nonumber\\
&&
(\eta _{{G^ \pm }}^{L(-1/3)})_{c,i }^\dag {(\eta _{{G^ \pm }}^{R(-1/3)})_{i ,t}} = \frac{{{e^2}}}{{2s_{\rm{w}}^2}}\sum\limits_{i = 1}^3 ( V_{CKM}^{(0)}{)_{ci}}(V_{CKM}^{(0)})_{it}^\dag \frac{{{m_c}{m_i}}}{{m_W^2}} + \frac{{{e^2}}}{{2s_{\rm{w}}^2}}\sum\limits_{i = 1}^3 {{{(\Upsilon _{i,2}^{(b)})}_{ct}}}  + O(\frac{{{\upsilon ^4}}}{{\Lambda _{KK}^4}})\;,\nonumber\\
&&
(\eta _{{G^ \pm }}^{R(-1/3)})_{c,i }^\dag {(\eta _{{G^ \pm }}^{L(-1/3)})_{i ,t}} = \frac{{{e^2}}}{{2s_{\rm{w}}^2}}\sum\limits_{i = 1}^3 ( V_{CKM}^{(0)}{)_{ci}}(V_{CKM}^{(0)})_{it}^\dag \frac{{{m_t}{m_i}}}{{m_W^2}} + \frac{{{e^2}}}{{2s_{\rm{w}}^2}}\sum\limits_{i = 1}^3 ( \Upsilon _{i,3}^{(b)}{)_{ct}} + O(\frac{{{\upsilon ^4}}}{{\Lambda _{KK}^4}})\;,\nonumber\\
&&
(\eta _{{G^ \pm }}^{R(-1/3)})_{c,i }^\dag {(\eta _{{G^ \pm }}^{R(-1/3)})_{i ,t}} = \frac{{{e^2}}}{{2s_{\rm{w}}^2}}\sum\limits_{i = 1}^3 ( V_{CKM}^{(0)}{)_{ci}}(V_{CKM}^{(0)})_{it}^\dag \frac{{m_i^2}}{{m_W^2}} + \frac{{{e^2}}}{{2s_{\rm{w}}^2}}\sum\limits_{i = 1}^3 {{{(\Upsilon _{i,4}^{(b)})}_{ct}}} + O(\frac{{{\upsilon ^4}}}{{\Lambda _{KK}^4}})\;,\nonumber\\
\label{eq1}
\end{eqnarray}
The first term of each equations are the SM case. And the short-cut
notation are

\begin{eqnarray}
&&
{(\Upsilon _{i,1}^{(b)})_{ct}} =  - {(V_{CKM}^{(0)})_{ci}}(V_{CKM}^{(0)})_{it}^\dag [\frac{{{m_t}}}{{m_{\rm{W}}^2}}{(\delta {M^u})_{ii}} + \frac{{{m_c}}}{{m_{\rm{W}}^2}}(\delta {M^u})_{ii}^*]\nonumber\\
&&\hspace{1.5cm}
 - \frac{{2\pi {m_t}{m_c}}}{{\Lambda _{KK}^2}}{(V_{CKM}^{(0)})_{ci}}(V_{CKM}^{(0)})_{it}^\dag \{ [\Sigma _{( +  + )}^G(1,1)] + [\Sigma _{( -  + )}^G(1,1)]\}\nonumber\\
&&\hspace{1.5cm}
 + \sum\limits_{j = 1}^3 {\frac{{{m_t}{m_c}}}{{m_{\rm{W}}^2}}} {(V_{CKM}^{(0)})_{ci}}[{(\delta Z_L^d)_{ij}} + (\delta Z_L^d)_{ij}^\dag ](V_{CKM}^{(0)})_{jt}^\dag \nonumber\\
&&\hspace{1.5cm}
 + \sum\limits_{j = 1}^3 {\frac{{m_{{d_i}}^2}}{{m_{\rm{W}}^2}}} [{(V_{CKM}^{(0)})_{ci}}(V_{CKM}^{(0)})_{ij}^\dag (\delta Z_R^u)_{it}^\dag  + {(\delta Z_R^u)_{cj}}{(V_{CKM}^{(0)})_{ji}}(V_{CKM}^{(0)})_{it}^\dag ]\nonumber\\
&&\hspace{1.5cm}
 + \frac{{{\upsilon ^2}}}{{4\Lambda _{KK}^2}}[\frac{{{m_c}}}{{{m_W}}}{(V_{CKM}^{(0)})_{ci}}{(\Delta _{{G^ \pm }}^L)_{it}} + \frac{{{m_t}}}{{{m_W}}}(\Delta _{{G^ \pm }}^L)_{ci}^\dag (V_{CKM}^{(0)})_{it}^\dag ]\;,\nonumber\\
&&
{(\Upsilon _{i,2}^{(b)})_{ct}} =  - {(V_{CKM}^{(0)})_{ci}}(V_{CKM}^{(0)})_{it}^\dag [\frac{{{m_{{d_i}}}}}{{m_{{W^ \pm }}^2}}{(\delta {M^u})_{ii}} + \frac{{{m_c}}}{{m_{\rm{W}}^2}}(\delta {M^d})_{ii}^*]\nonumber\\
&&\hspace{1.5cm}
 - \frac{\pi }{{\Lambda _{KK}^2}}({m_c}{m_{{d_i}}} + {m_c}{m_t}){(V_{CKM}^{(0)})_{ci}}(V_{CKM}^{(0)})_{it}^\dag \{ [\Sigma _{( +  + )}^G(1,1)] + [\Sigma _{( -  + )}^G(1,1)]\} \nonumber\\
&&\hspace{1.5cm}
 + \sum\limits_{j = 1}^3 {\frac{{{m_c}}}{{m_{\rm{W}}^2}}} {(V_{CKM}^{(0)})_{ci}}[{m_{{d_i}}}(\delta Z_L^d)_{ij}^\dag (V_{CKM}^{(0)})_{jt}^\dag  + {m_{{d_j}}}(V_{CKM}^{(0)})_{ij}^\dag (\delta Z_R^u)_{jt}^\dag ]\nonumber\\
&&\hspace{1.5cm}
 + \sum\limits_{j = 1}^3 [ \frac{{{m_{{d_i}}}{m_c}}}{{m_{\rm{W}}^2}}{(V_{CKM}^{(0)})_{cj}}{(\delta Z_L^d)_{ji}} + \frac{{{m_{{d_i}}}{m_{{u_j}}}}}{{m_{\rm{W}}^2}}{(\delta Z_R^u)_{cj}}{(V_{CKM}^{(0)})_{ji}}](V_{CKM}^{(0)})_{it}^\dag \nonumber\\
&&\hspace{1.5cm}
 + \frac{{{\upsilon ^2}}}{{4\Lambda _{KK}^2}}[\frac{{{m_c}}}{{{m_W}}}{(V_{CKM}^{(0)})_{ci}}{(\Delta _{{G^ \pm }}^R)_{it}} + \frac{{{m_{{d_i}}}}}{{{m_W}}}(\Delta _{{G^ \pm }}^L)_{ci}^\dag (V_{CKM}^{(0)})_{it}^\dag ]\;,\nonumber\\
&&
{(\Upsilon _{i,3}^{(b)})_{ct}} =  - {(V_{CKM}^{(0)})_{ci}}(V_{CKM}^{(0)})_{it}^\dag [\frac{{{m_{{d_i}}}}}{{m_{\rm{W}}^2}}(\delta {M^u})_{ii}^* + \frac{{{m_t}}}{{m_{\rm{W}}^2}}{(\delta {M^d})_{ii}}]\nonumber\\
&&\hspace{1.5cm}
 - \frac{\pi }{{\Lambda _{KK}^2}}({m_t}{m_{{d_i}}} + {m_c}{m_t}){(V_{CKM}^{(0)})_{ci}}(V_{CKM}^{(0)})_{it}^\dag \{ [\Sigma _{( +  + )}^G(1,1)] + [\Sigma _{( -  + )}^G(1,1)]\} \nonumber\\
&&\hspace{1.5cm}
 + \sum\limits_{j = 1}^3 {\frac{{{m_{{d_i}}}}}{{m_{\rm{W}}^2}}} {(V_{CKM}^{(0)})_{ci}}[{m_t}(\delta Z_L^d)_{ij}^\dag (V_{CKM}^{(0)})_{jt}^\dag  + {m_{{u_j}}}(V_{CKM}^{(0)})_{ij}^\dag (\delta Z_R^u)_{jt}^\dag ]\nonumber\\
&&\hspace{1.5cm}
 + \sum\limits_{j = 1}^3 [ \frac{{{m_t}{m_{{d_j}}}}}{{m_{\rm{W}}^2}}{(V_{CKM}^{(0)})_{cj}}{(\delta Z_L^d)_{ji}} + \frac{{{m_{{d_i}}}{m_t}}}{{m_{\rm{W}}^2}}{(\delta Z_R^u)_{cj}}{(V_{CKM}^{(0)})_{ji}}](V_{CKM}^{(0)})_{it}^\dag \nonumber\\
&&\hspace{1.5cm}
 + \frac{{{\upsilon ^2}}}{{4\Lambda _{KK}^2}}[\frac{{{m_{{d_i}}}}}{{{m_W}}}{(V_{CKM}^{(0)})_{ci}}{(\Delta _{{G^ \pm }}^L)_{it}} + \frac{{{m_t}}}{{{m_W}}}(\Delta _{{G^ \pm }}^R)_{ci}^\dag (V_{CKM}^{(0)})_{it}^\dag ]\;,\nonumber\\
&&
{(\Upsilon _{i,4}^{(b)})_{ct}} =  - {(V_{CKM}^{(0)})_{ci}}(V_{CKM}^{(0)})_{it}^\dag [\frac{{{m_{{d_i}}}}}{{m_{\rm{W}}^2}}{(\delta {M^d})_{ii}} + \frac{{{m_{{d_i}}}}}{{m_{\rm{W}}^2}}(\delta {M^d})_{ii}^*]\nonumber\\
&&\hspace{1.5cm}
 - \frac{\pi }{{\Lambda _{KK}^2}}({m_t}{m_{{d_i}}} + {m_c}{m_{{d_i}}}){(V_{CKM}^{(0)})_{ci}}(V_{CKM}^{(0)})_{it}^\dag \{ [\Sigma _{( +  + )}^G(1,1)] + [\Sigma _{( -  + )}^G(1,1)]\} \nonumber\\
&&\hspace{1.5cm}
 + \sum\limits_{j = 1}^3 {\frac{{{m_{{d_i}}}{m_{{d_j}}}}}{{m_{\rm{W}}^2}}} {(V_{CKM}^{(0)})_{ci}}[{(\delta Z_L^d)_{ij}} + (\delta Z_L^d)_{ij}^\dag ](V_{CKM}^{(0)})_{jt}^\dag \nonumber\\
&&\hspace{1.5cm}
 + \sum\limits_{j = 1}^3 {\frac{{m_{{d_i}}^2}}{{m_{\rm{W}}^2}}} [{(V_{CKM}^{(0)})_{ci}}(V_{CKM}^{(0)})_{ij}^\dag (\delta Z_R^u)_{it}^\dag  + {(\delta Z_R^u)_{cj}}{(V_{CKM}^{(0)})_{ji}}(V_{CKM}^{(0)})_{it}^\dag ]\nonumber\\
&&\hspace{1.5cm}
 + \frac{{{\upsilon ^2}}}{{4\Lambda _{KK}^2}}[\frac{{{m_{{d_i}}}}}{{{m_W}}}{(V_{CKM}^{(0)})_{ci}}{(\Delta _{{G^ \pm }}^R)_{it}} + \frac{{{m_{{d_i}}}}}{{{m_W}}}(\Delta _{{G^ \pm }}^R)_{ci}^\dag (V_{CKM}^{(0)})_{it}^\dag
 ]\;.
\label{eq1}
\end{eqnarray}

where $\delta Z_L^u, \delta Z_R^u, \delta Z_L^d, \delta Z_R^d$ are
presented in the equation (154) and (160) of Ref. \cite{ref39} too.

\subsubsection{Fig.(c)with SM propagators}

Similarly, the contributions from Fig.(c) with the zero mode of
neutral gauge bosons $Z,Z_{H_\alpha}$ and charge $2/3$ quarks are

\begin{eqnarray}
&&F_{TL}^{\gamma(c)}=\frac{{i{e^3}{Q^F}}}{{32{\pi
^2}{m_c}s_W^2c_W^2}}\Big(2 (\xi _Z^{R(2/3)})_{c,i }^\dag
{(\xi _Z^{L(2/3)})_{i ,t}} m_{i} \text{C}_1\nonumber\\
&&\hspace{2cm}+(\xi _Z^{R(2/3)})_{c,i }^\dag {(\xi _Z^{R(2/3)})_{i
,t}} m_t \text{C}_{12}-(\xi _Z^{L(2/3)})_{c,i }^\dag {(\xi
_Z^{L(2/3)})_{i ,t}} m_c
   (\text{C}_1-\text{C}_{11}+\text{C}_{12})\Big)\nonumber\\
&&F_{TR}^{\gamma(c)}=\frac{{i{e^3}{Q^F}}}{{32{\pi
^2}{m_t}s_W^2c_W^2}}\Big(2 (\xi _Z^{L(2/3)})_{c,i }^\dag
{(\xi _Z^{R(2/3)})_{i ,t}} m_{i} \text{C}_1\nonumber\\
&&\hspace{2cm}+(\xi _Z^{L(2/3)})_{c,i }^\dag {(\xi _Z^{L(2/3)})_{i
,t}} m_t \text{C}_{12}-(\xi _Z^{R(2/3)})_{c,i }^\dag {(\xi
_Z^{R(2/3)})_{i ,t}} m_c
   (\text{C}_1-\text{C}_{11}+\text{C}_{12})\Big)\nonumber\\\label{eq1}
\end{eqnarray}

with
$\text{C}_{ij}=\text{C}_{ij}(p^2,(2p-p')^2,(p-p')^2,m_{i}^2,m_Z^2,m_{i}^2)$

\begin{eqnarray}
&&F_{TL}^{g(c)}=\frac{{i{e^2}{g_S}{T^a}}}{{32{\pi
^2}{m_c}s_W^2c_W^2}}\Big(2 (\xi _Z^{R(2/3)})_{c,i }^\dag
{(\xi _Z^{L(2/3)})_{i ,t}} m_{i} \text{C}_1\nonumber\\
&&\hspace{2cm}+(\xi _Z^{R(2/3)})_{c,i }^\dag {(\xi _Z^{R(2/3)})_{i
,t}} m_t \text{C}_{12}-(\xi _Z^{L(2/3)})_{c,i }^\dag {(\xi
_Z^{L(2/3)})_{i ,t}} m_c
   (\text{C}_1-\text{C}_{11}+\text{C}_{12})\Big)\nonumber\\
&&F_{TR}^{g(c)}=\frac{{i{e^4}{g_S}{T^a}}}{{32{\pi
^2}{m_t}s_W^2c_W^2}}\Big(2 (\xi _Z^{L(2/3)})_{c,i }^\dag
{(\xi _Z^{R(2/3)})_{i ,t}} m_{i} \text{C}_1\nonumber\\
&&\hspace{2cm}+(\xi _Z^{L(2/3)})_{c,i }^\dag {(\xi _Z^{L(2/3)})_{i
,t}} m_t \text{C}_{12}-(\xi _Z^{R(2/3)})_{c,i }^\dag {(\xi
_Z^{R(2/3)})_{i ,t}} m_c
   (\text{C}_1-\text{C}_{11}+\text{C}_{12})\Big)\nonumber\\\label{eq1}
\end{eqnarray}

with
$\text{C}_{ij}=\text{C}_{ij}(p^2,(2p-p')^2,(p-p')^2,m_{i}^2,m_Z^2,m_{i}^2)$

and there is no contribution to the processes with zero mode of $\gamma$.

The relevant coefficients are approximate to the order ${\cal
O}(\upsilon^2/\Lambda_{KK}^2)$ as

\begin{eqnarray}
&&
(\xi _Z^{L(2/3)})_{c,i }^\dag {(\xi _Z^{L(2/3)})_{i ,t}} = {(\frac{{3 - 4s_{\rm{w}}^2}}{{6{s_{\rm{w}}}{{\rm{c}}_{\rm{w}}}}})^2}[{\delta _{ci }}((\delta Z_L^u)_{i t}^\dag  + {(\delta Z_L^u)_{i t}}) + ((\delta Z_L^u)_{ci }^\dag  + {(\delta Z_L^u)_{ci }}){\delta _{i t}}\nonumber\\
&&\hspace{3.3cm}
 + {\delta _{ci }}\frac{{{\upsilon ^2}}}{{2\Lambda _{KK}^2}}{(\Delta _Z^{L(2/3)})_{i t}} + \frac{{{\upsilon ^2}}}{{2\Lambda _{KK}^2}}(\Delta _Z^{L(2/3)})_{ci }^\dag {\delta _{i t}}] + O(\frac{{{\upsilon ^4}}}{{\Lambda _{KK}^4}})\;\nonumber\\
&&
(\xi _Z^{L(2/3)})_{c,i }^\dag {(\xi _Z^{R(2/3)})_{i ,t}} =  - \frac{2}{3}\frac{{{s_{\rm{w}}}}}{{{{\rm{c}}_{\rm{w}}}}}(\frac{{3 - 4s_{\rm{w}}^2}}{{6{s_{\rm{w}}}{{\rm{c}}_{\rm{w}}}}})[(\delta Z_L^u)_{ci }^\dag {\delta _{i t}} + {(\delta Z_L^u)_{ci }}{\delta _{i t}} + \frac{{{\upsilon ^2}}}{{2\Lambda _{KK}^2}}(\Delta _Z^{L(2/3)})_{ci }^\dag {\delta _{i t}}\nonumber\\
&&\hspace{3.3cm}
  + {\delta _{ci }}(\delta Z_R^u)_{i t}^\dag  + {\delta _{ci }}{(\delta Z_R^u)_{i t}} + {\delta _{ci }}\frac{{{\upsilon ^2}}}{{2\Lambda _{KK}^2}}{(\Delta _Z^{R(2/3)})_{i t}}] + O(\frac{{{\upsilon ^4}}}{{\Lambda _{KK}^4}})\;\nonumber\\
&&
(\xi _Z^{R(2/3)})_{c,i }^\dag {(\xi _Z^{R(2/3)})_{i ,t}} = \{ {( - \frac{2}{3}\frac{{{s_{\rm{w}}}}}{{{{\rm{c}}_{\rm{w}}}}})^2}[{\delta _{ci }}((\delta Z_R^u)_{i t}^\dag  + {(\delta Z_R^u)_{i t}}) + ((\delta Z_R^u)_{ci }^\dag  + {(\delta Z_R^u)_{ci }}){\delta _{i t}}\nonumber\\
&&\hspace{3.3cm}
 + {\delta _{ci }}\frac{{{\upsilon ^2}}}{{2\Lambda _{KK}^2}}{(\Delta _Z^{R(2/3)})_{i t}} + \frac{{{\upsilon ^2}}}{{2\Lambda _{KK}^2}}(\Delta _Z^{R(2/3)})_{ci }^\dag {\delta _{i t}}] + O(\frac{{{\upsilon ^4}}}{{\Lambda _{KK}^4}})\;\nonumber\\
&&
(\xi _Z^{R(2/3)})_{c,i }^\dag {(\xi _Z^{L(2/3)})_{i ,t}} =  - \frac{2}{3}\frac{{{s_{\rm{w}}}}}{{{{\rm{c}}_{\rm{w}}}}}(\frac{{3 - 4s_{\rm{w}}^2}}{{6{s_{\rm{w}}}{{\rm{c}}_{\rm{w}}}}})[{\delta _{ci }}(\delta Z_L^u)_{i t}^\dag  + {\delta _{ci }}{(\delta Z_L^u)_{i t}} + {\delta _{ci }}\frac{{{\upsilon ^2}}}{{2\Lambda _{KK}^2}}{(\Delta _Z^{L(2/3)})_{i t}}\nonumber\\
&&\hspace{3.3cm}
 + (\delta Z_R^u)_{ci }^\dag {\delta _{i t}} + {(\delta Z_R^u)_{ci }}{\delta _{i t}} + \frac{{{\upsilon ^2}}}{{2\Lambda _{KK}^2}}(\Delta _Z^{R(2/3)})_{ci }^\dag {\delta _{i t}}] + O(\frac{{{\upsilon ^4}}}{{\Lambda _{KK}^4}})\;\nonumber\\
\label{eq1}
\end{eqnarray}

\subsubsection{Fig.(d1) and (d2)  with SM propagators}

In Fig.(d1)(d2), the relevant coefficients ${(\xi
_{{g_{(n)}}}^{(2/3)})_{c,i}^\dag} (\xi _{{g_{(n)}}}^{(2/3)})_{i ,t}$
are at order $O(\frac{{{\upsilon ^4}}}{{\Lambda _{KK}^4}})$. So we
ignore them.

\subsubsection{Fig.(e)  with SM propagators}

Finally, the zero mode of neutral Higgs/Goldstone
$H_0,G_0$ and charge $2/3$ quarks contribution in Fig.(e) reads
\begin{eqnarray}
&&F_{TL}^{\gamma(e)}=\frac{{i{e}{Q^F}}}{{16{\pi ^2}{m_c}}}\Big((\eta
_{H_0/G_0}^{L(2/3)})_{c,i }^\dag {(\eta _{H_0/G_0}^{L(2/3)})_{i ,t}}
(\text{C}_0+\text{C}_1)
m_{i}\nonumber\\
&&\hspace{2cm}+(\eta _{H_0/G_0}^{R(2/3)})_{c,i }^\dag {(\eta
_{H_0/G_0}^{L(2/3)})_{i ,t}} (\text{C}_{12}-\text{C}_{11}) m_c-(\eta
_{H_0/G_0}^{L(2/3)})_{c,i }^\dag {(\eta _{H_0/G_0}^{R(2/3)})_{i ,t}}
   (\text{C}_{1}+\text{C}_{12}) m_t\Big)\nonumber\\
&&F_{TR}^{\gamma(e)}=\frac{{i{e}{Q^F}}}{{16{\pi ^2}{m_t}}}
\Big((\eta _{H_0/G_0}^{R(2/3)})_{c,i }^\dag {(\eta
_{H_0/G_0}^{R(2/3)})_{i ,t}}
(\text{C}_{0}+\text{C}_{1}) m_{i}\nonumber\\
&&\hspace{2cm}+(\eta _{H_0/G_0}^{L(2/3)})_{c,i }^\dag {(\eta
_{H_0/G_0}^{R(2/3)})_{i ,t}} (\text{C}_{12}-\text{C}_{11}) m_c-(\eta
_{H_0/G_0}^{R(2/3)})_{c,i }^\dag {(\eta _{H_0/G_0}^{L(2/3)})_{i ,t}}
   (\text{C}_{1}+\text{C}_{12}) m_t\Big)\nonumber\\\label{eq1}
\end{eqnarray}

with
$\text{C}_{ij}=\text{C}_{ij}(p^2,(2p-p')^2,(p-p')^2,m_{i}^2,m_{H_0/G_0}^2,m_{i}^2)$

\begin{eqnarray}
&&F_{TL}^{g(e)}=\frac{{i{g_S}{T^a}}}{{16{\pi ^2}{m_c}}}\Big((\eta
_{H_0/G_0}^{L(2/3)})_{c,i }^\dag {(\eta _{H_0/G_0}^{L(2/3)})_{i ,t}}
(\text{C}_0+\text{C}_1)
m_{i}\nonumber\\
&&\hspace{2cm}+(\eta _{H_0/G_0}^{R(2/3)})_{c,i }^\dag {(\eta
_{H_0/G_0}^{L(2/3)})_{i ,t}} (\text{C}_{12}-\text{C}_{11}) m_c-(\eta
_{H_0/G_0}^{L(2/3)})_{c,i }^\dag {(\eta _{H_0/G_0}^{R(2/3)})_{i ,t}}
   (\text{C}_{1}+\text{C}_{12}) m_t\Big)\nonumber\\
&&F_{TR}^{g(e)}=\frac{{i{g_S}{T^a}}}{{16{\pi ^2}{m_t}}}\Big((\eta
_{H_0/G_0}^{R(2/3)})_{c,i }^\dag {(\eta _{H_0/G_0}^{R(2/3)})_{i ,t}}
(\text{C}_{0}+\text{C}_{1}) m_{i}\nonumber\\
&&\hspace{2cm}+(\eta _{H_0/G_0}^{L(2/3)})_{c,i }^\dag {(\eta
_{H_0/G_0}^{R(2/3)})_{i ,t}} (\text{C}_{12}-\text{C}_{11}) m_c-(\eta
_{H_0/G_0}^{R(2/3)})_{c,i }^\dag {(\eta _{H_0/G_0}^{L(2/3)})_{i ,t}}
   (\text{C}_{1}+\text{C}_{12}) m_t\Big)\nonumber\\\label{eq1}
\end{eqnarray}

with
$\text{C}_{ij}=\text{C}_{ij}(p^2,(2p-p')^2,(p-p')^2,m_{i}^2,m_{H_0/G_0}^2,m_{i}^2)$

 The relevant coefficients can be expanded
according ${\cal O}(\upsilon^2/\Lambda_{KK}^2)$ as

\begin{eqnarray}
&&(\eta _{{H_0}}^{L(2/3)})_{c,i }^\dag {(\eta _{{H_0}}^{L(2/3)})_{i ,t}} =(\eta _{{G_0}}^{L(2/3)})_{c,i }^\dag {(\eta _{{G_0}}^{L(2/3)})_{i ,t}} = \frac{{{e^2}}}{{2s_{\rm{w}}^2}}\{ \frac{{{m_i }}}{{{m_{\rm{W}}}}}{\delta _{ci }}\frac{{{m_t}}}{{{m_{\rm{W}}}}}(\delta Z_R^u)_{i t}^\dag  + \frac{{{m_i }}}{{{m_{\rm{W}}}}}{\delta _{ci }}\frac{{{m_i }}}{{{m_{\rm{W}}}}}{(\delta Z_L^u)_{i t}}\nonumber\\
&&\hspace{3.3cm}
 + \frac{{{\upsilon ^2}}}{{4\Lambda _{KK}^2}}\frac{{{m_i }}}{{{m_{\rm{W}}}}}{\delta _{ci }}[\frac{{{m_t}}}{{{m_{\rm{W}}}}}{(\Delta _{{H_0}}^{(1)2/3})_{i t}} + \frac{{{m_i }}}{{{m_{\rm{W}}}}}{(\Delta _{{H_0}}^{(2)2/3})_{i t}}]\nonumber\\
&&\hspace{3.3cm}
 + \frac{{{m_c}}}{{{m_{\rm{W}}}}}{(\delta Z_R^u)_{ci }}\frac{{{m_t}}}{{{m_{\rm{W}}}}}{\delta _{i t}} + \frac{{{m_i }}}{{{m_{\rm{W}}}}}(\delta Z_L^u)_{ci }^\dag \frac{{{m_t}}}{{{m_{\rm{W}}}}}{\delta _{i t}} \nonumber\\
&&\hspace{3.3cm}+ \frac{{{\upsilon ^2}}}{{4\Lambda _{KK}^2}}\frac{{{m_t}}}{{{m_{\rm{W}}}}}{\delta _{i t}}[\frac{{{m_c}}}{{{m_{\rm{W}}}}}(\Delta _{{H_0}}^{(1)2/3})_{ci }^\dag  + \frac{{{m_i }}}{{{m_{\rm{W}}}}}(\Delta _{{H_0}}^{(2)2/3})_{ci }^\dag ]\}  + O(\frac{{{\upsilon ^3}}}{{\Lambda _{KK}^3}})\;\nonumber\\
&&
(\eta _{{H_0}}^{L(2/3)})_{c,i }^\dag {(\eta _{{H_0}}^{R(2/3)})_{i ,t}} =(\eta _{{G_0}}^{L(2/3)})_{c,i }^\dag {(\eta _{{G_0}}^{R(2/3)})_{i ,t}}= (\eta _{{H_0}}^{L(2/3)})_{c,i }^\dag (\eta _{{H_0}}^{L(2/3)})_{i ,t}^\dag  \nonumber\\
&&\hspace{3.3cm}= \frac{{{e^2}}}{{2s_{\rm{w}}^2}}\{ \frac{{{m_i }}}{{{m_{\rm{W}}}}}{\delta _{ci }}\frac{{{m_i }}}{{{m_{\rm{W}}}}}{(\delta Z_R^u)_{i t}} + \frac{{{m_i }}}{{{m_{\rm{W}}}}}{\delta _{ci }}\frac{{{m_t}}}{{{m_{\rm{W}}}}}(\delta Z_L^u)_{i t}^\dag \nonumber\\
&&\hspace{3.3cm}
 + \frac{{{\upsilon ^2}}}{{4\Lambda _{KK}^2}}\frac{{{m_i }}}{{{m_{\rm{W}}}}}{\delta _{ci }}[\frac{{{m_i }}}{{{m_{\rm{W}}}}}(\Delta _{{H_0}}^{(1)2/3})_{i t}^\dag  + \frac{{{m_t}}}{{{m_{\rm{W}}}}}(\Delta _{{H_0}}^{(2)2/3})_{i t}^\dag ]\nonumber\\
&&\hspace{3.3cm}
 + \frac{{{m_c}}}{{{m_{\rm{W}}}}}{(\delta Z_R^u)_{ci }}\frac{{{m_t}}}{{{m_{\rm{W}}}}}{\delta _{i t}} + \frac{{{m_i }}}{{{m_{\rm{W}}}}}(\delta Z_L^u)_{ci }^\dag \frac{{{m_t}}}{{{m_{\rm{W}}}}}{\delta _{i t}} \nonumber\\
&&\hspace{3.3cm}+ \frac{{{\upsilon ^2}}}{{4\Lambda _{KK}^2}}[\frac{{{m_c}}}{{{m_{\rm{W}}}}}(\Delta _{{H_0}}^{(1)2/3})_{ci }^\dag  + \frac{{{m_i }}}{{{m_{\rm{W}}}}}(\Delta _{{H_0}}^{(2)2/3})_{ci }^\dag ]\frac{{{m_t}}}{{{m_{\rm{W}}}}}{\delta _{i t}}\}  + O(\frac{{{\upsilon ^3}}}{{\Lambda _{KK}^3}})\;,\nonumber\\
&&
(\eta _{{H_0}}^{R(2/3)})_{c,i }^\dag {(\eta _{{H_0}}^{L(2/3)})_{i ,t}} =(\eta _{{G_0}}^{R(2/3)})_{c,i }^\dag {(\eta _{{G_0}}^{L(2/3)})_{i ,t}}= {(\eta _{{H_0}}^{L(2/3)})_{c,i }}{(\eta _{{H_0}}^{L(2/3)})_{i ,t}}\nonumber\\
&&\hspace{3.3cm} = \frac{{{e^2}}}{{2s_{\rm{w}}^2}}\{ \frac{{{m_i }}}{{{m_{\rm{W}}}}}{\delta _{ci }}\frac{{{m_t}}}{{{m_{\rm{W}}}}}(\delta Z_R^u)_{i t}^\dag  + \frac{{{m_i }}}{{{m_{\rm{W}}}}}{\delta _{ci }}\frac{{{m_i }}}{{{m_{\rm{W}}}}}{(\delta Z_L^u)_{i t}}\nonumber\\
&&\hspace{3.3cm}
 + \frac{{{\upsilon ^2}}}{{4\Lambda _{KK}^2}}\frac{{{m_i }}}{{{m_{\rm{W}}}}}{\delta _{ci }}[\frac{{{m_t}}}{{{m_{\rm{W}}}}}{(\Delta _{{H_0}}^{(1)2/3})_{i t}} + \frac{{{m_i }}}{{{m_{\rm{W}}}}}{(\Delta _{{H_0}}^{(2)2/3})_{i t}}]\nonumber\\
&&\hspace{3.3cm}
 + \frac{{{m_i }}}{{{m_{\rm{W}}}}}(\delta Z_R^u)_{ci }^\dag \frac{{{m_t}}}{{{m_{\rm{W}}}}}{\delta _{i t}} + \frac{{{m_c}}}{{{m_{\rm{W}}}}}{(\delta Z_L^u)_{ci }}\frac{{{m_t}}}{{{m_{\rm{W}}}}}{\delta _{i t}} \nonumber\\
&&\hspace{3.3cm}+ \frac{{{\upsilon ^2}}}{{4\Lambda _{KK}^2}}\frac{{{m_t}}}{{{m_{\rm{W}}}}}{\delta _{i t}}[\frac{{{m_i }}}{{{m_{\rm{W}}}}}{(\Delta _{{H_0}}^{(1)2/3})_{ci }} + \frac{{{m_c}}}{{{m_{\rm{W}}}}}{(\Delta _{{H_0}}^{(2)2/3})_{ci }}]\}  + O(\frac{{{\upsilon ^3}}}{{\Lambda _{KK}^3}})\;,\nonumber\\
&&
(\eta _{{H_0}}^{R(2/3)})_{c,i }^\dag {(\eta _{{H_0}}^{R(2/3)})_{i ,t}} =(\eta _{{G_0}}^{R(2/3)})_{c,i }^\dag {(\eta _{{G_0}}^{R(2/3)})_{i ,t}} = {(\eta _{{H_0}}^{L(2/3)})_{c,i }}(\eta _{{H_0}}^{L(2/3)})_{i ,t}^\dag \nonumber\\
&&\hspace{3.3cm} = \frac{{{e^2}}}{{2s_{\rm{w}}^2}}\{ \frac{{{m_i }}}{{{m_{\rm{W}}}}}{\delta _{ci }}\frac{{{m_i }}}{{{m_{\rm{W}}}}}{(\delta Z_R^u)_{i t}} + \frac{{{m_i }}}{{{m_{\rm{W}}}}}{\delta _{ci }}\frac{{{m_t}}}{{{m_{\rm{W}}}}}(\delta Z_L^u)_{i t}^\dag \nonumber\\
&&\hspace{3.3cm}
 + \frac{{{\upsilon ^2}}}{{4\Lambda _{KK}^2}}\frac{{{m_i }}}{{{m_{\rm{W}}}}}{\delta _{ci }}[\frac{{{m_i }}}{{{m_{\rm{W}}}}}(\Delta _{{H_0}}^{(1)2/3})_{i t}^\dag  + \frac{{{m_t}}}{{{m_{\rm{W}}}}}(\Delta _{{H_0}}^{(2)2/3})_{i t}^\dag ]\nonumber\\
&&\hspace{3.3cm}
 + \frac{{{m_i }}}{{{m_{\rm{W}}}}}(\delta Z_R^u)_{ci }^\dag \frac{{{m_t}}}{{{m_{\rm{W}}}}}{\delta _{i t}} + \frac{{{m_c}}}{{{m_{\rm{W}}}}}{(\delta Z_L^u)_{ci }}\frac{{{m_t}}}{{{m_{\rm{W}}}}}{\delta _{i t}}\nonumber\\
&&\hspace{3.3cm} + \frac{{{\upsilon ^2}}}{{4\Lambda _{KK}^2}}\frac{{{m_t}}}{{{m_{\rm{W}}}}}{\delta _{i t}}[\frac{{{m_i }}}{{{m_{\rm{W}}}}}{(\Delta _{{H_0}}^{(1)2/3})_{ci }} + \frac{{{m_c}}}{{{m_{\rm{W}}}}}{(\Delta _{{H_0}}^{(2)2/3})_{ci }}]\}  + O(\frac{{{\upsilon ^3}}}{{\Lambda _{KK}^3}})\;\nonumber\\
\label{eq1}
\end{eqnarray}

\subsection{The second case}

When the propagator is the KK mode, since the mass of the KK mode is
large then the top quark, we expand the amplitudes to the order
$m_t^2/\Lambda_{KK}^2$ for simplicity.

In a conventional form, the effective Hamilton is written as:
\begin{eqnarray}
&&H_{eff}=\sum_iC_i(\mu)\mathcal {O}_i\label{eq1}
\end{eqnarray}

with the magnetic and chromomagnetic dipole moment operators  are
defined through \cite{ref433}

\begin{eqnarray}
&&\mathcal
{O}_{7\gamma}=-{i\over2}m_t\overline{c}_\alpha\sigma^{\mu\nu} P_R
t_\alpha F_{\mu\nu} ,
\nonumber\\
&&\mathcal
{O}_{7\gamma}=-{i\over2}m_c\overline{c}_\alpha\sigma^{\mu\nu}
P_L t_\alpha F_{\mu\nu}, \nonumber\\
&&\mathcal {O}_{8g}=-{i\over2}m_t\overline{c}_\alpha
\sigma^{\mu\nu}P_R t_\beta G^a_{\mu\nu} , \nonumber\\
&&\mathcal {O}_{8g}=-{i\over2}m_c\overline{c}_\alpha
 \sigma^{\mu\nu}P_L t_\beta G^a_{\mu\nu} ,
\label{dipole-operators}
\end{eqnarray}

where $F_{\mu\nu}$ and $G_{\mu\nu}^a$ are the electromagnetic and
strong field strength tensors  respectively. And in the momentum
representation the operators have the same form as in equation (34).

\subsubsection{Fig.(a1)and (a2)  with KK mode propagators}

In Fig. (a1)and (a2), when one loop diagrams are composed by the KK
mode of  charged gauge bosons and the KK mode of charged $-1/3$
quarks, the corrections to the coefficients at the EW scale
$\mu_{_{\rm EW}}$ are formulated as
\begin{eqnarray}
&&C_{7\gamma}^{(a)}={ie^3\over16\pi^2\mu_{_{\rm EW}}^2 s_{_{\rm
W}}^2}\sum\limits_{\beta=1}^\infty\Big\{\Big(\xi_{W^\pm}^{L(-1/3)}\Big)_{_{c,\beta}}^\dagger
\Big(\xi_{W^\pm}^{L(-1/3)}\Big)_{_{\beta,t}}F_{1,\gamma}^{(a)}(x_{_{D_\beta}},x_{_{W^\pm}})
\nonumber\\
&&\hspace{3.0cm} +{m_{_{D_\beta}}\over
m_t}\Big(\xi_{W^\pm}^{L(-1/3)}\Big)_{_{c,\beta}}^\dagger\Big(\xi_{W^\pm}^{R(-1/3)}\Big)_{_{\beta,t}}
F_{2,\gamma}^{(a)}(x_{_{D_\beta}},x_{_{W^\pm}})
\nonumber\\
&&\hspace{3.0cm}
+\sum\limits_{\alpha=1}^\infty\Big(\xi_{W_{H_\alpha}^\pm}^{L(-1/3)}\Big)_{_{c,\beta}}^\dagger
\Big(\xi_{W_{H_\alpha}^\pm}^{L(-1/3)}\Big)_{_{\beta,t}}F_{1,\gamma}^{(a)}(x_{_{D_\beta}},x_{_{W_{H_\alpha}^\pm}})
\nonumber\\
&&\hspace{3.0cm} +{m_{_{D_\beta}}\over
m_t}\sum\limits_{\alpha=1}^\infty
\Big(\xi_{W_{H_\alpha}^\pm}^{L(-1/3)}\Big)_{_{c,\beta}}^\dagger\Big(\xi_{W_{H_\alpha}^\pm}^{R(-1/3)}\Big)_{_{\beta,t}}
F_{2,\gamma}^{(a)}(x_{_{D_\beta}},x_{_{W_{H_\alpha}^\pm}})\Big\}\;,
\nonumber\\
&&C_{8G}^{(a)}={ie^2g_sT^a\over16\pi^2\mu_{_{\rm EW}}^2 s_{_{\rm
W}}^2}\sum\limits_{\beta=1}^\infty\Big\{\Big(\xi_{W^\pm}^{L(-1/3)}\Big)_{_{c,\beta}}^\dagger
\Big(\xi_{W^\pm}^{L(-1/3)}\Big)_{_{\beta,t}}F_{1,g}^{(a)}(x_{_{D_\beta}},x_{_{W^\pm}})
\nonumber\\
&&\hspace{3.0cm} +{m_{_{D_\beta}}\over
m_t}\Big(\xi_{W^\pm}^{L(-1/3)}\Big)_{_{c,\beta}}^\dagger\Big(\xi_{W^\pm}^{R(-1/3)}\Big)_{_{\beta,t}}
F_{2,g}^{(a)}(x_{_{D_\beta}},x_{_{W^\pm}})
\nonumber\\
&&\hspace{3.0cm}
+\sum\limits_{\alpha=1}^\infty\Big(\xi_{W_{H_\alpha}^\pm}^{L(-1/3)}\Big)_{_{c,\beta}}^\dagger
\Big(\xi_{W_{H_\alpha}^\pm}^{L(-1/3)}\Big)_{_{\beta,t}}F_{1,g}^{(a)}
(x_{_{D_\beta}},x_{_{W_{H_\alpha}^\pm}})
\nonumber\\
&&\hspace{3.0cm} +{m_{_{D_\beta}}\over
m_t}\sum\limits_{\alpha=1}^\infty
\Big(\xi_{W_{H_\alpha}^\pm}^{L(-1/3)}\Big)_{_{c,\beta}}^\dagger\Big(\xi_{W_{H_\alpha}^\pm}^{R(-1/3)}\Big)_{_{\beta,t}}
F_{2,g}^{(a)}(x_{_{D_\beta}},x_{_{W_{H_\alpha}^\pm}})\Big\}\;,
\nonumber\\
&&\widetilde{C}_{7\gamma}^{(a)}=
C_{7\gamma}^{(a)}\Big(\xi_{W^\pm}^{L(-1/3)}\leftrightarrow
\xi_{W^\pm}^{R(-1/3)},\;\xi_{W_{H_\alpha}^\pm}^{L(-1/3)}\leftrightarrow\xi_{W_{H_\alpha}^\pm}^{R(-1/3)}\Big)\;,
\nonumber\\
&&\widetilde{C}_{8G}^{(a)}=C_{8G}^{(a)}
\Big(\xi_{W^\pm}^{L(-1/3)}\leftrightarrow
\xi_{W^\pm}^{R(-1/3)},\;\xi_{W_{H_\alpha}^\pm}^{L(-1/3)}\leftrightarrow\xi_{W_{H_\alpha}^\pm}^{R(-1/3)}\Big)
\label{Hamilton-a}
\end{eqnarray}
with $x_i=m_i^2/\mu_{_{\rm EW}}^2$, and $\beta=1,2...\infty$ is the
index of all the KK exciting modes and all the three quark
generations in each KK exciting modes. And the form factors are
explicitly given by
\begin{eqnarray}
&&F_{1,\gamma}^{(a)}(x,y)=\Big[{1\over18}{\partial^3\varrho_{_{3,1}}\over\partial
y^3}-{1\over4}{\partial^2\varrho_{_{2,1}}\over\partial y^2}
-{5\over6}{\partial\varrho_{_{1,1}}\over\partial y}\Big](x,y)\;,
\nonumber\\
&&F_{2,\gamma}^{(a)}(x,y)=
\Big[-{2\over3}{\partial^2\varrho_{_{2,1}}\over\partial y^2}
+{10\over3}{\partial\varrho_{_{1,1}}\over\partial y}\Big](x,y)\;,
\nonumber\\
&&F_{1,g}^{(a)}(x,y)=
\Big[{1\over12}{\partial^3\varrho_{_{3,1}}\over\partial
y^3}-{1\over2}{\partial\varrho_{_{1,1}}\over\partial y}\Big](x,y)\;,
\nonumber\\
&&F_{2,g}^{(a)}(x,y)= \Big[-{\partial^2\varrho_{_{2,1}}\over\partial
y^2} +2{\partial\varrho_{_{1,1}}\over\partial y}\Big](x,y)\;.
\label{form-(a)}
\end{eqnarray}
Here, the function $\varrho_{_{m,n}}(x,y)$ is defined through
\begin{eqnarray}
&&\varrho_{_{m,n}}(x,y)={x^m\ln^nx-y^m\ln^ny\over x-y}\;.
\label{varrh0}
\end{eqnarray}

 When  Fig.(a1)and (a2), are composed by the charged $5/3$ quarks, the
corrections to the coefficients at the EW scale $\mu_{_{\rm EW}}$
are formulated as

\begin{eqnarray}
&&C_{7\gamma}^{(a)\frac{5}{3}}={ie^3\over16\pi^2\mu_{_{\rm EW}}^2
s_{_{\rm
W}}^2}\sum\limits_{\beta=1}^\infty\Big\{\Big(\xi_{W^\pm}^{L(5/3)}\Big)_{_{c,\beta}}^\dagger
\Big(\xi_{W^\pm}^{L(5/3)}\Big)_{_{\beta,t}}F_{1,\gamma}^{(a)\frac{5}{3}}(x_{_{H_\beta}},x_{_{W^\pm}})
\nonumber\\
&&\hspace{3.0cm} +{m_{_{H_\beta}}\over
m_t}\Big(\xi_{W^\pm}^{L(5/3)}\Big)_{_{c,\beta}}^\dagger\Big(\xi_{W^\pm}^{R(5/3)}\Big)_{_{\beta,t}}
F_{2,\gamma}^{(a)\frac{5}{3}}(x_{_{H_\beta}},x_{_{W^\pm}})
\nonumber\\
&&\hspace{3.0cm}
+\sum\limits_{\alpha=1}^\infty\Big(\xi_{W_{H_\alpha}^\pm}^{L(5/3)}\Big)_{_{c,\beta}}^\dagger
\Big(\xi_{W_{H_\alpha}^\pm}^{L(5/3)}\Big)_{_{\beta,t}}F_{1,\gamma}^{(a)\frac{5}{3}}(x_{_{H_\beta}},x_{_{W_{H_\alpha}^\pm}})
\nonumber\\
&&\hspace{3.0cm} +{m_{_{H_\beta}}\over
m_t}\sum\limits_{\alpha=1}^\infty
\Big(\xi_{W_{H_\alpha}^\pm}^{L(5/3)}\Big)_{_{c,\beta}}^\dagger\Big(\xi_{W_{H_\alpha}^\pm}^{R(5/3)}\Big)_{_{\beta,t}}
F_{2,\gamma}^{(a)\frac{5}{3}}(x_{_{H_\beta}},x_{_{W_{H_\alpha}^\pm}})\Big\}\;,
\nonumber\\
&&C_{8G}^{(a)\frac{5}{3}}={ie^2g_sT^a\over16\pi^2\mu_{_{\rm EW}}^2
s_{_{\rm
W}}^2}\sum\limits_{\beta=1}^\infty\Big\{\Big(\xi_{W^\pm}^{L(5/3)}\Big)_{_{c,\beta}}^\dagger
\Big(\xi_{W^\pm}^{L(5/3)}\Big)_{_{\beta,t}}F_{1,g}^{(a)\frac{5}{3}}(x_{_{H_\beta}},x_{_{W^\pm}})
\nonumber\\
&&\hspace{3.0cm} +{m_{_{H_\beta}}\over
m_t}\Big(\xi_{W^\pm}^{L(5/3)}\Big)_{_{c,\beta}}^\dagger\Big(\xi_{W^\pm}^{R(5/3)}\Big)_{_{\beta,t}}
F_{2,g}^{(a)\frac{5}{3}}(x_{_{H_\beta}},x_{_{W^\pm}})
\nonumber\\
&&\hspace{3.0cm}
+\sum\limits_{\alpha=1}^\infty\Big(\xi_{W_{H_\alpha}^\pm}^{L(5/3)}\Big)_{_{c,\beta}}^\dagger
\Big(\xi_{W_{H_\alpha}^\pm}^{L(5/3)}\Big)_{_{\beta,t}}F_{1,g}^{(a)\frac{5}{3}}
(x_{_{H_\beta}},x_{_{W_{H_\alpha}^\pm}})
\nonumber\\
&&\hspace{3.0cm} +{m_{_{H_\beta}}\over
m_t}\sum\limits_{\alpha=1}^\infty
\Big(\xi_{W_{H_\alpha}^\pm}^{L(5/3)}\Big)_{_{c,\beta}}^\dagger\Big(\xi_{W_{H_\alpha}^\pm}^{R(5/3)}\Big)_{_{\beta,t}}
F_{2,g}^{(a)\frac{5}{3}}(x_{_{H_\beta}},x_{_{W_{H_\alpha}^\pm}})\Big\}\;,
\nonumber\\
&&\widetilde{C}_{7\gamma}^{(a)\frac{5}{3}}=
C_{7\gamma}^{(a)\frac{5}{3}}\Big(\xi_{W^\pm}^{L(5/3)}\leftrightarrow
\xi_{W^\pm}^{R(5/3)},\;\xi_{W_{H_\alpha}^\pm}^{L(5/3)}\leftrightarrow\xi_{W_{H_\alpha}^\pm}^{R(5/3)}\Big)\;,
\nonumber\\
&&\widetilde{C}_{8G}^{(a)\frac{5}{3}}=C_{8G}^{(a)\frac{5}{3}}
\Big(\xi_{W^\pm}^{L(5/3)}\leftrightarrow
\xi_{W^\pm}^{R(5/3)},\;\xi_{W_{H_\alpha}^\pm}^{L(5/3)}\leftrightarrow\xi_{W_{H_\alpha}^\pm}^{R(5/3)}\Big)
\label{Hamilton-a}
\end{eqnarray}
And the form factors are

\begin{eqnarray}
&&F_{1,\gamma}^{(a)\frac{5}{3}}(x,y)=\Big[{5\over36}{\partial^3\varrho_{_{3,1}}\over\partial
y^3}-{1\over4}{\partial^2\varrho_{_{2,1}}\over\partial y^2}
-{4\over3}{\partial\varrho_{_{1,1}}\over\partial y}\Big](x,y)\;,
\nonumber\\
&&F_{2,\gamma}^{(a)\frac{5}{3}}(x,y)=
\Big[-{5\over3}{\partial^2\varrho_{_{2,1}}\over\partial y^2}
+{16\over3}{\partial\varrho_{_{1,1}}\over\partial y}\Big](x,y)\;,
\nonumber\\
&&F_{1,g}^{(a)\frac{5}{3}}(x,y)=
\Big[{1\over12}{\partial^3\varrho_{_{3,1}}\over\partial
y^3}-{1\over2}{\partial\varrho_{_{1,1}}\over\partial y}\Big](x,y)\;,
\nonumber\\
&&F_{2,g}^{(a)\frac{5}{3}}(x,y)=
\Big[-{\partial^2\varrho_{_{2,1}}\over\partial y^2}
+2{\partial\varrho_{_{1,1}}\over\partial y}\Big](x,y)\;.
\label{11}
\end{eqnarray}

Using $\xi^{L,R} $ in appendix, we could see that in Fig.(a1)(a2),
 the relevant coefficients $C_{7\gamma}^{(a)}$ and $C_{8g}^{(a)}$ are
at order $O(\frac{{{\upsilon ^4}}}{{\Lambda _{KK}^4}})$, so we
ignore them.

\subsubsection{Fig.(b1)and (b2)  with KK mode propagators}

Similarly, we can write down the corrections to Wilson coefficients
at the EW scale $\mu_{_{\rm EW}}$ from Fig.(b1),(b2), which are
composed by the KK mode of  charged Goldstone and the KK mode of
charged $-1/3$ quarks
\begin{eqnarray}
&&C_{7\gamma}^{(b)}={ie\over8\pi^2\mu_{_{\rm EW}}^2}
\sum\limits_{\beta=1}^\infty\Big\{\Big(\eta_{G^\pm}^{L(-1/3)}\Big)_{_{c,\beta}}^\dagger
\Big(\eta_{G^\pm}^{L(-1/3)}\Big)_{_{\beta,t}}F_{1,\gamma}^{(b)}(x_{_{D_\beta}},x_{_{W^\pm}})
\nonumber\\
&&\hspace{3.0cm} +{m_{_{D_\beta}}\over
m_t}\Big(\eta_{G^\pm}^{L(-1/3)}\Big)_{_{c,\beta}}^\dagger\Big(\eta_{G^\pm}^{R(-1/3)}\Big)_{_{\beta,t}}
F_{2,\gamma}^{(b)}(x_{_{D_\beta}},x_{_{W^\pm}})\Big\}\;,
\nonumber\\
&&C_{8G}^{(b)}={ig_sT^a\over8\pi^2\mu_{_{\rm EW}}^2}
\sum\limits_{\beta=1}^\infty\Big\{\Big(\eta_{G^\pm}^{L(-1/3)}\Big)_{_{c,\beta}}^\dagger
\Big(\eta_{G^\pm}^{L(-1/3)}\Big)_{_{\beta,t}}F_{1,g}^{(b)}(x_{_{D_\beta}},x_{_{W^\pm}})
\nonumber\\
&&\hspace{3.0cm} +{m_{_{D_\beta}}\over
m_t}\Big(\eta_{G^\pm}^{L(-1/3)}\Big)_{_{c,\beta}}^\dagger\Big(\eta_{G^\pm}^{R(-1/3)}\Big)_{_{\beta,t}}
F_{2,g}^{(b)}(x_{_{D_\beta}},x_{_{W^\pm}})\Big\}\;,
\nonumber\\
&&\widetilde{C}_{7\gamma}^{(b)}=
C_{7\gamma}^{(b)}\Big(\eta_{G^\pm}^{L(-1/3)}\leftrightarrow
\eta_{G^\pm}^{R(-1/3)}\Big)\;,
\nonumber\\
&&\widetilde{C}_{8G}^{(b)}=C_{8G}^{(b)}\Big(\eta_{G^\pm}^{L(-1/3)}\leftrightarrow\eta_{G^\pm}^{R(-1/3)}\Big)\;,
\label{Hamilton-b}
\end{eqnarray}
and those form factors are given by
\begin{eqnarray}
&&F_{1,\gamma}^{(b)}(x,y)=\Big[{1\over36}{\partial^3\varrho_{_{3,1}}\over\partial
y^3}-{7\over24}{\partial^2\varrho_{_{2,1}}\over\partial y^2}
+{5\over12}{\partial\varrho_{_{1,1}}\over\partial y}\Big](x,y)\;,
\nonumber\\
&&F_{2,\gamma}^{(b)}(x,y)=
\Big[-{1\over6}{\partial^2\varrho_{_{2,1}}\over\partial y^2}
+{1\over3}{\partial\varrho_{_{1,1}}\over\partial y}
-{5\over6}{\partial\varrho_{_{1,1}}\over\partial x}\Big](x,y)\;,
\nonumber\\
&&F_{1,g}^{(b)}(x,y)=
\Big[{1\over24}{\partial^3\varrho_{_{3,1}}\over\partial y^3}
-{1\over4}{\partial^2\varrho_{_{2,1}}\over\partial y^2}
+{1\over4}{\partial\varrho_{_{1,1}}\over\partial y}\Big](x,y)\;,
\nonumber\\
&&F_{2,g}^{(b)}(x,y)=\Big[-{1\over4}{\partial^2\varrho_{_{2,1}}\over\partial
y^2} +{1\over2}{\partial\varrho_{_{1,1}}\over\partial y}
-{1\over2}{\partial\varrho_{_{1,1}}\over\partial x}\Big](x,y)\;.
\label{form-(b)}
\end{eqnarray}

And the coefficients in Fig.(b1),(b2) composed by the KK mode of
charged $5/3$ quarks are

\begin{eqnarray}
&&C_{7\gamma}^{(b)\frac{5}{3}}={ie\over8\pi^2\mu_{_{\rm EW}}^2}
\sum\limits_{\beta=1}^\infty\Big\{\Big(\eta_{G^\pm}^{L(5/3)}\Big)_{_{c,\beta}}^\dagger
\Big(\eta_{G^\pm}^{L(5/3)}\Big)_{_{\beta,t}}F_{1,\gamma}^{(b)\frac{5}{3}}(x_{_{H_\beta}},x_{_{W^\pm}})
\nonumber\\
&&\hspace{3.0cm} +{m_{_{H_\beta}}\over
m_t}\Big(\eta_{G^\pm}^{L(5/3)}\Big)_{_{c,\beta}}^\dagger\Big(\eta_{G^\pm}^{R(5/3)}\Big)_{_{\beta,t}}
F_{2,\gamma}^{(b)\frac{5}{3}}(x_{_{H_\beta}},x_{_{W^\pm}})\Big\}\;,
\nonumber\\
&&C_{8G}^{(b)\frac{5}{3}}={ig_sT^a\over8\pi^2\mu_{_{\rm EW}}^2}
\sum\limits_{\beta=1}^\infty\Big\{\Big(\eta_{G^\pm}^{L(5/3)}\Big)_{_{c,\beta}}^\dagger
\Big(\eta_{G^\pm}^{L(5/3)}\Big)_{_{\beta,t}}F_{1,g}^{(b)\frac{5}{3}}(x_{_{H_\beta}},x_{_{W^\pm}})
\nonumber\\
&&\hspace{3.0cm} +{m_{_{H_\beta}}\over
m_t}\Big(\eta_{G^\pm}^{L(5/3)}\Big)_{_{c,\beta}}^\dagger\Big(\eta_{G^\pm}^{R(5/3)}\Big)_{_{\beta,t}}
F_{2,g}^{(b)\frac{5}{3}}(x_{_{H_\beta}},x_{_{W^\pm}})\Big\}\;,
\nonumber\\
&&\widetilde{C}_{7\gamma}^{(b)\frac{5}{3}}=
C_{7\gamma}^{(b)\frac{5}{3}}\Big(\eta_{G^\pm}^{L(5/3)}\leftrightarrow
\eta_{G^\pm}^{R(5/3)}\Big)\;,
\nonumber\\
&&\widetilde{C}_{8G}^{(b)\frac{5}{3}}=C_{8G}^{(b)\frac{5}{3}}\Big(\eta_{G^\pm}^{L(5/3)}\leftrightarrow\eta_{G^\pm}^{R(5/3)}\Big)\;,
\label{Hamilton-b}
\end{eqnarray}
and those form factors are given by
\begin{eqnarray}
&&F_{1,\gamma}^{(b)\frac{5}{3}}(x,y)=\Big[{5\over72}{\partial^3\varrho_{_{3,1}}\over\partial
y^3}-{13\over24}{\partial^2\varrho_{_{2,1}}\over\partial y^2}
+{2\over3}{\partial\varrho_{_{1,1}}\over\partial y}\Big](x,y)\;,
\nonumber\\
&&F_{2,\gamma}^{(b)\frac{5}{3}}(x,y)=
\Big[-{5\over12}{\partial^2\varrho_{_{2,1}}\over\partial y^2}
+{5\over6}{\partial\varrho_{_{1,1}}\over\partial y}
-{4\over3}{\partial\varrho_{_{1,1}}\over\partial x}\Big](x,y)\;,
\nonumber\\
&&F_{1,g}^{(b)\frac{5}{3}}(x,y)=
\Big[{1\over24}{\partial^3\varrho_{_{3,1}}\over\partial y^3}
-{1\over4}{\partial^2\varrho_{_{2,1}}\over\partial y^2}
+{1\over4}{\partial\varrho_{_{1,1}}\over\partial y}\Big](x,y)\;,
\nonumber\\
&&F_{2,g}^{(b)\frac{5}{3}}(x,y)=\Big[-{1\over4}{\partial^2\varrho_{_{2,1}}\over\partial
y^2} +{1\over2}{\partial\varrho_{_{1,1}}\over\partial y}
-{1\over2}{\partial\varrho_{_{1,1}}\over\partial x}\Big](x,y)\;.
\label{form}
\end{eqnarray}

The corrections to relevant Wilson coefficients with
the KK exciting modes of virtual charged $-1/3$ quarks, are
analogously formulated to the order ${\cal
O}(\upsilon^2/\Lambda_{KK}^2)$ as
\begin{eqnarray}
&&C_{7\gamma}^{(b)}=
{11ie\over576\pi^2\Lambda_{KK}^2}\sum\limits_{i,j,k=1}^3(
U_L^{(0)})_{ci}^\dag [f_{( +  + )}^{L,c_B^i}(0,1)]Y_{ik}^d[\Sigma
_{( \mp  \mp )}^{R,c_T^k}(1,1)]Y_{kj}^{d\dag }[f_{( +  +
)}^{L,c_B^j}(0,1)]{(U_L^{(0)})_{jt}} +{\cal
O}({\upsilon^4\over\Lambda_{KK}^4})\;,
\nonumber\\
&&C_{8G}^{(b)}={ig_sT^a\over192\pi^2\Lambda_{KK}^2}\sum\limits_{i,j,k=1}^3(
U_L^{(0)})_{ci}^\dag [f_{( +  + )}^{L,c_B^i}(0,1)]Y_{ik}^d[\Sigma
_{( \mp  \mp )}^{R,c_T^k}(1,1)]Y_{kj}^{d\dag }[f_{( +  +
)}^{L,c_B^j}(0,1)]{(U_L^{(0)})_{jt}} +{\cal
O}({\upsilon^4\over\Lambda_{KK}^4})\;,
\nonumber\\
&&\widetilde{C}_{7\gamma}^{(b)}={11ie\over576\pi^2\Lambda_{KK}^2}\sum\limits_{i,j,k=1}^3
( U_R^{(0)})_{ci}^\dag [f_{( +  + )}^{R,c_S^i}(0,1)]Y_{ik}^{u\dag
}[\Sigma _{( \pm  \pm  )}^{L,c_B^k}(1,1)]Y_{kj}^u[f_{( +  +
)}^{R,c_S^j}(0,1)]{(U_R^{(0)})_{jt}} +{\cal
O}({\upsilon^4\over\Lambda_{KK}^4})\;,
\nonumber\\
&&\widetilde{C}_{8G}^{(b)}={ig_sT^a\over192\pi^2\Lambda_{KK}^2}\sum\limits_{i,j,k=1}^3
( U_R^{(0)})_{ci}^\dag [f_{( +  + )}^{R,c_S^i}(0,1)]Y_{ik}^{u\dag
}[\Sigma _{( \pm  \pm  )}^{L,c_B^k}(1,1)]Y_{kj}^u[f_{( +  +
)}^{R,c_S^j}(0,1)]{(U_R^{(0)})_{jt}} +{\cal
O}({\upsilon^4\over\Lambda_{KK}^4})\;.
\nonumber\\\label{Hamilton-b-E}
\end{eqnarray}

and for charge $5/3$ quarks
\begin{eqnarray}
&&C_{7\gamma}^{(b)\frac{5}{3}}=
{7ie\over288\pi^2\Lambda_{KK}^2}\sum\limits_{i,j,k=1}^3 (
U_L^{(0)})_{ci}^\dag [f_{( +  + )}^{L,c_B^i}(0,1)]Y_{ik}^{d
}[2\Sigma _{( \pm  \mp  )}^{R,c_T^k}(1,1)]Y_{kj}^{d\dag}[f_{( +  +
)}^{L,c_B^j}(0,1)]{(U_L^{(0)})_{jt}} +{\cal
O}({\upsilon^4\over\Lambda_{KK}^4})\;,
\nonumber\\
&&C_{8G}^{(b)\frac{5}{3}}={ig_sT^a\over192\pi^2\Lambda_{KK}^2}\sum\limits_{i,j,k=1}^3
( U_L^{(0)})_{ci}^\dag [f_{( +  + )}^{L,c_B^i}(0,1)]Y_{ik}^{d
}[2\Sigma _{( \pm  \mp  )}^{R,c_T^k}(1,1)]Y_{kj}^{d\dag}[f_{( +  +
)}^{L,c_B^j}(0,1)]{(U_L^{(0)})_{jt}} +{\cal
O}({\upsilon^4\over\Lambda_{KK}^4})\;,
\nonumber\\
&&\widetilde{C}_{7\gamma}^{(b)\frac{5}{3}}={7ie\over288\pi^2\Lambda_{KK}^2}\sum\limits_{i,j,k=1}^3
( U_R^{(0)})_{ci}^\dag [f_{( +  + )}^{R,c_S^i}(0,1)]Y_{ik}^{u\dag
}[\Sigma _{( \mp  \pm )}^{L,c_B^k}(1,1)]Y_{kj}^{u}[f_{( +  +
)}^{R,c_S^j}(0,1)]{(U_R^{(0)})_{jt}} +{\cal
O}({\upsilon^4\over\Lambda_{KK}^4})\;,
\nonumber\\
&&\widetilde{C}_{8G}^{(b)\frac{5}{3}}={ig_sT^a\over192\pi^2\Lambda_{KK}^2}\sum\limits_{i,j,k=1}^3
( U_R^{(0)})_{ci}^\dag [f_{( +  + )}^{R,c_S^i}(0,1)]Y_{ik}^{u\dag
}[\Sigma _{( \mp  \pm )}^{L,c_B^k}(1,1)]Y_{kj}^{u}[f_{( +  +
)}^{R,c_S^j}(0,1)]{(U_R^{(0)})_{jt}} +{\cal
O}({\upsilon^4\over\Lambda_{KK}^4})\;.\nonumber\\
\label{Hamilton-b-E}
\end{eqnarray}

\subsubsection{Fig.(c)  with KK mode propagators}

For the Feynman diagrams drawn in Fig.(c), intermediate virtual
particles involve the KK mode of neutral gauge bosons
$Z,\;Z_{H_\alpha},\;\gamma_{(n)}$, and  the KK mode of charged $2/3$
quarks, the corresponding corrections to Wilson coefficients at
electroweak scale are expressed as
\begin{eqnarray}
&&C_{7\gamma}^{(c)}={ie^3\over48\pi^2\mu_{_{\rm EW}}^2 s_{_{\rm
W}}^2c_{_{\rm
W}}^2}\sum\limits_{\beta=1}^\infty\Big\{\Big(\xi_{Z}^{L(2/3)}\Big)_{_{c,\beta}}^\dagger
\Big(\xi_{Z}^{L(2/3)}\Big)_{_{\beta,t}}F_{1,g}^{(a)}(x_{_{U_\beta}},x_{_Z})
\nonumber\\
&&\hspace{3.0cm} +{m_{_{U_\beta}}\over
m_t}\Big(\xi_{Z}^{L(2/3)}\Big)_{_{c,\beta}}^\dagger\Big(\xi_{Z}^{R(2/3)}\Big)_{_{\beta,t}}
F_{2,g}^{(a)}(x_{_{U_\beta}},x_{_{Z}})
\nonumber\\
&&\hspace{3.0cm}
+\sum\limits_{\alpha=1}^\infty\Big(\xi_{Z_{H_\alpha}}^{L(2/3)}\Big)_{_{c,\beta}}^\dagger
\Big(\xi_{Z_{H_\alpha}}^{L(2/3)}\Big)_{_{\beta,t}}F_{1,g}^{(a)}(x_{_{U_\beta}},x_{_{Z_{H_\alpha}}})
\nonumber\\
&&\hspace{3.0cm} +{m_{_{U_\beta}}\over
m_t}\sum\limits_{\alpha=1}^\infty
\Big(\xi_{Z_{H_\alpha}}^{L(2/3)}\Big)_{_{c,\beta}}^\dagger\Big(\xi_{Z_{H_\alpha}}^{R(2/3)}\Big)_{_{\beta,t}}
F_{2,g}^{(a)}(x_{_{U_\beta}},x_{_{Z_{H_\alpha}}})\Big\}
\nonumber\\
&&\hspace{3.0cm} {2e^2\over27\mu_{_{\rm
EW}}^2}\sum\limits_{n=1}^\infty\sum\limits_{\beta=1}^\infty\Big\{
\Big(\xi_{\gamma_{(n)}}^{L(2/3)}\Big)_{_{c,\beta}}^\dagger
\Big(\xi_{\gamma_{(n)}}^{L(2/3)}\Big)_{_{\beta,t}}F_{1,g}^{(a)}(x_{_{U_\beta}},x_{_{\gamma_{(n)}}})
\nonumber\\
&&\hspace{3.0cm} +{m_{_{U_\beta}}\over
m_t}\Big(\xi_{\gamma_{(n)}}^{L(2/3)}\Big)_{_{c,\beta}}^\dagger\Big(\xi_{\gamma_{(n)}}^{R(2/3)}\Big)_{_{\beta,t}}
F_{2,g}^{(a)}(x_{_{U_\beta}},x_{_{\gamma_{(n)}}})\Big\}\;,
\nonumber\\
&&C_{8G}^{(c)}={3g_sT^a\over2e}C_{7\gamma}^{(c)}\;,
\nonumber\\
&&\widetilde{C}_{7\gamma}^{(c)}=
C_{7\gamma}^{(c)}\Big(\xi_{Z}^{L(2/3)}\leftrightarrow
\xi_{Z}^{R(2/3)},\;\xi_{Z_{H_\alpha}}^{L(2/3)}\leftrightarrow\xi_{Z_{H_\alpha}}^{R(2/3)},\;
\xi_{\gamma_{(n)}}^{L(2/3)}\leftrightarrow\xi_{\gamma_{(n)}}^{R(2/3)}\Big)\;,
\nonumber\\
&&\widetilde{C}_{8G}^{(c)}=C_{8G}^{(c)}
\Big(\xi_{Z}^{L(2/3)}\leftrightarrow\xi_{Z}^{R(2/3)},\;\xi_{Z_{H_\alpha}}^{L(2/3)}\leftrightarrow\xi_{Z_{H_\alpha}}^{R(2/3)},\;
\xi_{\gamma_{(n)}}^{L(2/3)}\leftrightarrow\xi_{\gamma_{(n)}}^{R(2/3)}\Big)\;.
\label{Hamilton-c}
\end{eqnarray}

The corrections to relevant Wilson coefficients from Fig.(c) are
analogously formulated to the order ${\cal
O}(\upsilon^2/\Lambda_{KK}^2)$ as

\begin{eqnarray}
&&C_{7\gamma}^{(c)}={32ie^3\over9\pi\Lambda_{KK}^2(kr\epsilon)^2}\sum\limits_{i,j,k=1}^3\bigg\{
{m_{_{u_k}}\over m_t}(U_L^{(0)})_{ci}^\dag {(U_L^{(0)})_{ik}}(U_R^{(0)})_{kj}^\dag {(U_R^{(0)})_{jt}}\nonumber\\
&&\hspace{2cm}
 \times \int_{\epsilon}^1 d t\int_{\epsilon}^1 d {t^\prime }\Big(\frac{1}{{{{\rm{c}}^2_{\rm{w}}}}}[\Sigma _{( +  + )}^G(t,{t^\prime })] +   \frac{{3 - 2s_{\rm{w}}^2}}{{{{\rm{c}}^2_{\rm{w}}}( {1 - 2s_{\rm{w}}^2}) }} [\Sigma _{( -  + )}^G(t,{t^\prime })]\Big)\nonumber\\
&&\hspace{2cm}
 \times{[f_{( +  + )}^{L,c_B^i}(0,t)]^2}{[f_{( +  + )}^{R,c_S^j}(0,{t^\prime })]^2}\bigg\}
+{\cal O}({\upsilon^4\over\Lambda_{KK}^4})\;,
\nonumber\\
&&C_{8G}^{(c)}=\frac{3g_sT^a}{2e}C_{7\gamma}^{(c)}\;,
\nonumber\\
&&\widetilde{C}_{7\gamma}^{(c)}={32ie^3\over9\pi\Lambda_{KK}^2(kr\epsilon)^2}
\sum\limits_{i,j,k=1}^3\bigg\{ {m_{_{u_k}}\over m_t} (U_R^{(0)})_{ci}^\dag {(U_R^{(0)})_{ik}}(U_L^{(0)})_{kj}^\dag {(U_L^{(0)})_{jt}}\nonumber\\
&&\hspace{2cm}
 \times \int_{\epsilon}^1 d t\int_{\epsilon}^1 d {t^\prime }\Big(\frac{1}{{{{\rm{c}}^2_{\rm{w}}}}}[\Sigma _{( +  + )}^G(t,{t^\prime })] +   \frac{{3 - 2s_{\rm{w}}^2}}{{{{\rm{c}}^2_{\rm{w}}}( {1 - 2s_{\rm{w}}^2}) }} [\Sigma _{( -  + )}^G(t,{t^\prime })]\Big)\nonumber\\
&&\hspace{2cm}
 \times{[f_{( +  + )}^{L,c_B^i}(0,t)]^2}{[f_{( +  + )}^{R,c_S^j}(0,{t^\prime })]^2}\bigg\}
+{\cal O}({\upsilon^4\over\Lambda_{KK}^4})\;,
\nonumber\\
&&\widetilde{C}_{8G}^{(c)}=\frac{3g_sT^a}{2e}\widetilde{C}_{7\gamma}^{(c)}\;.
\label{Hamilton-c-E}
\end{eqnarray}

\subsubsection{Fig.(d1) and (d2) with KK mode propagators}

Correspondingly, the contributions to Wilson coefficients at
electroweak scale from Fig.(d) with the KK mode of gluon, and  the
KK mode of charged $2/3$ quarks are
\begin{eqnarray}
&&C_{7\gamma}^{(d)}= {ieg_s^2\over9\pi^2\mu_{_{\rm
EW}}^2}\sum\limits_{n=1}^\infty\sum\limits_{\beta=1}^\infty\Big\{
\Big(\xi_{g_{(n)}}^{L(2/3)}\Big)_{_{c,\beta}}^\dagger
\Big(\xi_{g_{(n)}}^{L(2/3)}\Big)_{_{\beta,t}}F_{1,g}^{(a)}(x_{_{U_\beta}},x_{_{g_{(n)}}})
\nonumber\\
&&\hspace{3.0cm} +{m_{_{U_\beta}}\over
m_t}\Big(\xi_{g_{(n)}}^{L(2/3)}\Big)_{_{c,\beta}}^\dagger\Big(\xi_{g_{(n)}}^{R(2/3)}\Big)_{_{\beta,t}}
F_{2,g}^{(a)}(x_{_{U_\beta}},x_{_{g_{(n)}}})\Big\}\;,
\nonumber\\
&&C_{8G}^{(d)}= {ig_s^3T^a\over8\pi^2\mu_{_{\rm
EW}}^2}\sum\limits_{n=1}^\infty\sum\limits_{\beta=1}^\infty\Big\{
\Big(\xi_{g_{(n)}}^{L(2/3)}\Big)_{_{c,\beta}}^\dagger
\Big(\xi_{g_{(n)}}^{L(2/3)}\Big)_{_{\beta,t}}F_{1,g}^{(d)}(x_{_{U_\beta}},x_{_{g_{(n)}}})
\nonumber\\
&&\hspace{3.0cm} +{m_{_{U_\beta}}\over
m_t}\Big(\xi_{g_{(n)}}^{L(2/3)}\Big)_{_{c,\beta}}^\dagger\Big(\xi_{g_{(n)}}^{R(2/3)}\Big)_{_{\beta,t}}
F_{2,g}^{(d)}(x_{_{U_\beta}},x_{_{g_{(n)}}})\Big\}\;,
\nonumber\\
&&\widetilde{C}_{7\gamma}^{(d)}= C_{7\gamma}^{(d)}\Big(
\xi_{g_{(n)}}^{L(2/3)}\leftrightarrow\xi_{g_{(n)}}^{R(2/3)}\Big)\;,
\nonumber\\
&&\widetilde{C}_{8G}^{(d)}=C_{8G}^{(d)}
\Big(\xi_{g_{(n)}}^{L(2/3)}\leftrightarrow\xi_{g_{(n)}}^{R(2/3)}\Big)\;,
\label{Hamilton-d}
\end{eqnarray}
where the form factors are defined as
\begin{eqnarray}
&&F_{1,g}^{(d)}(x,y)=\Big[-{5\over36}{\partial^3\varrho_{_{3,1}}
\over\partial y^3}-{3\over8}{\partial^2\varrho_{_{2,1}}\over\partial
y^2} +{1\over12}{\partial\varrho_{_{1,1}}\over\partial
y}\Big](x,y)\;,
\nonumber\\
&&F_{2,g}^{(d)}(x,y)=
\Big[{5\over3}{\partial^2\varrho_{_{2,1}}\over\partial y^2}
-{1\over3}{\partial\varrho_{_{1,1}}\over\partial y}\Big](x,y)\;.
\label{form-(d)}
\end{eqnarray}

Using the coefficients in the appendix , the Wilson coefficients
could be approximated by

\begin{eqnarray}
&&C_{7\gamma}^{(d)}=
{32ieg_s^2\over9\pi\Lambda_{KK}^2(kr\epsilon)^2}
\sum\limits_{i,j,k=1}^3\bigg\{ {m_{_{u_k}}\over m_t}(U_L^{(0)})_{ci}^\dag {(U_L^{(0)})_{ik}}(U_R^{(0)})_{kj}^\dag {(U_R^{(0)})_{jt}}\nonumber\\
&&\hspace{3.3cm}
 \times \int_{\epsilon}^1 d t\int_{\epsilon}^1 d {t^\prime }[\Sigma _{( +  + )}^G(t,{t^\prime })]{[f_{( +  + )}^{L,c_B^i}(0,t)]^2}{[f_{( +  + )}^{R,c_S^j}(0,{t^\prime })]^2}\bigg\}
+{\cal O}({\upsilon^4\over\Lambda_{KK}^4})\;,
\nonumber\\
&&C_{8G}^{(d)}= {16ieg_s^3T^a\over3\pi\Lambda_{KK}^2(kr\epsilon)^2}
\sum\limits_{i,j,k=1}^3\bigg\{ {m_{_{u_k}}\over m_t}(U_L^{(0)})_{ci}^\dag {(U_L^{(0)})_{ik}}(U_R^{(0)})_{kj}^\dag {(U_R^{(0)})_{jt}}\nonumber\\
&&\hspace{3.3cm}
 \times \int_{\epsilon}^1 d t\int_{\epsilon}^1 d {t^\prime }[\Sigma _{( +  + )}^G(t,{t^\prime })]{[f_{( +  + )}^{L,c_B^i}(0,t)]^2}{[f_{( +  + )}^{R,c_S^j}(0,{t^\prime })]^2}\bigg\}
+{\cal O}({\upsilon^4\over\Lambda_{KK}^4})\;,
\nonumber\\
&&\widetilde{C}_{7\gamma}^{(d)}=
{32ieg_s^2\over9\pi\Lambda_{KK}^2(kr\epsilon)^2}
\sum\limits_{i,j,k=1}^3\bigg\{ {m_{_{u_k}}\over m_t}(U_R^{(0)})_{ci}^\dag {(U_R^{(0)})_{ik}}(U_L^{(0)})_{kj}^\dag {(U_L^{(0)})_{jt}}\nonumber\\
&&\hspace{3.3cm}
 \times \int_{\epsilon}^1 d t\int_{\epsilon}^1 d {t^\prime }[\Sigma _{( +  + )}^G(t,{t^\prime })]{[f_{( +  + )}^{R,c_S^j}(0,{t^\prime })]^2}{[f_{( +  + )}^{L,c_B^i}(0,t)]^2}
\bigg\}+{\cal O}({\upsilon^4\over\Lambda_{KK}^4})\;,
\nonumber\\
&&\widetilde{C}_{8G}^{(d)}=
{16ieg_s^3T^a\over3\pi\Lambda_{KK}^2(kr\epsilon)^2}
\sum\limits_{i,j,k=1}^3\bigg\{ {m_{_{u_k}}\over m_t}(U_R^{(0)})_{ci}^\dag {(U_R^{(0)})_{ik}}(U_L^{(0)})_{kj}^\dag {(U_L^{(0)})_{jt}}\nonumber\\
&&\hspace{3.3cm}
 \times \int_{\epsilon}^1 d t\int_{\epsilon}^1 d {t^\prime }[\Sigma _{( +  + )}^G(t,{t^\prime })]{[f_{( +  + )}^{R,c_S^j}(0,{t^\prime })]^2}{[f_{( +  + )}^{L,c_B^i}(0,t)]^2}
\bigg\}+{\cal O}({\upsilon^4\over\Lambda_{KK}^4})\;.
\label{Hamilton-d-E}
\end{eqnarray}

\subsubsection{Fig.(e)  with KK mode propagators}

As intermediate virtual particles in Fig.(e) are the KK mode of neutral
Higgs/Goldstone, and  the KK mode of charged $2/3$ quarks, the
corresponding corrections to relevant Wilson coefficients can be
written as
\begin{eqnarray}
&&C_{7\gamma}^{(e)}={ie\over12\pi^2\mu_{_{\rm EW}}^2}
\sum\limits_{\beta=1}^\infty\Big\{\Big(\eta_{H_0}^{(2/3)}\Big)_{_{c,\beta}}^\dagger
\Big(\eta_{H_0}^{(2/3)}\Big)_{_{\beta,t}}F_{1,g}^{(b)}(x_{_{U_\beta}},x_{_{H_0}})
\nonumber\\
&&\hspace{3.0cm} +{m_{_{U_\beta}}\over
m_t}\Big(\eta_{H_0}^{(2/3)}\Big)_{_{c,\beta}}^\dagger
\Big(\eta_{H_0}^{(2/3)}\Big)_{_{\beta,t}}^\dagger F_{2,g}^{(b)}
(x_{_{U_\beta}},x_{_{H_0}})
\nonumber\\
&&\hspace{3.0cm} +\Big(\eta_{G_0}^{(2/3)}\Big)_{_{c,\beta}}^\dagger
\Big(\eta_{G_0}^{(2/3)}\Big)_{_{\beta,t}}F_{1,g}^{(b)}(x_{_{U_\beta}},x_{_{Z}})
\nonumber\\
&&\hspace{3.0cm} -{m_{_{U_\beta}}\over
m_t}\Big(\eta_{G_0}^{(2/3)}\Big)_{_{c,\beta}}^\dagger
\Big(\eta_{G_0}^{(2/3)}\Big)_{_{\beta,t}}^\dagger
F_{2,g}^{(b)}(x_{_{U_\beta}},x_{_{Z}}) \Big\}\;,
\nonumber\\
&&C_{8G}^{(e)}={3g_sT^a\over2e}C_{7\gamma}^{(e)}
\;,\nonumber\\
&&\widetilde{C}_{7\gamma}^{(e)}=
C_{7\gamma}^{(e)}\Big(\eta_{H_0}^{(2/3)}\leftrightarrow
(\eta_{H_0}^{(2/3)})^\dagger,\;\eta_{G_0}^{(2/3)}\leftrightarrow-(\eta_{G_0}^{(2/3)})^\dagger\Big)\;,
\nonumber\\
&&\widetilde{C}_{8G}^{(e)}=C_{8G}^{(e)}
\Big(\eta_{H_0}^{(2/3)}\leftrightarrow(\eta_{H_0}^{(2/3)})^\dagger,\;\eta_{G_0}^{(2/3)}\leftrightarrow-(\eta_{G_0}^{(2/3)})^\dagger\Big)\;.
\label{Hamilton-e}
\end{eqnarray}

Using the coefficients in the appendix,we can approximate
$C_{7\gamma}$ and $C_{8g}$ as

\begin{eqnarray}
&&C_{7\gamma}^{(e)}={ie\over144\pi^2\Lambda_{KK}^2}
\sum\limits_{i,j,k=1}^3\bigg\{ (U_L^{(0)})_{ci}^\dag f_{( +  +
)}^{L,c_B^i}(0,1)Y_{ik}^u([\Sigma _{( \mp  \mp
)}^{R,c_S^k}(1,1)])\nonumber\\
&&\hspace{3.3cm}\times Y_{kj}^{{u^\dag }}f_{( +  +
)}^{L,c_B^j}(0,1){(U_L^{(0)})_{jt}}\bigg\}+{\cal
O}({\upsilon^4\over\Lambda_{KK}^4})\;,
\nonumber\\
&&C_{8G}^{(e)}=\frac{3g_sT^a}{2e}C_{7\gamma}^{(e)}
\;,\nonumber\\
&&\widetilde{C}_{7\gamma}^{(e)}={ie\over144\pi^2\Lambda_{KK}^2}
\sum\limits_{i,j,k=1}^3\bigg\{ (U_R^{(0)})_{ci}^\dag f_{( +  +
)}^{R,c_S^i}(0,1)Y_{ik}^{{u^\dag }}([\Sigma _{( \pm  \pm
)}^{L,c_B^k}(1,1)] + [\Sigma _{( \mp  \pm
)}^{L,c_B^k}(1,1)])\nonumber\\
&&\hspace{3.3cm}\times Y_{kj}^uf_{( +  +
)}^{R,c_S^j}(0,1){(U_R^{(0)})_{jt}}\bigg\} +{\cal
O}({\upsilon^4\over\Lambda_{KK}^4})\;,\nonumber\\
&&\widetilde{C}_{8G}^{(e)}=\frac{3g_sT^a}{2e}\widetilde{C}_{7\gamma}^{(e)}
\;. \label{Hamilton-e-E}
\end{eqnarray}

\section{Numerical analysis\label{sec3}}
\indent\indent

 In general case, the partial widths of these processes are \cite{ref1}

\begin{eqnarray}
&&\Gamma(t\rightarrow c\gamma)=\frac{m_t^3
}{16\pi}(m_c^2|F_{TL}^{\gamma}|^2+m_t^2|F_{TR}^{\gamma}|^2) \nonumber\\
&&\Gamma(t\rightarrow cg)=\frac{m_t^3C_F
}{16\pi}(m_c^2|F_{TL}^{g}|^2+m_t^2|F_{TR}^{g}|^2)\label{eq1}
\end{eqnarray}
with $C_F=4/3$ is a colour factor.
Applying Eq(34),(53),(54), we could get that the partial widths both
 composed by the SM particles and their KK exciting modes has the form,

\begin{eqnarray}
&&\Gamma(t\rightarrow c\gamma)=\frac{m_t^3
}{16\pi}(m_c^2|F_{TL}^{\gamma}+\widetilde{C}_{7\gamma}|^2+m_t^2|F_{TR}^{\gamma}+C_{7\gamma}|^2) \nonumber\\
&&\Gamma(t\rightarrow cg)=\frac{m_t^3C_F
}{16\pi}(m_c^2|F_{TL}^{g}+\widetilde{C}_{8g}|^2+m_t^2|F_{TR}^{g}+C_{8g}|^2)\label{eq1}
\end{eqnarray}

We compute the branching ratio by taking the SM charged-current
two-body decay $t\rightarrow bW$ to be the dominant $t$-quark decay
mode, which is $\Gamma(t\rightarrow bW^+)=1.466|V_{tb}|^2$ for $m_t=172$GeV, $m_W=80.399$GeV. We will then approximate
the branching ratio by
\begin{eqnarray}
&&Br(t\rightarrow c\gamma)=\frac{\Gamma(t\rightarrow c\gamma)}{\Gamma(t\rightarrow bW^+)} \nonumber\\
&&Br(t\rightarrow cg)=\frac{\Gamma(t\rightarrow
cg)}{\Gamma(t\rightarrow bW^+)}\label{eq1}
\end{eqnarray}

The input parameters which we are going to use in the numerical
computations are : $m_W=80.399$GeV, $m_Z=90.19$GeV, $m_c=1.27$GeV,
$m_t=172$GeV. We choose the Yukawa couplings $Y^u_{ij}=0.01 \;
(i\neq j,\; i, j=1,2,3),\;   Y^d_{21}=Y^d_{31}=Y^d_{32}=0.01$. And
the Wolfenstein parameters of the CKM matrix are
$\lambda=0.22,\;A=0.81,\;\overline{\rho}=0.13,\;\overline{\eta}=0.34.$

Assuming anarchic Yukawa couplings, i.e., complex-valued matrices
$Y^u,\;Y^d$ with random elements, we can reproduce the up- and
down-type quark mass hierarchies with the ansatz for hierarchical
structures of the profiles of zero modes on IR brane\cite{ref28}:
\begin{eqnarray}
&&\Big[f_{_{(++)}}^{L,c_B^1}(0,1)\Big]<\Big[f_{_{(++)}}^{L,c_B^2}(0,1)\Big]
<\Big[f_{_{(++)}}^{L,c_B^3}(0,1)\Big]\;,
\nonumber\\
&&\Big[f_{_{(++)}}^{R,c_T^1}(0,1)\Big]<\Big[f_{_{(++)}}^{R,c_T^2}(0,1)\Big]
<\Big[f_{_{(++)}}^{R,c_T^3}(0,1)\Big]\;,
\nonumber\\
&&\Big[f_{_{(++)}}^{R,c_S^1}(0,1)\Big]<\Big[f_{_{(++)}}^{R,c_S^2}(0,1)\Big]
<\Big[f_{_{(++)}}^{R,c_S^3}(0,1)\Big]\;.
\label{Froggatt-Nielsen1}
\end{eqnarray}
Since when $c>c'$, we have
$f_{_{(++)}}^{L,c'}(0,1)<f_{_{(++)}}^{L,c}(0,1)$, and when $c>c'$,
we have $f_{_{(++)}}^{R,c'}(0,1)<f_{_{(++)}}^{R,c}(0,1)$,  so the
bulk masses $c_B^i,c_T^i,c_S^i$ must satisfy the relation:

\begin{eqnarray}
&&c_B^1>c_B^2>c_B^3\;,
\nonumber\\
&&c_T^1<c_T^2<c_T^3\;,
\nonumber\\
&&c_S^1<c_S^2<c_S^3\;.
\label{1}
\end{eqnarray}

\begin{figure}[H]\small
  \centering
   \includegraphics[width=6in]{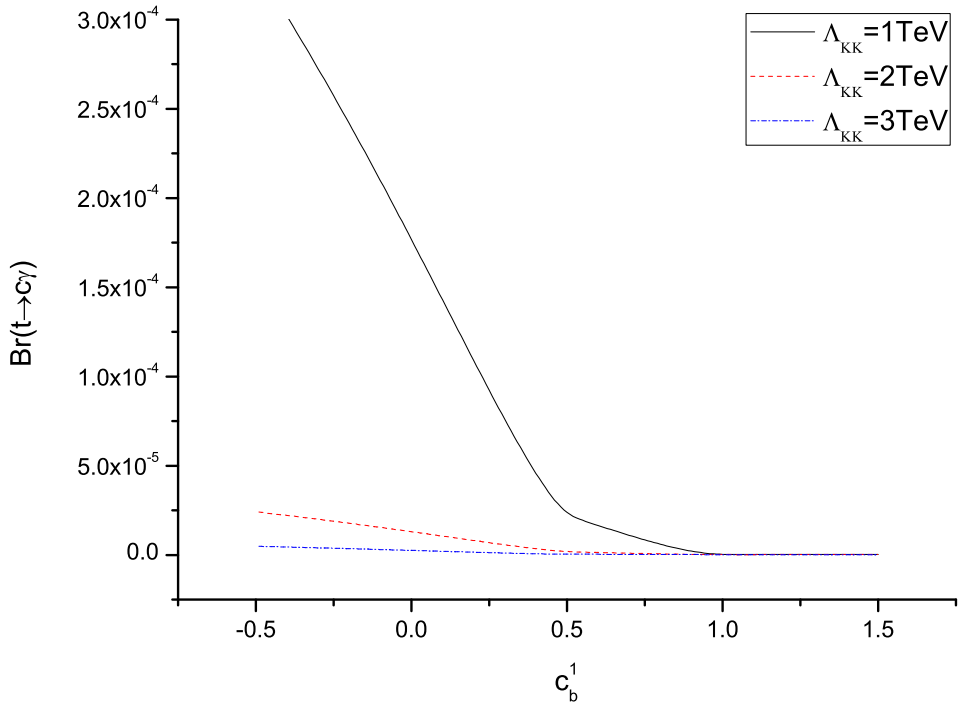}
     \caption{The branching ratio of  $t\rightarrow
c\gamma$ varying with the bulk mass $c_B^1$. Here we assuming that $c_T^1=c_S^1=-0.75,\; c_T^2=c_S^2=-0.55,\; c_T^3=c_S^3=-0.35,$
 and $c_B^2=-0.5+c_B^1,\; c_B^3=-1+c_B^1$. The solid line corresponds
to the numerical result with $\Lambda_{KK}=1$TeV, dash line corresponds
to the numerical result with $\Lambda_{KK}=2$TeV,
and dash-dot line corresponds to the numerical result with $\Lambda_{KK}=3$TeV, respectively}
    \label{Graph1}
\end{figure}

In order to reduce the number of free parameters in our analysis, we
assume
\begin{eqnarray}
&&c_T^1=c_S^1=-0.75,\; c_T^2=c_S^2=-0.55,\;
c_T^3=c_S^3=-0.35;\nonumber\\
&&c_B^2=-0.5+c_B^1,\; c_B^3=-1+c_B^1.
\label{100}
\end{eqnarray}
Actually, the assumption on the bulk masses guarantees the profiles
of zero modes on IR brane satisfying the hierarchical structures
(78). In Fig. \ref{Graph1}, we plot Br$(t\rightarrow c\gamma)$
varying with the bulk mass $c_B^1$ under above assumption on the
parameter space with the mass scale of low-lying KK states
$\Lambda_{KK}=1$TeV(solid line), $\Lambda_{KK}=2$TeV(dash line),
$\Lambda_{KK}=3$TeV(dash-dot line). As $c_B^1<0.5$, the
contributions from new physics to Br$(t\rightarrow c\gamma)$
decrease steeply, for low-lying KK states $\Lambda_{KK}=1,2,3$TeV
respectively.  Since the left-handed SM top and charm quarks are
contained in bidoublet, so the profiles of  $t$ and $c$ quarks on IR
brane are determined by the bulk mass of  $c_B^i$. As $c_B^1>1.0$,
the new physics corrections to Br$(t\rightarrow c\gamma)$ depend on
$c_B^1$ mildly, and line for $\Lambda_{KK}=1,2,3$TeV are coincide
with each other almost.

To investigate the dependence of Br$(t\rightarrow c\gamma)$ on bulk
mass of triplet quark $c_T^1$, we choose

\begin{eqnarray}
&&c_S^1=-0.75, \;c_S^2=-0.55, \;c_S^3=-0.35;\nonumber\\
&&c_B^1=0.55,\; c_B^2=0.25,\; c_B^3=-0.05;\nonumber\\
&&c_T^2=0.5+c_T^1,\; c_T^3=1+c_T^1,
\label{101}
\end{eqnarray}

\begin{figure}[H]\small
  \centering
   \includegraphics[width=6in]{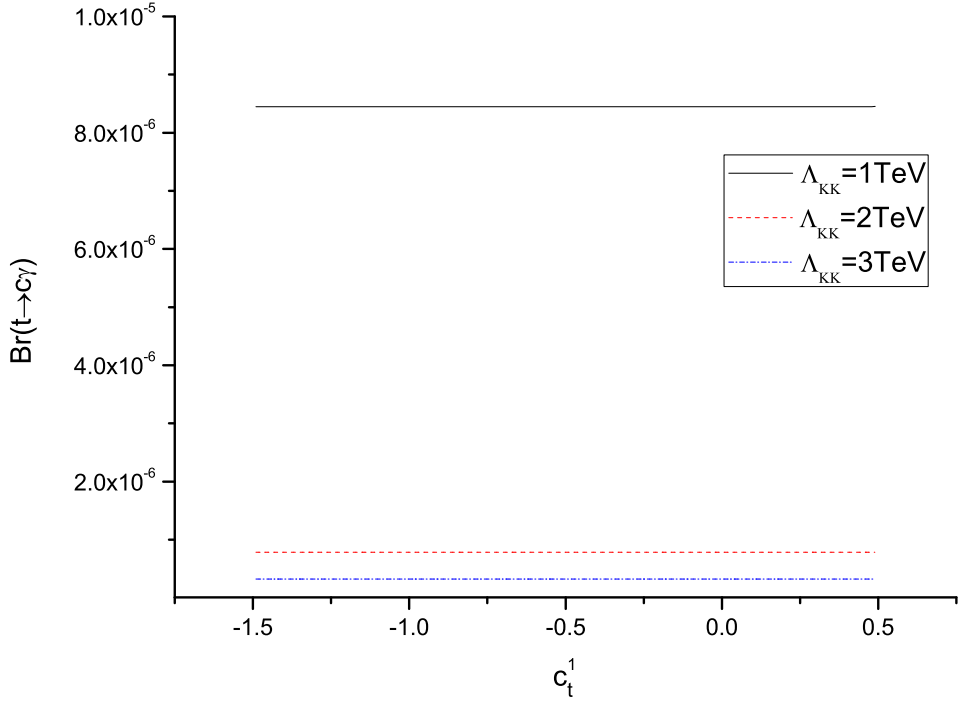}
     \caption{The branching ratio of  $t\rightarrow
c\gamma$ varying with the bulk mass $c_T^1$. Here we assuming that
$c_S^1=-0.75, \;c_S^2=-0.55, \;c_S^3=-0.35,\; c_B^1=0.55,\;
c_B^2=0.25,\; c_B^3=-0.05,$
 and $c_T^2=0.5+c_T^1,\; c_T^3=1+c_T^1$. The solid line corresponds
to the numerical result with $\Lambda_{KK}=1$TeV, dash line corresponds
to the numerical result with $\Lambda_{KK}=2$TeV,
and dash-dot line corresponds to the numerical result with $\Lambda_{KK}=3$TeV, respectively}
    \label{Graph2}
\end{figure}

The bulk masses satisfying Eq.(81) can guarantee the profiles of
zero modes on IR brane satisfy the hierarchical structures(78).
Under this choice on parameter space, we draw Br$(t\rightarrow
c\gamma)$ varying with the bulk mass $c_T^1$  in Fig.\ref{Graph2}.
Because the profiles of $t$ and $c$ quarks on IR brane do not depend
on $c_T^1$, the dependence of Br$(t\rightarrow c\gamma)$ on $c^1_T$
is very mildly, for $\Lambda_{KK}= 1, 2, 3$TeV respectively.

\begin{figure}[H]\small
  \centering
   \includegraphics[width=6in]{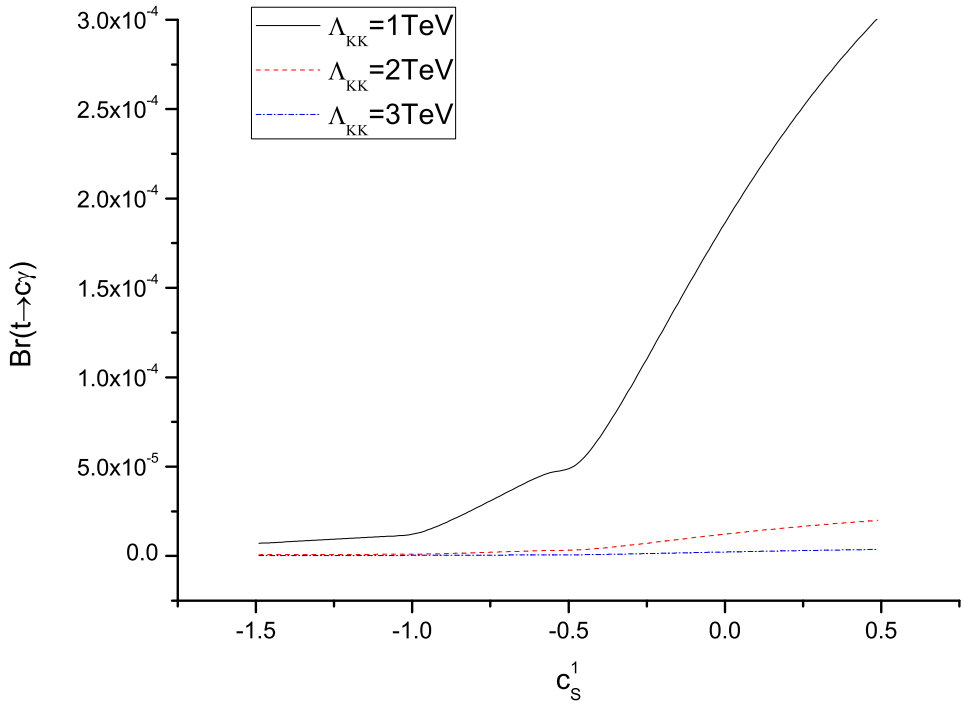}
     \caption{The branching ratio of  $t\rightarrow
c\gamma$ varying with the bulk mass $c_S^1$. Here we assuming that
$c_T^1=-0.75, \;c_T^2=-0.55, \;c_T^3=-0.35,\; c_B^1=0.55,
\;c_B^2=0.25, \;c_B^3=-0.05,$
 and $c_S^2=0.5+c_S^1,\; c_S^3=1+c_S^1$. The solid line corresponds
to the numerical result with $\Lambda_{KK}=1$TeV, dash line corresponds
to the numerical result with $\Lambda_{KK}=2$TeV,
and dash-dot line corresponds to the numerical result with $\Lambda_{KK}=3$TeV, respectively}
    \label{Graph3}
\end{figure}

Adopting

\begin{eqnarray}
&&c_T^1=-0.75, \;c_T^2=-0.55, \;c_T^3=-0.35,\; \nonumber\\
&&c_B^1=0.55, \;c_B^2=0.25, \;c_B^3=-0.05,\nonumber\\
&&c_S^2=0.5+c_S^1,\; c_S^3=1+c_S^1,
\label{100}
\end{eqnarray}
to guarantee the profiles of zero modes on IR brane satisfy the
hierarchical structures (78). We show the Br$(t\rightarrow c\gamma)$
varying with $c_S^1$ in Fig. \ref{Graph3} for
$\Lambda_{KK}=1$TeV(solid line), $\Lambda_{KK}=2$TeV(dash line),
$\Lambda_{KK}=3$TeV(dash-dot line), respectively. With the
assumption on parameter space, the bulk mass $c_S^1$ affect the new
physics corrections on Br$(t\rightarrow c\gamma)$ strongly, since
the profiles of right-handed top and charm quarks on IR brane are
determined by the bulk mass of singlet $c_S^i$.

In order to appreciate the size of the 3TeV curve, we redraw
 Fig.\ref{Graph1} and Fig. \ref{Graph3} in logarithmic coordinate in Fig.\ref{gammalog}:

\begin{figure}[H]
\centering
\subfigure[×Ó±êÌâ]{\includegraphics[width=0.45\textwidth]{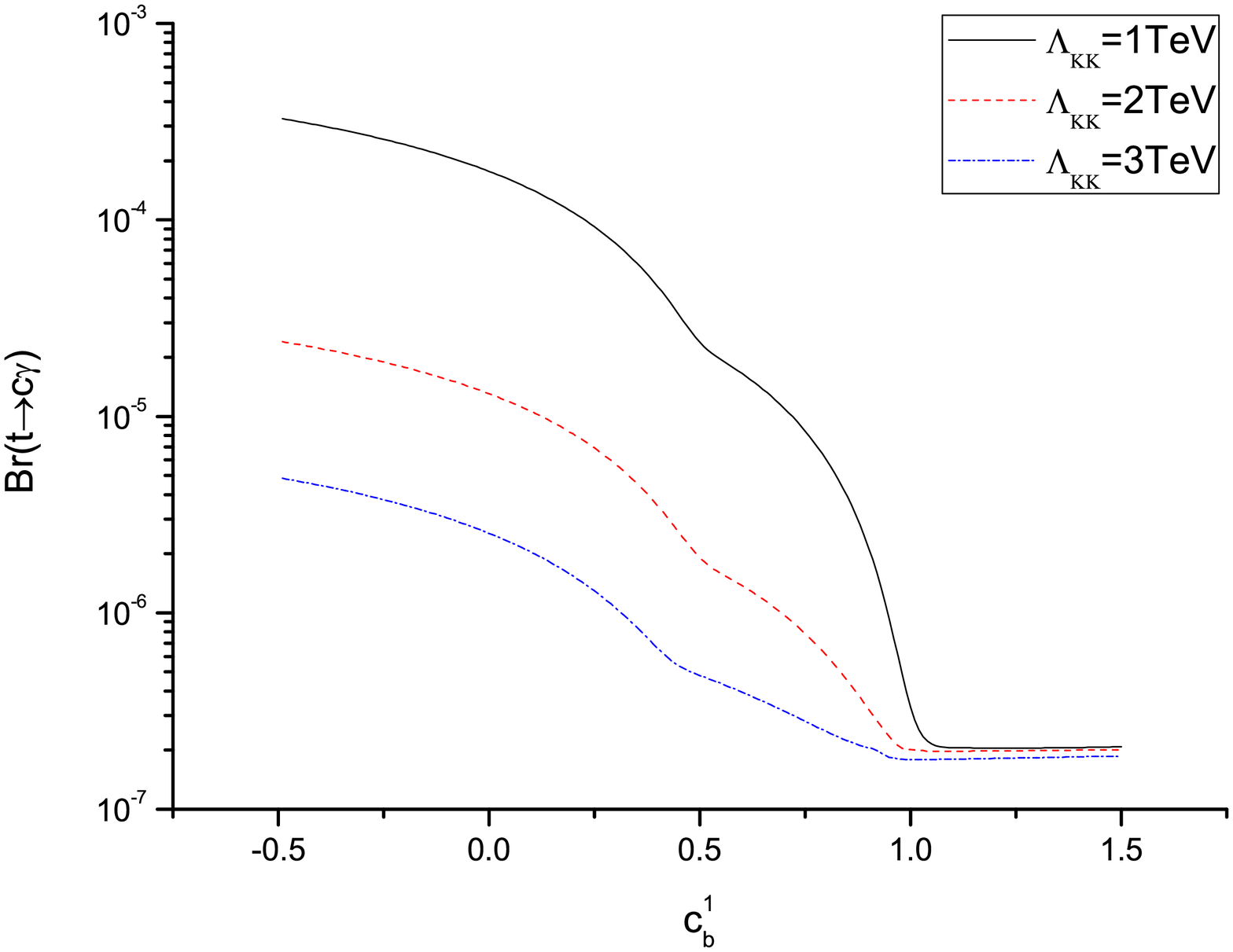}}
\subfigure[×Ó±êÌâ]{\includegraphics[width=0.45\textwidth]{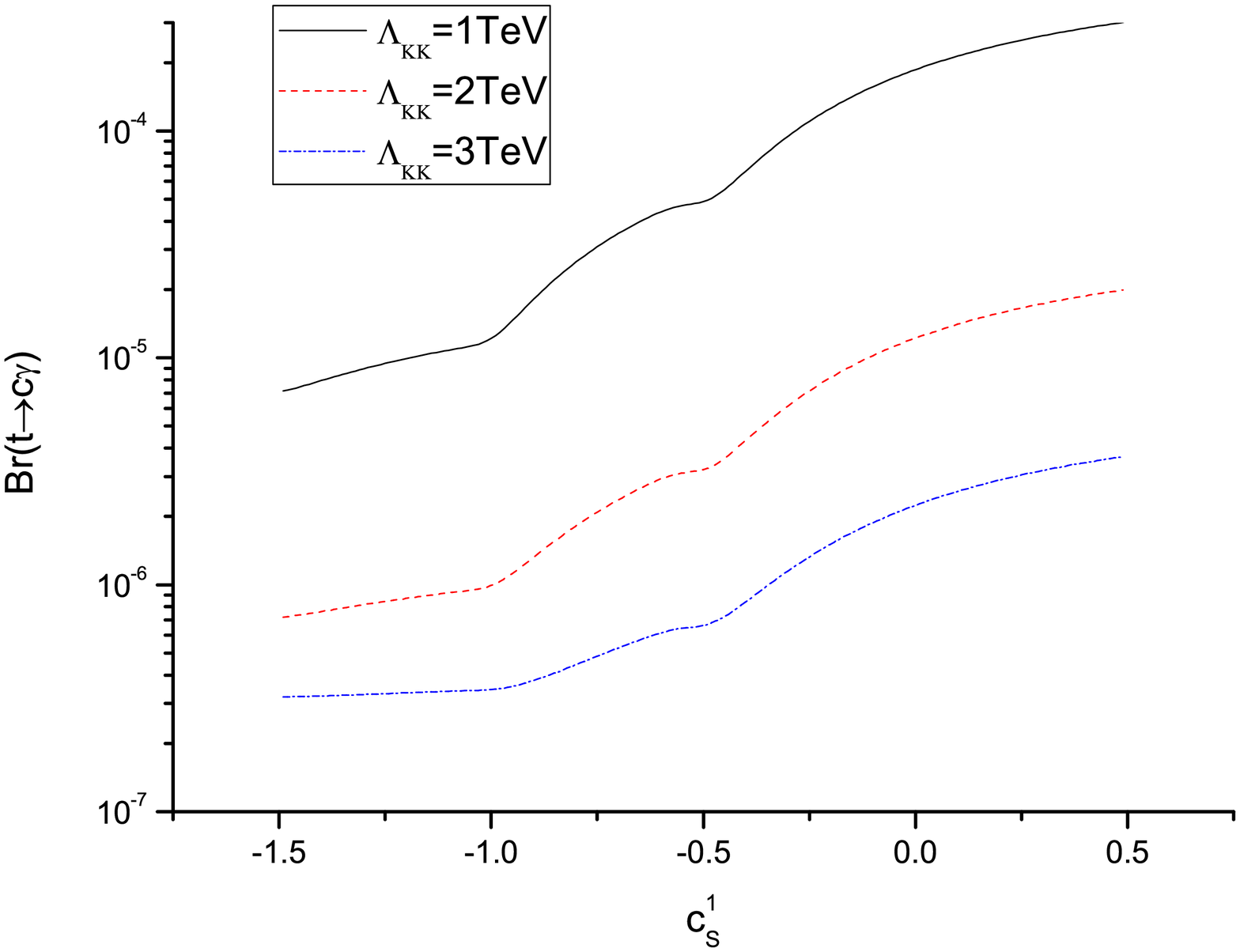}}
\caption{(a) Redraw Fig.\ref{Graph1} in logarithmic coordinate,
 (b)Redraw Fig.\ref{Graph3} in logarithmic coordinate.}
\label{gammalog}
\end{figure}
%

In Fig.\ref{gammalog}(a), when $c_B^1\geq1$, the three curves become
very close, but not coincide with each other,  and both in Fig.
\ref{gammalog}(a) and (b), the branching ratio decreases as
$\Lambda_{KK}$ runs from 1TeV to 3TeV.

\begin{figure}[H]\small
  \centering
   \includegraphics[width=6in]{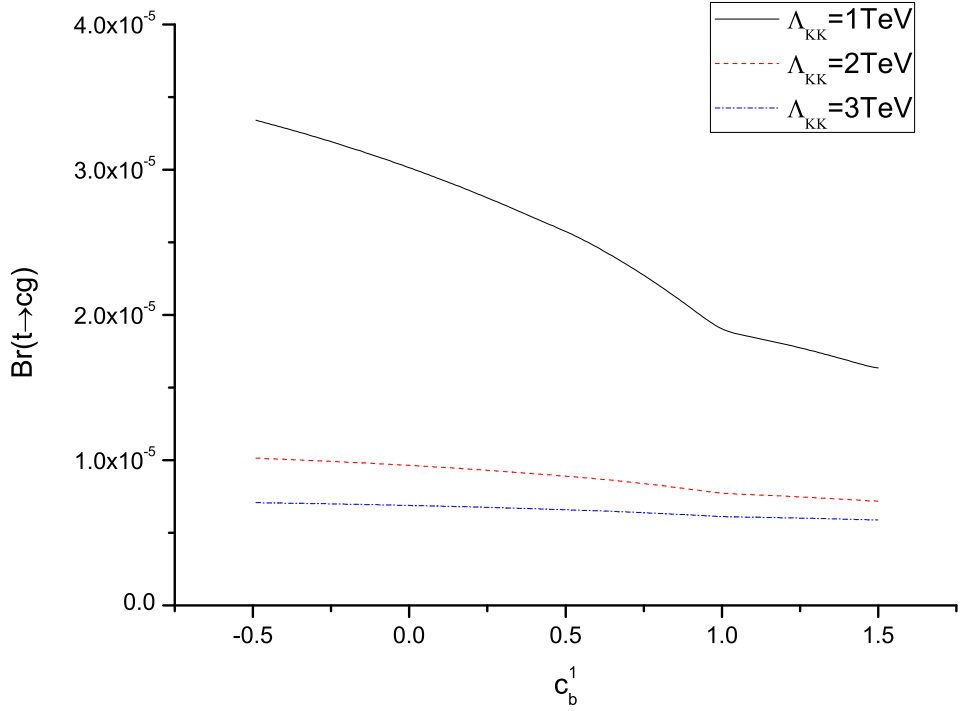}
     \caption{The branching ratio of  $t\rightarrow
cg$ varying with the bulk mass $c_B^1$. Here we assuming that $c_T^1=c_S^1=-0.75, \;c_T^2=c_S^2=-0.55, \;c_T^3=c_S^3=-0.35,$
 and $c_B^2=-0.5+c_B^1,\; c_B^3=-1+c_B^1$. The solid line corresponds
to the numerical result with $\Lambda_{KK}=1$TeV, dash line corresponds
to the numerical result with $\Lambda_{KK}=2$TeV,
and dash-dot line corresponds to the numerical result with $\Lambda_{KK}=3$TeV, respectively}
    \label{Graph4}
\end{figure}

Similarly assuming
 \begin{eqnarray}
&&c_T^1=c_S^1=-0.75,\; c_T^2=c_S^2=-0.55,\;
c_T^3=c_S^3=-0.35;\nonumber\\
&&c_B^2=-0.5+c_B^1,\; c_B^3=-1+c_B^1,
\label{100}
\end{eqnarray}
to guarantee that the profiles of zero modes on IR brane satisfy the
hierarchical structures (78), we plot
 Br$(t\rightarrow cg)$ varying with the bulk
mass $c_B^1$ in Fig.\ref{Graph4} under above assumption on the
parameter space as the mass scale $\Lambda_{KK}=1$TeV(solid line),
$\Lambda_{KK}=2$TeV(dash line), $\Lambda_{KK}=3$TeV(dash-dot line).
The contributions from new physics to the branching ratio of
$t\rightarrow cg$ decrease, but slower than which in Fig.
\ref{Graph1}.

\begin{figure}[H]\small
  \centering
   \includegraphics[width=6in]{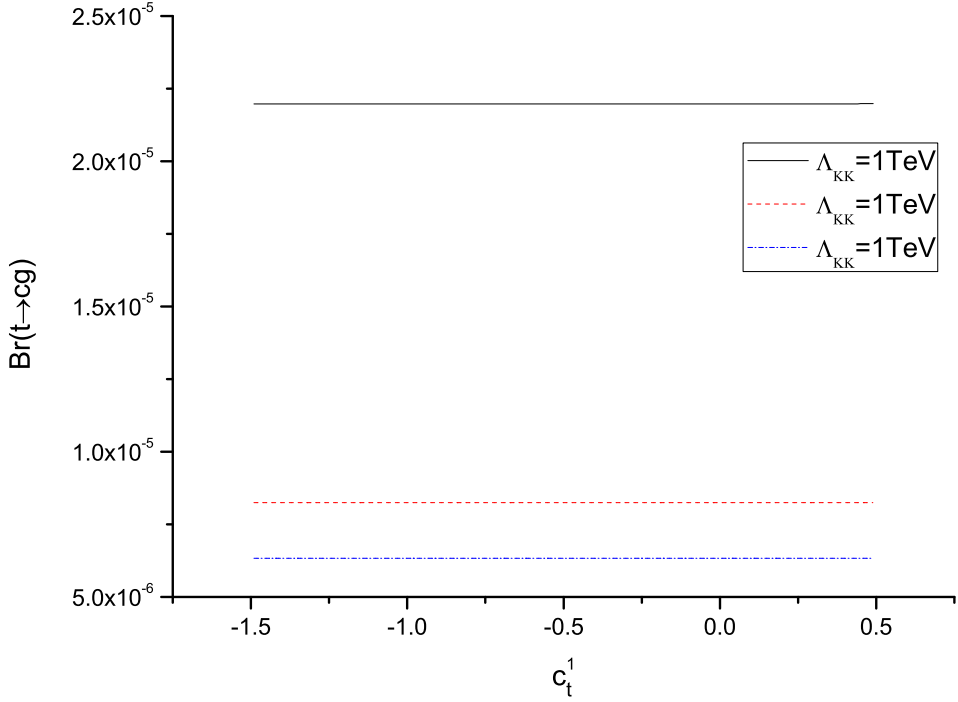}
     \caption{The branching ratio of  $t\rightarrow
cg$ varying with the bulk mass $c_T^1$. Here we assuming that $c_S^1=-0.75, \;c_S^2=-0.55, \;c_S^3=-0.35,
c_B^1=0.55,\; c_B^2=0.25,\; c_B^3=-0.05,$
 and $c_T^2=0.5+c_T^1,\; c_T^3=1+c_T^1$. The solid line corresponds
to the numerical result with $\Lambda_{KK}=1$TeV, dash line corresponds
to the numerical result with $\Lambda_{KK}=2$TeV,
and dash-dot line corresponds to the numerical result with $\Lambda_{KK}=3$TeV, respectively}
    \label{Graph5}
\end{figure}

Taking
\begin{eqnarray}
&&c_S^1=-0.75, \;c_S^2=-0.55, \;c_S^3=-0.35;\nonumber\\
&&c_B^1=0.55,\; c_B^2=0.25,\; c_B^3=-0.05;\nonumber\\
&&c_T^2=0.5+c_T^1,\; c_T^3=1+c_T^1,
\label{101}
\end{eqnarray}
to guarantee that the profiles of zero modes on IR brane satisfy the
hierarchical structures (78). We present Br$(t\rightarrow cg)$
varying with the bulk mass $c_T^1$ in Fig.\ref{Graph5}, for
$\Lambda_{KK}=1$TeV(solid line), $\Lambda_{KK}=2$TeV(dash line),
$\Lambda_{KK}=3$TeV(dash-dot line), respectively. Similarly as in
Fig.\ref{Graph2}, the dependence of the $t\rightarrow cg$ process on
the bulk mass $c^1_T$ is very mild also, since the profiles of $t$
and $c$ quarks on IR brane do not depend on $c_T^1$,

\begin{figure}[H]\small
  \centering
   \includegraphics[width=6in]{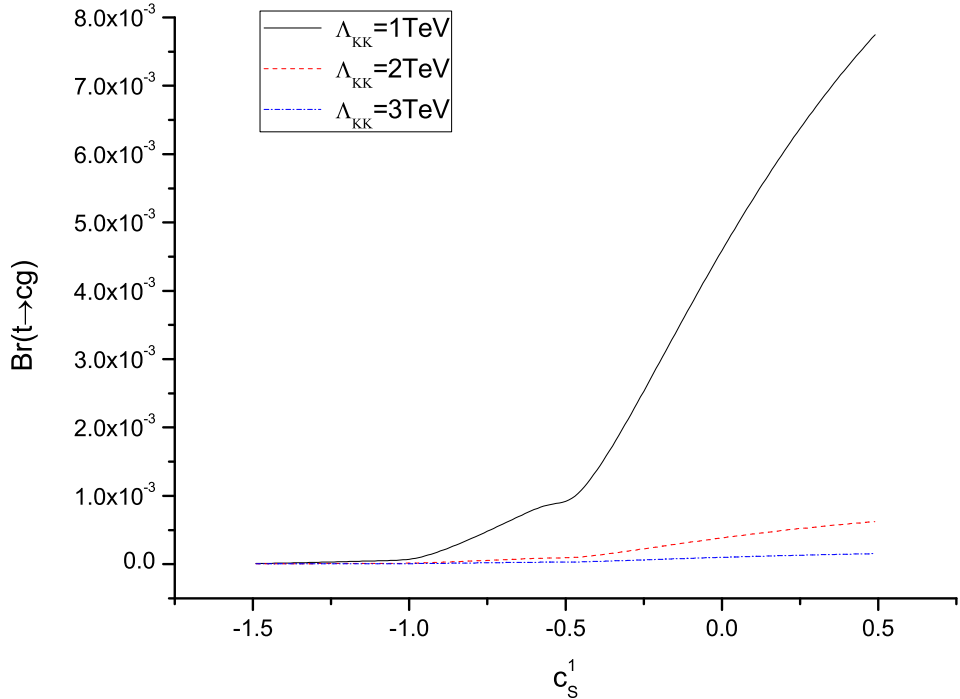}
     \caption{The branching ratio of  $t\rightarrow
cg$ varying with the bulk mass $c_S^1$. Here we assuming that $c_T^1=-0.75, \;c_T^2=-0.55, \;c_T^3=-0.35,
c_B^1=0.55, \;c_B^2=0.25, \;c_B^3=-0.05,$
 and $c_S^2=0.5+c_S^1,\; c_S^3=1+c_S^1$. The solid line corresponds
to the numerical result with $\Lambda_{KK}=1$TeV, dash line corresponds
to the numerical result with $\Lambda_{KK}=2$TeV,
and dash-dot line corresponds to the numerical result with $\Lambda_{KK}=3$TeV, respectively}
    \label{Graph6}
\end{figure}

Taking
\begin{eqnarray}
&&c_T^1=-0.75, \;c_T^2=-0.55, \;c_T^3=-0.35,\; \nonumber\\
&&c_B^1=0.55, \;c_B^2=0.25, \;c_B^3=-0.05,\nonumber\\
&&c_S^2=0.5+c_S^1,\; c_S^3=1+c_S^1,
\label{100}
\end{eqnarray}
to guarantee that the profiles of zero modes on IR brane satisfy the
hierarchical structures (78), and we present Br$(t\rightarrow cg)$
varying with the bulk mass $c_S^1$ in Fig.\ref{Graph6},. We could
see that the contributions from new physics to the $t\rightarrow cg$
increase quickly when $c_S^1>0.5$, because of the reason mentioned
above, which is similar to that for Fig.\ref{Graph3}.

In order to appreciate the size of the 3TeV curve, we redraw
 Fig.\ref{Graph4} and Fig. \ref{Graph6} in logarithmic coordinate in Fig. \ref{gammalog2}:

\begin{figure}[H]
\centering
\subfigure[×Ó±êÌâ]{\includegraphics[width=0.45\textwidth]{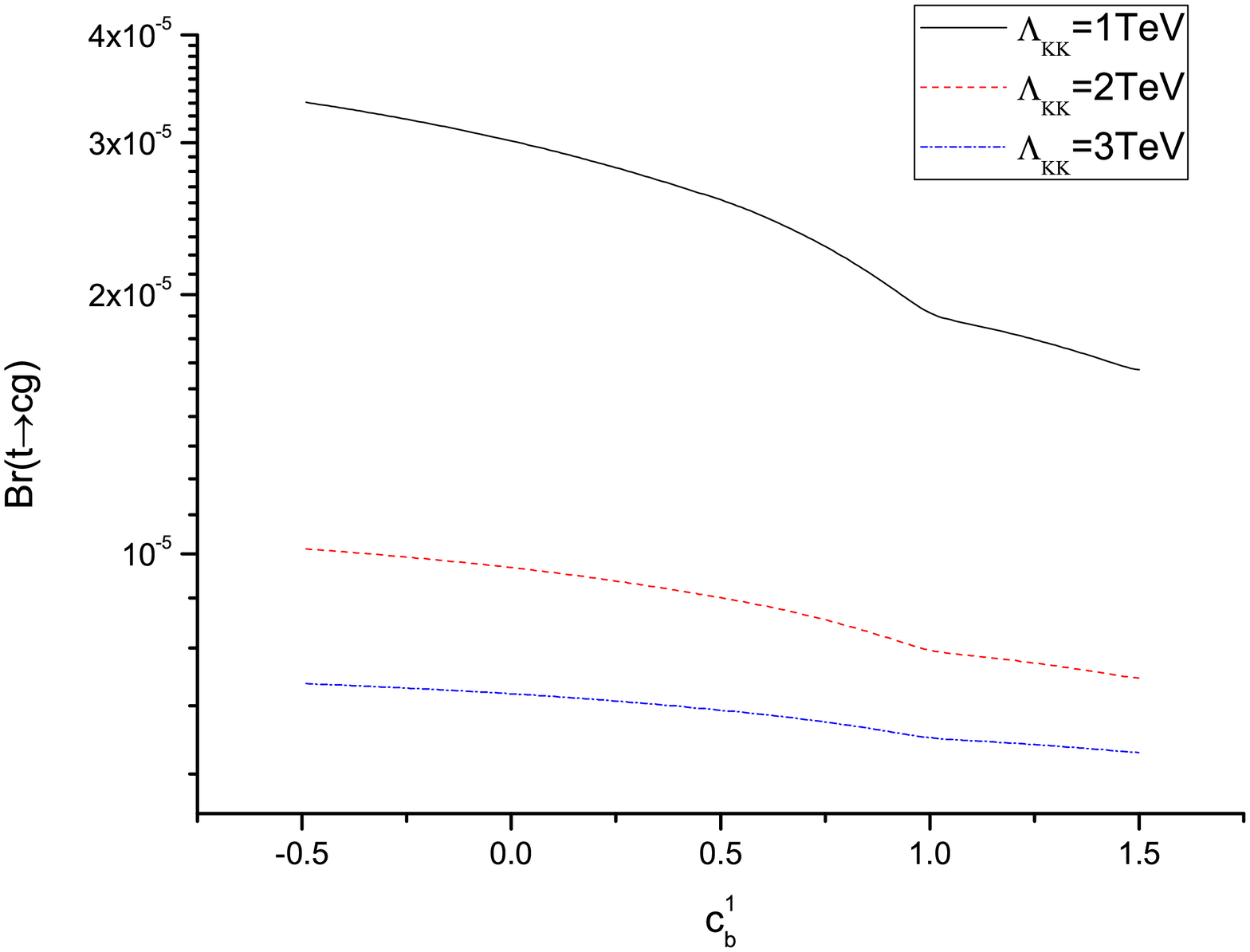}}
\subfigure[×Ó±êÌâ]{\includegraphics[width=0.45\textwidth]{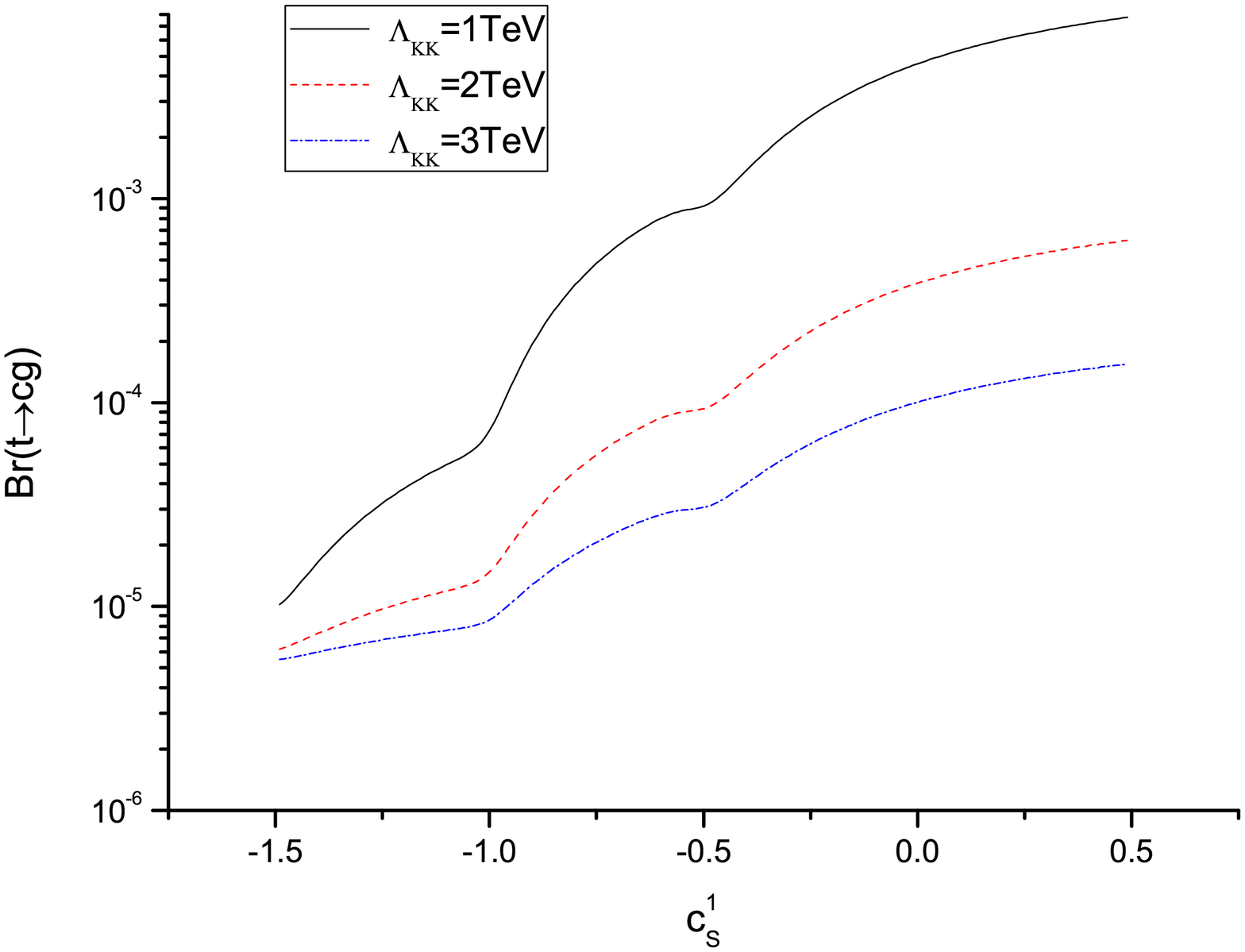}}
\caption{(a) Redraw Fig.\ref{Graph1} in logarithmic coordinate,
 (b)Redraw Fig.\ref{Graph3} in logarithmic coordinate.}
\label{gammalog2}
\end{figure}
%

Being similar to the case of $t\rightarrow c\gamma$, the branching
ratio decreases as $\Lambda_{KK}$ runs from 1TeV to 3TeV.

As we could see above, the branching ratio
of $t\rightarrow c\gamma$ and $t\rightarrow cg$  varying with the bulk mass $c_B^1$  in
Fig. \ref{Graph1} and Fig. \ref{Graph4} are all decrease, since the dominating corrections  to the branching
ratio of  $t\rightarrow c\gamma$ and $t\rightarrow cg$  depend on the bulk masses $c_B^i
(i=1,2,3)$ in terms of
 $[f^{L,c_B^i}_{(++)}(0,t)][f^{L,c_B^j}_{(++)}(0,t)]$, with $f^{L,c}_{(++)}(0,t)$ have the form(Eq. 36 and Eq. 109 of Ref.\cite{ref39}):

\begin{eqnarray}
&&f^{L,c}_{(++)}(0,t)={t^{-c}\over2}\sqrt{-2(1-2c)\epsilon\ln\epsilon\over\pi(1-\epsilon^{1-2c})}
\;.
\label{1}
\end{eqnarray}

\begin{figure}[H]
\centering
\subfigure[×Ó±êÌâ]{\includegraphics[width=0.45\textwidth]{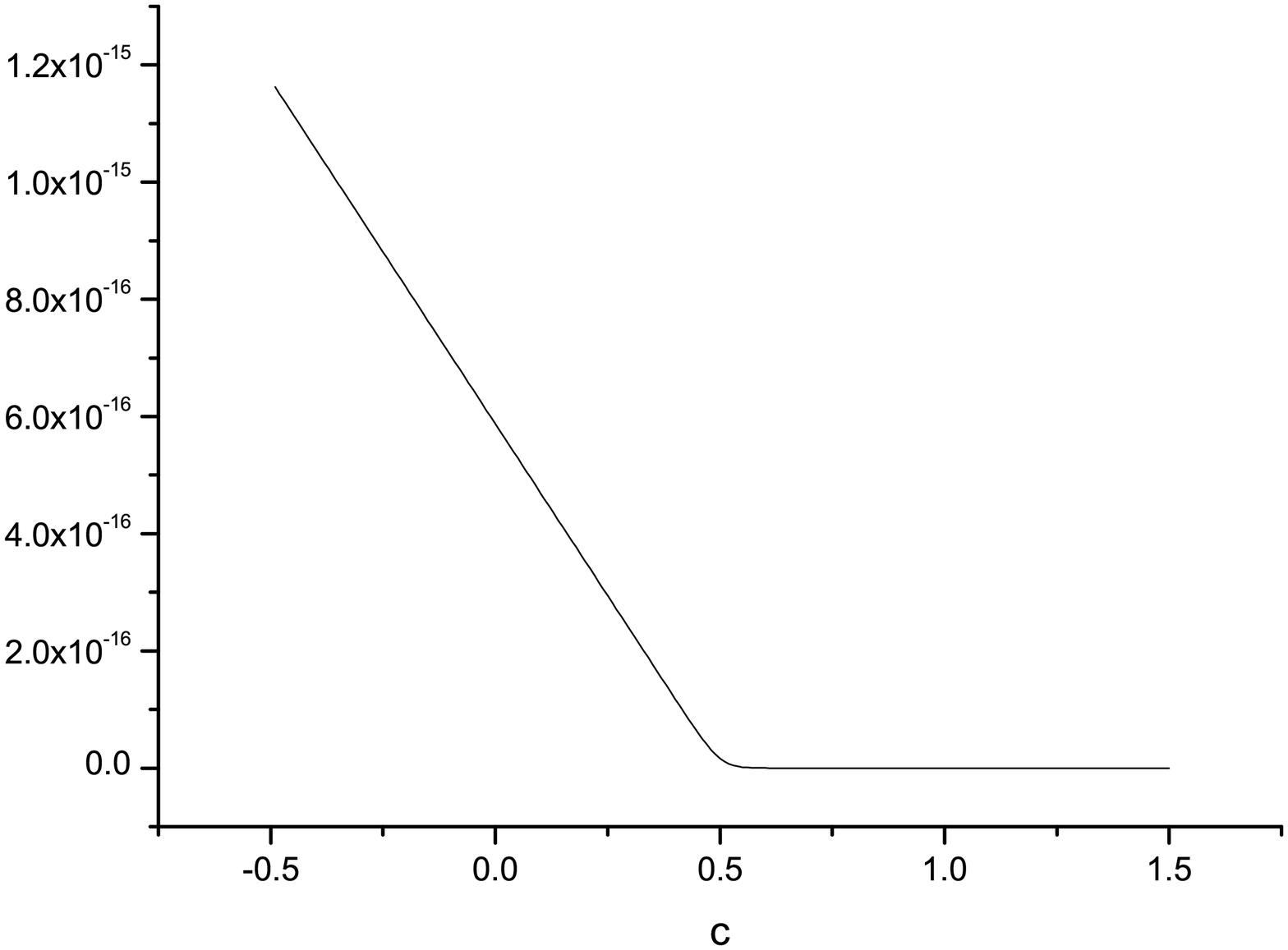}}
\subfigure[×Ó±êÌâ]{\includegraphics[width=0.45\textwidth]{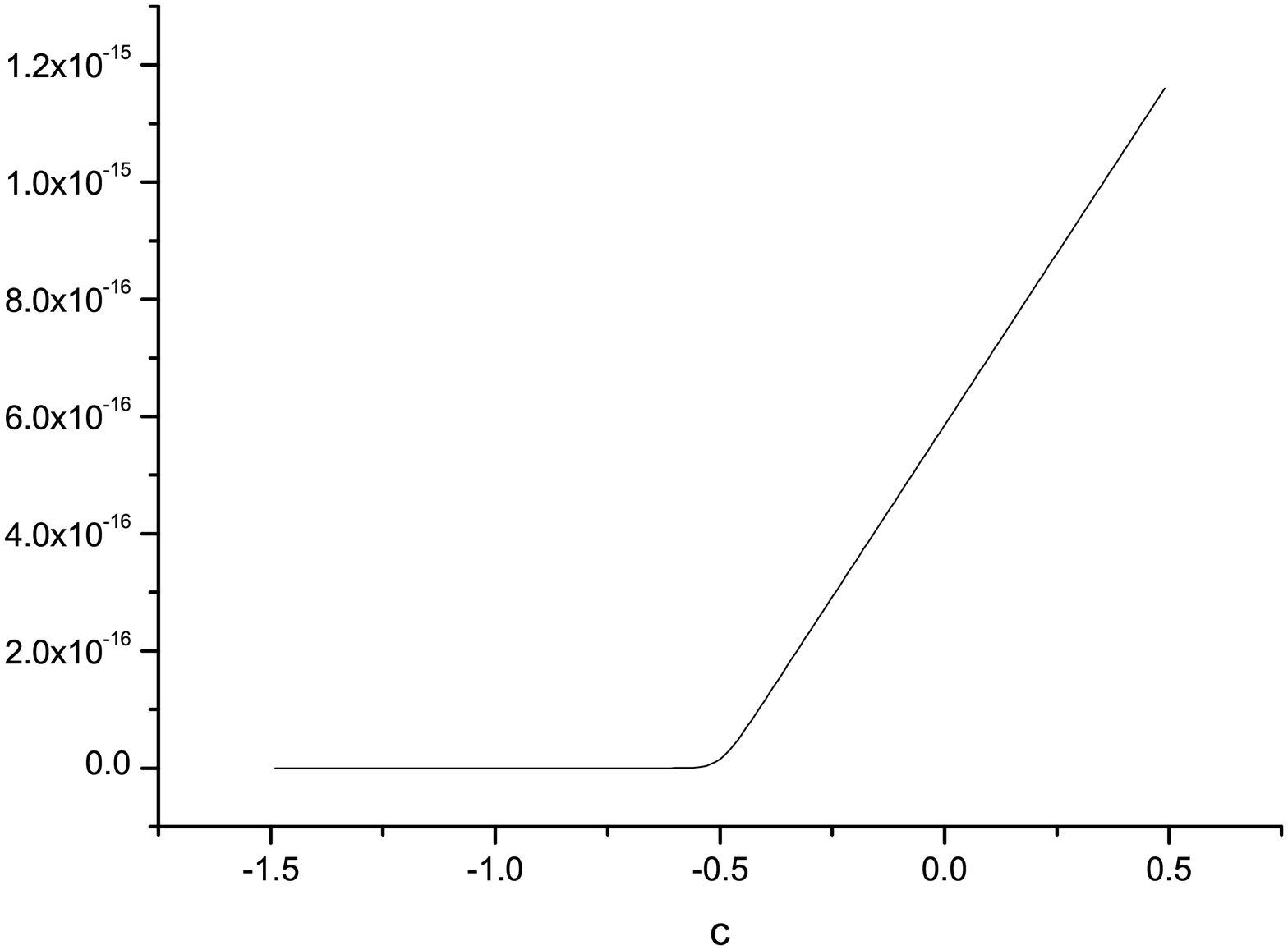}}
\caption{(a)$[f^{L,c}_{(++)}(0,1)][f^{L,c}_{(++)}(0,1)]$ varying with the bulk mass $c$,
 (b)$[f^{R,c}_{(++)}(0,1)][f^{R,c}_{(++)}(0,1)]$ varying with the bulk mass $c$}
\label{fLR}
\end{figure}

In the $t=1$ case, the relation between bulk mass $c$ and $[f^{L,c}_{(++)}(0,1)][f^{L,c}_{(++)}(0,1)]$ are drawn in Fig. \ref{fLR}(a),
we could  see that  the curve is similarly as in Fig. \ref{Graph1} and Fig. \ref{Graph4}.

%

Similarly, the curves in Fig. \ref{Graph3} and Fig. \ref{Graph6}  increase quickly when $c_S^1>0.5$, since the dominating corrections  to the branching
ratio of $t\rightarrow c\gamma$ and $t\rightarrow cg$ depend on the bulk masses $c_S^i
(i=1,2,3)$ in terms of
 $[f^{R,c_S^i}_{(++)}(0,t)][f^{R,c_S^j}_{(++)}(0,t)]$,  with $f^{R,c}_{(++)}(0,t)$ have the form:

\begin{eqnarray}
&&f^{R,c}_{(++)}(0,t)=f^{L,-c}_{(++)}(0,t)={t^{c}\over2}\sqrt{-2(1+2c)\epsilon\ln\epsilon\over\pi(1-\epsilon^{1+2c})}
\;.
\label{1}
\end{eqnarray}

In the $t=1$ case, the relation between bulk mass $c$ and $[f^{R,c}_{(++)}(0,1)][f^{R,c}_{(++)}(0,1)]$ are drawn in Fig. \ref{fLR}(b),
the curve increase quickly when $c_S^1>0.5$, which is similarly as in Fig. \ref{Graph3} and Fig. \ref{Graph6}.

\section{Conclusion\label{sec4}}
\indent\indent  In the SM, the rare FCNC top decay $t\rightarrow
c\gamma$ and $t\rightarrow cg$  are suppressed strongly
and\cite{ref8}:

\begin{eqnarray}
&&\rm{Br}(t\rightarrow c\gamma)\approx 4.6\times10^{-14},
\nonumber\\
&&\rm{Br}(t\rightarrow cg)\approx 4.6\times10^{-12},
\label{1}
\end{eqnarray}
which cannot be detected in near future experiment. Considering the
constraints from the precise electroweak observations, we
investigate the radiative corrections to $t\rightarrow c\gamma$ and
$t\rightarrow cg$ in warped extra dimensions with the custodial
symmetry $SU(3)_c\times SU(2)_L\times SU(2)_R\times U(1)_X\times
P_{LR}$. Since the fifth dimensional profiles depend on the bulk
masses $c_B^i,\;c_T^i,\;c_S^i\;(i=1,2,3)$ strongly, we mainly
analyze those bulk masses how affecting the corrections to
Br$(t\rightarrow c\gamma)$ and Br$(t\rightarrow cg)$ from new
physics. Numerical results indicate that

\begin{eqnarray}
&&\rm{Br}(t\rightarrow c\gamma)\sim10^{-6},
\nonumber\\
&&\rm{Br}(t\rightarrow cg)\sim10^{-5},
\label{1}
\end{eqnarray}
under our assumption on parameter space. With an integrated
luminosity of 100 fb$^{-1}$, the estimated precision of LHC to the
Br of $t \rightarrow c\gamma$ is $1.2\times10^{-5}$\cite{ref44} and
that of the  Br to $t \rightarrow cg$ is about $2.7 \times10^{-5}$
\cite{ref45}, respectively. In addition, the TESLA precision to the
Br of $t \rightarrow c\gamma$ can reach $3.6\times10^{-6}$
\cite{ref46} with a center of mass energy of 800GeV and an integrate
luminosity of 500 fb$^{-1}$. We can expert to detect those FCNC
processes in near future hopefully.

\begin{acknowledgments}
\indent\indent The work has been supported by the National Natural
Science Foundation of China (NNSFC) with Grant No. 10975027, No.
11275036, No. 11047002 and Natural Science Fund of Hebei University
with Grant No. 2011JQ05, No. 2012-242.
\end{acknowledgments}

\section{The couplings between bosons and quarks at the order ${\cal
O}(\upsilon^2/\Lambda_{KK}^2)$\label{app1}}

The relevant nontrivial couplings in Fig.(a1),(a2) with charged
$-1/3$ quarks could be approached to the order ${\cal
O}(\upsilon^2/\Lambda_{KK}^2)$ as:
\begin{eqnarray}
&&
{(\xi _{{W^ \pm }}^{L(-1/3)})_{b,t}} = (V_{CKM}^{(0)})_{bt}^\dag  + (V_{CKM}^{(0)\dag }\delta Z_L^{u})_{bt}  + (V_{CKM}^{(0)}\delta Z_L^d)_{bt}^\dag  - \frac{{{\upsilon ^2}}}{{4\Lambda _{KK}^2}}(\Delta _{{W^ \pm }}^L)_{bt}\nonumber\\
&&\hspace{2.3cm}
 + O(\frac{{{\upsilon ^3}}}{{\Lambda _{KK}^3}})\;,\nonumber\\
&&
{(\xi _{{W^ \pm }}^{R(-1/3)})_{b,t}} = \frac{{{\upsilon ^2}}}{{2\Lambda _{KK}^2}}(\Delta _{{W^ \pm }}^R)_{bt} + O(\frac{{{\upsilon ^3}}}{{\Lambda _{KK}^3}})\;,\nonumber\\
&&
{(\xi _{W_{{H_{(2n - 1)}}}^ \pm }^{L(-1/3)})_{b,t}} = \frac{{4\sqrt {2\pi } }}{{kr\epsilon}}\sum\limits_{i = 1}^3 {(D_L^{(0)})_{bi}^\dag } \int_{\epsilon}^1 d t\chi _{( +  + )}^{{W_L}}(y_{( +  + )}^{{W_L}(n)},t){[f_{( +  + )}^{L,c_B^i}(0,t)]^2}{(U_L^{(0)})_{it}}\nonumber\\
&&\hspace{2.3cm}
 + O(\frac{{{\upsilon ^2}}}{{\Lambda _{KK}^2}})\;,\;\;(\alpha  \ge 4)\;,\nonumber\\
&&
{(\xi _{W_{{H_{(2n - 1)}}}^ \pm }^{L(-1/3)})_{\alpha ,t}} = \frac{{4\sqrt {2\pi } }}{{kr\epsilon}}\sum\limits_{i = 1}^3 {\sum\limits_{{n^\prime } = 1}^\infty  {{\delta _{\alpha (9{n^\prime } - 6 + i)}}} } \int_{\epsilon}^1 d t\chi _{( +  + )}^{{W_L}}(y_{( +  + )}^{{W_L}(n)},t)\nonumber\\
&&\hspace{2.3cm}
 \times [f_{( +  + )}^{L,c_B^i}(y_{( \pm  \pm )}^{c_B^i({n^\prime })},t)][f_{( +  + )}^{L,c_B^i}(0,t)]{(U_L^{(0)})_{it}} + O(\frac{{{\upsilon ^2}}}{{\Lambda _{KK}^2}})\;,\;\;(\alpha  \ge 4)\;
\label{eq1}
\end{eqnarray}

with
\begin{eqnarray}
&& {(\Delta _{{W^ \pm }}^L)_{bt}} =  - \sum\limits_{i,j,k = 1}^3 ( D_L^{(0)})_{bi}^\dag f_{( +  + )}^{L,c_B^i}(0,1)\{ 2Y_{ik}^{d\dag }[\Sigma _{( \pm  \mp )}^{R,c_T^k}(1,1)]Y_{kj}^d\nonumber\\
&&\hspace{2cm}
 + Y_{ik}^{d\dag }[\Sigma _{( \mp  \mp )}^{R,c_T^k}(1,1)]Y_{kj}^d + Y_{ik}^{u\dag }[\Sigma _{( \mp  \mp )}^{R,c_S^k}(1,1)]Y_{kj}^u\} f_{( +  + )}^{L,c_B^j}(0,1){(U_L^{(0)})_{jt}}\nonumber\\
&&\hspace{2cm}
 - \frac{{8\pi {e^2}}}{{s_{\rm{W}}^2}}\sum\limits_{i = 1}^3 ( D_L^{(0)})_{ti}^\dag \{ \frac{1}{{kr\epsilon}}\int_{\epsilon}^1 d t{[f_{( +  + )}^{L,c_B^i}(0,t)]^2}[\Sigma _{( +  + )}^G(t,1)]{(U_L^{(0)})_{ib}}\} \;,\nonumber\\
&& {(\Delta _{{W^ \pm }}^R)_{bt}} = \sum\limits_{i,j,k = 1}^3 (
D_R^{(0)})_{bi}^\dag f_{( +  + )}^{R,c_S^i}(0,1)Y_{ik}^d[\Sigma _{(
\pm  \pm )}^{L,c_B^k}(1,1)]Y_{kj}^{d\dag }f_{( +  +
)}^{R,c_T^j}(0,1){(U_R^{(0)})_{jt}}\;. \label{eq1}
\end{eqnarray}
Similarly, the nontrivial couplings involving in Fig.(a1),(a2) with
charged $5/3$ quarks are approximated by

\begin{eqnarray}
&& {(\xi _{W_{{H_{(2n)}}}^ \pm }^{L(5/3)})_{\alpha ,t}} = \frac{{4\sqrt {2\pi } }}{{kr\epsilon}}\sum\limits_{i = 1}^3 {\sum\limits_{{n^\prime } = 1}^\infty  {{\delta _{\alpha (9(n' - 1) - 3 + i)}}} } \int_{\epsilon}^1 d t\chi _{( -  + )}^{{W_R}}(y_{( -  + )}^{{W_R}(n)},t)\nonumber\\
&&\hspace{2.0cm}
 \times [f_{( -  + )}^{L,c_B^i}(y_{( \mp  \pm )}^{c_B^i({n^\prime })},t)][f_{( +  + )}^{L,c_B^i}(0,t)]{(U_L^{(0)})_{it}} + O(\frac{{{\upsilon ^2}}}{{\Lambda _{KK}^2}})\;,\;\;(\alpha  \ge 4)\;,
\label{eq1}
\end{eqnarray}
The nontrivial couplings involving in Fig.(b1),(b2) with charged
$-1/3$ quarks are approximated by

\begin{eqnarray}
&&
{(\eta _{{G^ \pm }}^{L(-1/3)})_{b,t}} = \frac{e}{{\sqrt 2 {s_{\rm{w}}}}}\{ \frac{{{m_t}}}{{{m_{\rm{W}}}}}(V_{CKM}^{(0)})_{bt}^\dag  - \frac{{(\delta {M^u})_{33}^*}}{{{m_{\rm{W}}}}}(V_{CKM}^{(0)})_{bt}^\dag  - \frac{{\pi {m_t}{m_{\rm{W}}}}}{{\Lambda _{KK}^2}}[\{ \Sigma _{( +  + )}^G(1,1)\} \nonumber\\
&&\hspace{1.5cm}
 + \{ \Sigma _{( -  + )}^G(1,1)\} ](V_{CKM}^{(0)})_{bt}^\dag  + \sum\limits_{i = 1}^3 [ \frac{{{m_{{u_i}}}}}{{{m_{\rm{W}}}}}(\delta Z_R^uV_{CKM}^{(0)})_{bt}^\dag\nonumber\\
&&\hspace{1.5cm}
 + \frac{{{m_t}}}{{{m_{\rm{W}}}}}(V_{CKM}^{(0)}\delta Z_L^d)_{bt}^\dag ] + \frac{{{\upsilon ^2}}}{{4\Lambda _{KK}^2}}(\Delta _{{G^ \pm }}^L)_{bt}\}  + O(\frac{{{\upsilon ^3}}}{{\Lambda _{KK}^3}})\;\nonumber\\
&&
{(\eta _{{G^ \pm }}^{R(-1/3)})_{b,t}} = \frac{e}{{\sqrt 2 {s_{\rm{w}}}}}\{ (V_{CKM}^{(0)})_{bt}^\dag \frac{{{m_b}}}{{{m_{\rm{W}}}}} - \frac{{(\delta {M^d})_{33}^*}}{{{m_{\rm{W}}}}}(V_{CKM}^{(0)})_{bt}^\dag  - \frac{{\pi {m_t}{m_{\rm{W}}}}}{{\Lambda _{KK}^2}}[\{ \Sigma _{( +  + )}^G(1,1)\} \nonumber\\
&&\hspace{1.5cm}
 + \{ \Sigma _{( -  + )}^G(1,1)\} ](V_{CKM}^{(0)})_{bt}^\dag  + \sum\limits_{i = 1}^3 [ \frac{{{m_b}}}{{{m_{\rm{W}}}}}(\delta Z_R^uV_{CKM}^{(0)})_{bt}^\dag \nonumber\\
&&\hspace{1.5cm}
 + \frac{{{m_{{d_i}}}}}{{{m_{\rm{W}}}}}(V_{CKM}^{(0)}\delta Z_L^d)_{bt}^\dag ] + \frac{{{\upsilon ^2}}}{{4\Lambda _{KK}^2}}(\Delta _{{G^ \pm }}^R)_{bt} \}  + O(\frac{{{\upsilon ^3}}}{{\Lambda _{KK}^3}})\;\nonumber\\
&& {(\eta _{{G^ \pm }}^{L(-1/3)})_{\alpha ,t}} =\sum\limits_{i,j =
1}^3 {\sum\limits_{n = 1}^\infty  {{\delta _{\alpha ,(9n + i)}}} }
f_{( +  + )}^{R,c_T^i}(y_{( \mp  \mp )}^{c_T^i(n)},1)Y_{ij}^{d\dag
}f_{( +  + )}^{L,c_B^j}(0,1){(U_L^{(0)})_{j,t}}
 + O(\frac{{{\upsilon ^2}}}{{\Lambda _{KK}^2}}) \nonumber\\
&& {(\eta _{{G^ \pm }}^{R(-1/3)})_{\alpha ,t}} = \sum\limits_{i,j =
1}^3 {\sum\limits_{n = 1}^\infty    } {\delta _{\alpha ,(9n - 6 +
i)}}f_{( +  + )}^{L,c_B^i}(y_{( \pm  \pm )}^{c_B^i(n)},1)Y_{ij}^u
f_{( +  + )}^{R,c_S^j}(0,1){(U_R^{(0)})_{j,t}} + O(\frac{{{\upsilon
^2}}}{{\Lambda _{KK}^2}})\; \label{eq1}
\end{eqnarray}
with
\begin{eqnarray}
&&{(\Delta _{{G^ \pm }}^L)_{bt}} =  - \sum\limits_{i,j,k,l = 1}^3 {\frac{{{m_{{u_l}}}}}{{{m_{\rm{W}}}}}} (V_{CKM}^{(0)})_{bl}^\dag (U_R^{(0)})_{li}^\dag f_{( +  + )}^{R,c_S^i}(0,1)Y_{ik}^u[\{ \Sigma _{( \pm  \pm )}^{L,c_B^k}(1,1)\}  + \{ \Sigma _{( \mp  \pm )}^{L,c_B^k}(1,1)\} ]\nonumber\\
&&\hspace{1.5cm}
 \times Y_{kj}^{u\dag }f_{( +  + )}^{R,c_S^j}(0,1){(U_R^{(0)})_{jt}}\nonumber\\
&&\hspace{1.5cm}
 - \sum\limits_{i,j,k,l = 1}^3 {\frac{{{m_t}}}{{{m_{\rm{W}}}}}} (V_{CKM}^{(0)})_{bl}^\dag (U_L^{(0)})_{li}^\dag f_{( +  + )}^{L,c_B^i}(0,1)Y_{ik}^{d\dag }[\{ \Sigma _{( \pm  \mp )}^{R,c_T^k}(1,1)\}  + \{ \Sigma _{( \mp  \mp )}^{R,c_T^k}(1,1)\} ]\nonumber\\
&&\hspace{1.5cm}
 \times Y_{kj}^df_{( +  + )}^{L,c_B^j}(0,1){(U_L^{(0)})_{jt}},\nonumber\\
&&
{(\Delta _{{G^ \pm }}^R)_{bt}} =  - \sum\limits_{i,j,k,l = 1}^3 {\frac{{{m_{{d_i}}}}}{{{m_{\rm{W}}}}}} (D_R^{(0)})_{bk}^\dag f_{( +  + )}^{R,c_T^k}(0,1)Y_{kl}^d[\Sigma _{( \pm  \pm )}^{L,c_B^l}(1,1)]\nonumber\\
&&\hspace{1.5cm}
 \times Y_{lj}^{d\dag }f_{( +  + )}^{R,c_S^j}(0,1){(D_R^{(0)})_{ji}}(V_{CKM}^{(0)})_{it}^\dag \nonumber\\
&&\hspace{1.5cm}
 - \sum\limits_{i,j,k,l = 1}^3 {\frac{{{m_b}}}{{{m_{\rm{W}}}}}} (D_L^{(0)})_{bk}^\dag f_{( +  + )}^{L,c_B^k}(0,1)[Y_{kl}^{d\dag }\{ \Sigma _{( \pm  \mp )}^{R,c_T^l}(1,1)\} Y_{lj}^d\nonumber\\
&&\hspace{1.5cm}
 + Y_{kl}^{u\dag }\{ \Sigma _{( \mp  \mp )}^{R,c_T^l}(1,1)\} Y_{lj}^u]f_{( +  + )}^{L,c_B^j}(0,1){(D_L^{(0)})_{ji}}\;(V_{CKM}^{(0)})_{it}^\dag .
\label{eq1}
\end{eqnarray}

The nontrivial couplings involving in Fig.(b1),(b2) with charged
$5/3$ quarks are approximated by

\begin{eqnarray}
&& {(\eta _{{G^ \pm }}^{L(5/3)})_{\alpha ,t}} = \sum\limits_{i,j =
1}^3 {\sum\limits_{n = 1}^\infty  \{  } {\delta _{\alpha ,(9n - 3 +
i)}}f_{( -  + )}^{R,c_T^i}(y_{( \pm  \mp
)}^{c_T^i(n)},1)Y_{ij}^{d\dag }
 + {\delta _{\alpha ,(9n - 6 + i)}} f_{( -  + )}^{R,c_T^i}(y_{( \pm  \mp )}^{c_T^i(n)},1)Y_{ij}^{d\dag }\}\nonumber\\
&&\hspace{1.7cm}\times f_{( +  + )}^{L,c_B^j}(0,1){(U_L^{(0)})_{j,t}} + O(\frac{{{\upsilon ^2}}}{{\Lambda _{KK}^2}})\;\nonumber\\
&& {(\eta _{{G^ \pm }}^{R(5/3)})_{\alpha ,t}} =  - \sum\limits_{i,j
= 1}^3 {\sum\limits_{n = 1}^\infty  {{\delta _{\alpha ,(9(n - 1) +
i)}}} } f_{( -  + )}^{L,c_B^i}(y_{( \mp  \pm
)}^{c_B^i(n)},1)Y_{ij}^uf_{( +  +
)}^{R,c_S^j}(0,1){(U_R^{(0)})_{j,t}}
 + O(\frac{{{\upsilon ^2}}}{{\Lambda _{KK}^2}})\nonumber\\
\label{eq1}
\end{eqnarray}

As the FCNC transitions are mediated by the massive neutral gauge
bosons $Z,\; Z_{_{H_\alpha}},\;\gamma_{(n)}$ in Fig.(c), relevant
couplings are expanded according $\upsilon^2/\Lambda_{KK}^2$ as

\begin{eqnarray}
&&{(\xi _Z^{L(2/3)})_{c,t}} = (\frac{{3 -
4s_{\rm{w}}^2}}{{6{s_{\rm{w}}}{{\rm{c}}_{\rm{w}}}}})\Big[{\delta
_{ct}} + (\delta Z_L^u)_{ct}^\dag  + {(\delta Z_L^u)_{ct}} +
\frac{{{\upsilon ^2}}}{{2\Lambda _{KK}^2}}{(\Delta
_Z^{L2/3})_{ct}}\Big] + O(\frac{{{\upsilon ^3}}}{{\Lambda
_{KK}^3}})\;,\nonumber\\
&& {(\xi _Z^{R(2/3)})_{c,t}} =  -
\frac{2}{3}\frac{{{s_{\rm{w}}}}}{{{{\rm{c}}_{\rm{w}}}}}\Big[{\delta
_{ct}} + (\delta Z_R^u)_{ct}^\dag  + {(\delta Z_R^u)_{ct}} +
\frac{{{\upsilon ^2}}}{{2\Lambda _{KK}^2}}{(\Delta
_Z^{R2/3})_{ct}}\Big] + O(\frac{{{\upsilon ^3}}}{{\Lambda
_{KK}^3}})\;,\nonumber\\
&&{(\xi _{{Z_{{H_{(2n - 1)}}}}}^{L(2/3)})_{c,t}} = (\frac{{3 -
4s_{\rm{w}}^2}}{{6{s_{\rm{w}}}{{\rm{c}}_{\rm{w}}}}})\frac{{4\sqrt
{2\pi } }}{{kr\epsilon}}\sum\limits_{i = 1}^3 ( U_L^{(0)})_{ci}^\dag
\int_{\epsilon}^1 d t\chi _{( +  + )}^Z(y_{( +  +
)}^{Z(n)},t)\nonumber\\
&&\hspace{2.5cm}\times {[f_{( +  + )}^{L,c_B^i}(0,t)]^2}{(U_L^{(0)})_{it}} + O(\frac{\upsilon }{{{\Lambda _{KK}}}})\;,\nonumber\\
&&{(\xi _{{Z_{{H_{(2n)}}}}}^{L(2/3)})_{c,t}} =  - \frac{{3 -
2s_{\rm{w}}^2}}{{6{s_{\rm{w}}}{{\rm{c}}_{\rm{w}}}\sqrt {1 -
2s_{\rm{w}}^2} }}
\frac{{4\sqrt {2\pi } }}{{kr\epsilon}}\sum\limits_{i = 1}^3 ( U_L^{(0)})_{ci}^\dag \int_{\epsilon}^1 d t\chi _{( -  + )}^{{Z_X}}(y_{( -  + )}^{{Z_X}(n)},t)\nonumber\\
&& \hspace{2.5cm}\times {[f_{( +  + )}^{L,c_B^i}(0,t)]^2}{(U_L^{(0)})_{it}} + O(\frac{\upsilon }{{{\Lambda _{KK}}}})\;,\nonumber\\
&&{(\xi _{{Z_{{H_{(2n - 1)}}}}}^{R(2/3)})_{c,t}} =  -
\frac{2}{3}\frac{{{s_{\rm{w}}}}}{{{{\rm{c}}_{\rm{w}}}}}
\frac{{4\sqrt {2\pi } }}{{kr\epsilon}}\sum\limits_{i = 1}^3 ( U_R^{(0)})_{ci}^\dag \int_{\epsilon}^1 d t\chi _{( +  + )}^Z(y_{( +  + )}^{Z(n)},t)\nonumber\\
&&\hspace{2.5cm} \times {[f_{( +  +
)}^{R,c_S^i}(0,t)]^2}{(U_R^{(0)})_{it}} +
O(\frac{\upsilon }{{{\Lambda _{KK}}}})\;,\nonumber\\
&&{(\xi _{{Z_{{H_{(2n)}}}}}^{R(2/3)})_{c,t}} =  -
\frac{{2{s_{\rm{w}}}}}{{3{{\rm{c}}_{\rm{w}}}\sqrt {1 -
2s_{\rm{w}}^2} }}
\frac{{4\sqrt {2\pi } }}{{kr\epsilon}}\sum\limits_{i = 1}^3 ( U_R^{(0)})_{ci}^\dag \int_{\epsilon}^1 d t\chi _{( -  + )}^{{Z_X}}(y_{( -  + )}^{{Z_X}(n)},t)\nonumber\\
&& \hspace{2.5cm}\times {[f_{( +  + )}^{R,c_S^i}(0,t)]^2}{(U_R^{(0)})_{it}} + O(\frac{\upsilon }{{{\Lambda _{KK}}}})\;,\nonumber\\
&&{(\xi _{{Z_{{H_{(2n - 1)}}}}}^{L(2/3)})_{\alpha ,t}} = (\frac{{3 -
4s_{\rm{w}}^2}}{{6{s_{\rm{w}}}{{\rm{c}}_{\rm{w}}}}}) \frac{{4\sqrt
{2\pi } }}{{kr\epsilon}}\sum\limits_{i = 1}^3
{\sum\limits_{{n^\prime } = 1}^\infty  {{\delta _{\alpha (15n' - 12
+ i)}}} }
\int_{\epsilon}^1 d t\chi _{( +  + )}^Z(y_{( +  + )}^{Z(n)},t)\nonumber\\
&&\hspace{2.5cm} \times [f_{( +  + )}^{L,c_B^i}(0,t)][f_{( +  +
)}^{L,c_B^i}(y_{( \pm  \pm )}^{c_B^i({n^\prime
})},t)]{(U_L^{(0)})_{it}}
+ O(\frac{\upsilon }{{{\Lambda _{KK}}}})\;,\nonumber\\
&&{(\xi _{{Z_{{H_{(2n)}}}}}^{L(2/3)})_{\alpha ,t}} =  - \frac{{3 -
2s_{\rm{w}}^2}}{{6{s_{\rm{w}}}{{\rm{c}}_{\rm{w}}}\sqrt {1 -
2s_{\rm{w}}^2} }} \frac{{4\sqrt {2\pi }
}}{{kr\epsilon}}\sum\limits_{i = 1}^3 {\sum\limits_{{n^\prime } =
1}^\infty  {{\delta _{\alpha (15n' - 12 + i)}}} }
\int_{\epsilon}^1 d t\chi _{( -  + )}^{{Z_X}}(y_{( -  + )}^{{Z_X}(n)},t)\nonumber\\
&& \hspace{2.5cm}\times [f_{( +  + )}^{L,c_B^i}(0,t)][f_{( +  +
)}^{L,c_B^i}(y_{( \pm  \pm )}^{c_B^i({n^\prime
})},t)]{(U_L^{(0)})_{it}}
+ O(\frac{\upsilon }{{{\Lambda _{KK}}}})\;,\nonumber\\
&&{(\xi _{{Z_{{H_{(2n - 1)}}}}}^{R(2/3)})_{\alpha ,t}} =  -
\frac{2}{3}\frac{{{s_{\rm{w}}}}}{{{{\rm{c}}_{\rm{w}}}}}\frac{{4\sqrt
{2\pi } }} {{kr\epsilon}}\sum\limits_{i = 1}^3
{\sum\limits_{{n^\prime } = 1}^\infty  {{\delta _{\alpha (15n' +
i)}}} }
\int_{\epsilon}^1 d t\chi _{( +  + )}^Z(y_{( +  + )}^{Z(n)},t)\nonumber\\
&& \hspace{2.5cm}\times [f_{( +  + )}^{R,c_S^i}(0,t)][f_{( +  +
)}^{R,c_S^i}(y_{( \pm  \pm )}^{c_S^i({n^\prime
})},t)]{(U_R^{(0)})_{it}}
+ O(\frac{\upsilon }{{{\Lambda _{KK}}}})\;,\nonumber\\
&&{(\xi _{{Z_{{H_{(2n)}}}}}^{R(2/3)})_{\alpha ,t}} =  -
\frac{{2{s_{\rm{w}}}}}{{3{c_{\rm{w}}}\sqrt {1 - 2s_{\rm{w}}^2} }}
\frac{{4\sqrt {2\pi } }}{{kr\epsilon}}\sum\limits_{i = 1}^3
{\sum\limits_{{n^\prime } = 1}^\infty  {{\delta _{\alpha (15n' +
i)}}} }
\int_{\epsilon}^1 d t\chi _{( -  + )}^{{Z_X}}(y_{( -  + )}^{{Z_X}(n)},t)\nonumber\\
&& \hspace{2.5cm}\times [f_{( +  + )}^{R,c_S^i}(0,t)][f_{( +  +
)}^{R,c_S^i}(y_{( \pm  \pm )}^{c_S^i({n^\prime
})},t)]{(U_R^{(0)})_{it}} + O(\frac{\upsilon }{{{\Lambda
_{KK}}}})\;,\nonumber \label{eq1}
\end{eqnarray}

\begin{eqnarray}
&& {(\xi _{{\gamma _{(n)}}}^{L(2/3)})_{c,t}} =
\frac{2}{3}\frac{{4\sqrt {2\pi } }} {{kr\epsilon}}\sum\limits_{i =
1}^3 ( U_L^{(0)})_{c,i}^\dag \int_{\epsilon}^1 d t\chi _{( +  +
)}^A(y_{( +  + )}^{A(n)},t)
{[f_{( +  + )}^{L,c_B^i}(0,t)]^2}{(U_L^{(0)})_{i,t}}\nonumber\\
&& \hspace{2.5cm}
 + O(\frac{\upsilon }{{{\Lambda _{KK}}}})\;,\nonumber\\
&& {(\xi _{{\gamma _{(n)}}}^{R(2/3)})_{c,t}} =
\frac{2}{3}\frac{{4\sqrt {2\pi } }} {{kr\epsilon}}\sum\limits_{i =
1}^3 ( U_R^{(0)})_{c,i}^\dag \int_{\epsilon}^1 d t\chi _{( +  +
)}^A(y_{( +  + )}^{A(n)},t) {[f_{( +  +
)}^{R,c_S^i}(0,t)]^2}{(U_R^{(0)})_{i,t}}
 + O(\frac{\upsilon }{{{\Lambda _{KK}}}})\;,\nonumber\\
&&  {(\xi _{{\gamma _{(n)}}}^{L(2/3)})_{\alpha ,t}} =
\frac{2}{3}\frac{{4\sqrt {2\pi } }} {{kr\epsilon}}\sum\limits_{i =
1}^3 {\sum\limits_{{n^\prime } = 1}^\infty  {{\delta _{\alpha ,(15n'
- 12 + i)}}} }
\int_{\epsilon}^1 d t\chi _{( +  + )}^A(y_{( +  + )}^{A(n)},t)\{ f_{( +  + )}^{L,c_B^i}(0,t)\nonumber\\
&& \hspace{2.5cm}
 \times f_{( +  + )}^{L,c_B^i}(y_{( \pm  \pm )}^{c_B^i({n^\prime })},t)\} {(U_L^{(0)})_{i,t}} + O(\frac{\upsilon }{{{\Lambda _{KK}}}})\;,\nonumber\\
&& {(\xi _{{\gamma _{(n)}}}^{R(2/3)})_{\alpha ,t}} =
\frac{2}{3}\frac{{4\sqrt {2\pi } }} {{kr\epsilon}}\sum\limits_{i =
1}^3 {\sum\limits_{{n^\prime } = 1}^\infty  {{\delta _{\alpha ,(15n'
+ i)}}} }
 \int_{\epsilon}^1 d t\chi _{( +  + )}^A(y_{( +  + )}^{A(n)},t)\{ f_{( +  + )}^{R,c_S^i}(0,t)\nonumber\\
&& \hspace{2.5cm}
 \times f_{( +  + )}^{R,c_S^i}(y_{( \mp  \mp )}^{c_S^i({n^\prime })},t)\} {(U_R^{(0)})_{i,b}} + O(\frac{\upsilon }{{{\Lambda _{KK}}}})\;.
\label{eq1}
\end{eqnarray}

Here the short-cut notations $\Delta_{Z}^L,\;\Delta_{Z}^R$ are
defined by

\begin{eqnarray}
&&{(\Delta _Z^{L2/3})_{ct}} = \sum\limits_{i,j,k = 1}^3 (
U_L^{(0)})_{ci}^\dag f_{( +  + )}^{L,c_B^i}(0,1)\{
Y_{ik}^d(\frac{6}{{4s_{\rm{w}}^2 - 3}}
[\Sigma _{( \pm  \mp )}^{R,c_T^k}(1,1)])Y_{kj}^{d\dag } \nonumber\\
&& \hspace{2cm}+ Y_{ik}^u(\frac{3}{{4s_{\rm{w}}^2 - 3}}[\Sigma _{(
\mp  \mp )}^{R,c_S^k}(1,1)])Y_{kj}^{u\dag }\} \times f_{( +  +
)}^{L,c_B^j}(0,1){(U_L^{(0)})_{jt}} \nonumber\\
&& \hspace{2cm}- \frac{{4\pi
{e^2}}}{{s_{\rm{w}}^2c_{\rm{w}}^2kr\epsilon}}\sum\limits_{i = 1}^3
( U_L^{(0)})_{ci}^\dag (\int_{\epsilon}^1 d t\{ [\Sigma _{( +  + )}^G(t,1)]\nonumber\\
&&\hspace{2cm} + \frac{{2s_{\rm{w}}^2 - 3}}{{3 - 4s_{\rm{w}}^2}}[\Sigma _{( -  + )}^G(t,1)]\} {[f_{( +  + )}^{L,c_B^i}(0,t)]^2}){(U_L^{(0)})_{it}}\;,\nonumber\\
&& {(\Delta _Z^{R2/3})_{ct}} =
\frac{3}{{4s_{\rm{w}}^2}}\sum\limits_{i,j,k = 1}^3 (
U_R^{(0)})_{ci}^\dag f_{( +  + )}^{R,c_S^i}(0,1)Y_{ik}^{u\dag }\{
 [\Sigma _{( \mp  \pm )}^{L,c_B^k}(1,1)]\nonumber\\
&& \hspace{2cm} - [\Sigma _{( \pm  \pm )}^{L,c_B^k}(1,1)]\} Y_{kj}^uf_{( +  + )}^{R,c_S^j}(0,1){(U_R^{(0)})_{jt}}\nonumber\\
&&\hspace{2cm} - \frac{{4\pi
{e^2}}}{{s_{\rm{w}}^2c_{\rm{w}}^2kr\epsilon}}\sum\limits_{i,j = 1}^3
( U_R^{(0)})_{ci}^\dag (\int_{\epsilon}^1 d t\{
[\Sigma _{( +  + )}^G(t,1)] + {s_{\rm{w}}}[\Sigma _{( -  + )}^G(t,1)]\} \nonumber\\
&&\hspace{2cm} \times {[f_{( +  +
)}^{L,c_S^i}(0,t)]^2}){(U_R^{(0)})_{it}}\;\label{eq1}
\end{eqnarray}

In Fig.(d1) and (d2), the KK exciting modes of gluon also mediate
the FCNC transitions and the relevant couplings are approximated by

\begin{eqnarray}
&& {(\xi _{{g_{(n)}}}^{L(2/3)})_{c,t}} = \frac{{4\sqrt {2\pi }
}}{{kr\epsilon}}\sum\limits_{i = 1}^3 ( U_L^{(0)})_{c,i}^\dag
\int_{\epsilon}^1 d t\chi _{( +  + )}^g(y_{( +  + )}^{g(n)},t){[f_{(
+  + )}^{L,c_B^i}(0,t)]^2}{(U_L^{(0)})_{i,t}}
 + O(\frac{\upsilon }{{{\Lambda _{KK}}}})\;,\nonumber\\
&& {(\xi _{{g_{(n)}}}^{R(2/3)})_{c,t}} = \frac{{4\sqrt {2\pi }
}}{{kr\epsilon}}\sum\limits_{i = 1}^3 ( U_R^{(0)})_{c,i}^\dag
\int_{\epsilon}^1 d t\chi _{( +  + )}^g(y_{( +  + )}^{g(n)},t){[f_{(
+  + )}^{R,c_S^i}(0,t)]^2}{(U_R^{(0)})_{i,t}}
 + O(\frac{\upsilon }{{{\Lambda _{KK}}}})\;,\nonumber\\
&&{(\xi _{{g_{(n)}}}^{L(2/3)})_{\alpha ,t}} = \frac{{4\sqrt {2\pi }
}}{{kr\epsilon}}\sum\limits_{i = 1}^3 {\sum\limits_{{n^\prime } =
1}^\infty  {{\delta _{\alpha ,(15n' - 12 + i)}}} }
 \int_{\epsilon}^1 d t\chi _{( +  + )}^g(y_{( +  + )}^{g(n)},t)\{ f_{( +  + )}^{L,c_B^i}(0,t)\nonumber\\
&&\hspace{2cm}
 \times f_{( +  + )}^{L,c_B^i}(y_{( \pm  \pm )}^{c_B^i({n^\prime })},t)\} {(U_L^{(0)})_{i,t}} + O(\frac{\upsilon }{{{\Lambda _{KK}}}})\;,\nonumber\\
&& {(\xi _{{g_{(n)}}}^{R(2/3)})_{\alpha ,t}} = \frac{{4\sqrt {2\pi }
}}{{kr\epsilon}}\sum\limits_{i = 1}^3 {\sum\limits_{{n^\prime } =
1}^\infty  {{\delta _{\alpha ,(15n' + i)}}} } \int_{\epsilon}^1 d
t\chi _{( +  + )}^g(y_{( +  + )}^{g(n)},t)\{ f_{( +  +
)}^{R,c_S^i}(0,t)
 \nonumber\\
&&\hspace{2cm}\times f_{( +  + )}^{R,c_S^i}(y_{( \mp  \mp
)}^{c_S^i({n^\prime })},t)\} {(U_R^{(0)})_{i,b}} + O(\frac{\upsilon
}{{{\Lambda _{KK}}}})\;.\label{eq1}
\end{eqnarray}

Finally, In Fig.(e), the relevant FCNC couplings mediated by neutral
Higgs and Goldstone are

\begin{eqnarray}
&&
{(\eta _{{H_0}}^{L(2/3)})_{c,t}} =  - \frac{e}{{\sqrt 2 {s_{\rm{w}}}}}\{ \frac{{{m_t}}}{{{m_{\rm{W}}}}}{\delta _{ct}} - \frac{{{{(\delta {M^u})}_{33}}}}{{{m_{\rm{W}}}}}{\delta _{ct}} + \frac{{{m_t}}}{{{m_{\rm{W}}}}}(\delta Z_R^u)_{ct}^\dag  + \frac{{{m_c}}}{{{m_{\rm{W}}}}}{(\delta Z_L^u)_{ct}}\nonumber\\
&&\hspace{2cm}
 - {\delta _{ct}}\frac{{\pi {m_t}{m_{\rm{W}}}}}{{\Lambda _{KK}^2}}[\{ \Sigma _{( +  + )}^G(1,1)\}  + \{ \Sigma _{( -  + )}^G(1,1)\} ]\nonumber\\
&&\hspace{2cm}
 + \frac{{{\upsilon ^2}}}{{4\Lambda _{KK}^2}}[\frac{{{m_t}}}{{{m_{\rm{W}}}}}{(\Delta _{{H_0}}^{(1)2/3})_{ct}} + \frac{{{m_c}}}{{{m_{\rm{W}}}}}{(\Delta _{{H_0}}^{(2)2/3})_{ct}}]\}  + O(\frac{{{\upsilon ^3}}}{{\Lambda _{KK}^3}})\;,\nonumber\\
&&
{(\eta _{{H_0}}^{R(2/3)})_{c,t}} = (\eta _{{H_0}}^{L(2/3)})_{c,t}^\dag \;,\nonumber\\
&&
{(\eta _{{H_0}}^{L(2/3)})_{\alpha ,t}} = \sum\limits_{i,j = 1}^3 {\sum\limits_{n = 1}^\infty  \{  }  - {\delta _{\alpha ,(15n + i)}}f_{( +  + )}^{R,c_S^i}(y_{( \mp  \mp )}^{c_S^i(n)},1)Y_{ij}^{u\dag }f_{( +  + )}^{L,c_B^j}(0,1){(U_L^{(0)})_{j,t}}\nonumber\\
&&\hspace{2cm}
 + \frac{1}{{\sqrt 2 }}{\delta _{\alpha ,(15n - 9 + i)}}f_{( -  + )}^{R,c_T^i}(y_{( \pm  \mp )}^{c_T^i(n)},1)Y_{ij}^{d\dag }f_{( +  + )}^{L,c_B^j}(0,1){(U_L^{(0)})_{j,t}}\nonumber\\
&&\hspace{2cm}
 - \frac{1}{{\sqrt 2 }}{\delta _{\alpha ,(15n - 6 + i)}}f_{( -  + )}^{R,c_T^i}(y_{( \pm  \mp )}^{c_T^i(n)},1)Y_{ij}^{d\dag }f_{( +  + )}^{L,c_B^j}(0,1){(U_L^{(0)})_{j,t}}\}  + O(\frac{\upsilon }{{{\Lambda _{KK}}}})\;\nonumber\\
&&
{(\eta _{{H_0}}^{R(2/3)})_{\alpha ,t}} = \sum\limits_{i,j = 1}^3 {\sum\limits_{n = 1}^\infty  {\{  - {\delta _{\alpha ,(15n - 12 + i)}}} } f_{( +  + )}^{L,c_B^i}(y_{( \pm  \pm )}^{c_B^i(n)},1)Y_{ij}^uf_{( +  + )}^{R,c_S^j}(0,1){(U_R^{(0)})_{j,t}}\nonumber\\
&&\hspace{2cm}
 + \sum\limits_{n = 1}^\infty  {{\delta _{\alpha ,(15n - 3 + i)}}f_{( -  + )}^{L,c_B^i}(y_{( \mp  \pm )}^{c_B^i(n)},1)Y_{ij}^uf_{( +  + )}^{R,c_S^j}(0,1){{(U_R^{(0)})}_{j,t}}} \}  + O(\frac{\upsilon }{{{\Lambda _{KK}}}})\;,\nonumber\\
&&
{(\eta _{{G_0}}^{L(2/3)})_{c,t}} =  {(\eta _{{H_0}}^{L(2/3)})_{c,t}}\;,\nonumber\\
&&
{(\eta _{{G_0}}^{R(2/3)})_{c,t}} =  (\eta _{{G_0}}^{L(2/3)})_{c,t}^\dag \;,\nonumber\\
&&
{(\eta _{{G_0}}^{L(2/3)})_{\alpha ,t}} = \sum\limits_{i,j = 1}^3 \{  \sum\limits_{n = 1}^\infty  (  - {\delta _{\alpha ,(15n + i)}}f_{( +  + )}^{R,c_S^i}(y_{( \mp  \mp )}^{c_S^i(n)},1)Y_{ij}^{u\dag }f_{( +  + )}^{L,c_B^j}(0,1){(U_L^{(0)})_{j,t}}\nonumber\\
&&\hspace{2cm}
 + \sum\limits_{n = 1}^\infty  ( \frac{1}{{\sqrt 2 }}{\delta _{\alpha ,(15n - 9 + i)}}f_{( -  + )}^{R,c_T^i}(y_{( \pm  \mp )}^{c_T^i(n)},1)Y_{ij}^{d\dag }f_{( +  + )}^{L,c_B^j}(0,1){(U_L^{(0)})_{j,t}}\nonumber\\
&&\hspace{2cm}
 -\sum\limits_{n = 1}^\infty  ( \frac{1}{{\sqrt 2 }}{\delta _{\alpha ,(15n - 6 + i)}}f_{( -  + )}^{R,c_T^i}(y_{( \pm  \mp )}^{c_T^i(n)},1)Y_{ij}^{d\dag }f_{( +  + )}^{L,c_B^j}(0,1){(U_L^{(0)})_{j,t}}\}  + O(\frac{\upsilon }{{{\Lambda _{KK}}}})\;,\nonumber\\
&&
{(\eta _{{G_0}}^{R(2/3)})_{\alpha ,t}} = \sum\limits_{i,j = 1}^3 {\sum\limits_{n = 1}^\infty  {\{  - {\delta _{\alpha ,(15n - 3 + i)}}} } f_{( -  + )}^{L,c_B^i}(y_{( \mp  \pm )}^{c_B^i(n)},1)Y_{ij}^uf_{( +  + )}^{R,c_S^j}(0,1){(U_R^{(0)})_{j,t}}\nonumber\\
&&\hspace{2cm}
 -{\delta _{\alpha ,(15n - 12 + i)}}f_{( +  + )}^{L,c_B^i}(y_{( \pm  \pm )}^{c_B^i(n)},1)Y_{ij}^uf_{( +  + )}^{R,c_S^j}(0,1){(U_R^{(0)})_{j,t}}\}  + O(\frac{\upsilon }{{{\Lambda
 _{KK}}}})\;.
\label{eq1}
\end{eqnarray}

where the abbreviations are given by

\begin{eqnarray}
&&
{(\Delta _{{H_0}}^{(1)2/3})_{ct}} =  - \sum\limits_{i,j,k,l = 1}^3 ( U_R^{(0)})_{ci}^\dag f_{( +  + )}^{R,c_S^i}(0,1)Y_{ik}^{u\dag }[\Sigma _{( \pm  \pm )}^{L,c_B^k}(1,1)]Y_{kj}^uf_{( +  + )}^{R,c_S^j}(0,1){(U_L^{(0)})_{jt}}\;\nonumber\\
&&\hspace{2cm}
 -\sum\limits_{i,j,k,l = 1}^3 ( U_R^{(0)})_{ci}^\dag f_{( +  + )}^{R,c_S^i}(0,1)Y_{ik}^{u\dag }[\Sigma _{( \mp  \pm )}^{L,c_B^k}(1,1)]Y_{kj}^uf_{( +  + )}^{R,c_S^j}(0,1){(U_L^{(0)})_{jt}}\;\nonumber\\
&&
{(\Delta _{{H_0}}^{(2)2/3})_{ct}} =  - \sum\limits_{i,j,k,l = 1}^3 ( U_R^{(0)})_{ci}^\dag f_{( +  + )}^{L,c_B^i}(0,1)\{ Y_{ik}^d[2\{ \Sigma _{( \pm  \mp )}^{R,c_T^k}(1,1)\} ]Y_{kj}^{d\dag } + Y_{ik}^u[\{ \Sigma _{( \mp  \mp )}^{R,c_S^k}(1,1)\} ]Y_{kj}^{u\dag }\} \nonumber\\
&&\hspace{2cm}
 \times f_{( +  + )}^{L,c_B^j}(0,1){(U_L^{(0)})_{jt}}\;. \label{eq1}
\end{eqnarray}

\end{document}